\documentclass[10pt,final,journal,letterpaper,oneside,twocolumn]{IEEEtran}
\usepackage[english]{babel}
\usepackage[utf8]{inputenc}
\usepackage[T1]{fontenc}
\usepackage{amsmath,amssymb,amsfonts,mathtools,bm}
\usepackage{graphicx}
\usepackage[table]{xcolor}
\usepackage{tabularx,array,booktabs,multirow,makecell}
\usepackage{ragged2e}
\usepackage{caption,subcaption}
\usepackage{cite}
\usepackage{pifont}
\usepackage{placeins}
\usepackage{stfloats}
\usepackage{enumitem}
\usepackage{tikz}
 
\usetikzlibrary{arrows.meta,calc,positioning,fit,shapes.geometric,decorations.pathreplacing,chains,backgrounds}
\usepackage[most]{tcolorbox}
\usepackage[hidelinks]{hyperref}
\graphicspath{{figs/}}

\definecolor{SIMHeader}{RGB}{210,225,245}
\definecolor{SIMFirstCol}{RGB}{226,241,225}
\definecolor{SIMBody}{RGB}{252,246,232}
\definecolor{SIMLine}{RGB}{150,150,150}
\definecolor{pastelgray}{RGB}{235,235,235}
\definecolor{pastelpurple}{RGB}{240,230,250}
\definecolor{pastelblue}{RGB}{220,240,255}
\definecolor{lessonbg}{RGB}{245,248,240}
\definecolor{lessonframe}{RGB}{90,130,70}
\definecolor{phbg}{RGB}{250,250,250}
\definecolor{phframe}{RGB}{180,60,60}
\definecolor{atomfill}{RGB}{200,220,240}
\definecolor{lightgreen}{RGB}{210,235,215}

\newcommand{\cmark}{\ding{51}}
\newcommand{\xmark}{\ding{55}}
\newcommand{\pmark}{$\circ$}   

\newcommand{\bC}{\mathbb{C}}
\newcommand{\bmm}[1]{\mathbf{#1}}
\DeclareMathOperator{\diag}{diag}
\DeclareMathOperator{\blkdiag}{blkdiag}

\newtcolorbox{lessons}[1][Lessons learned]{
  breakable, enhanced, colback=lessonbg, colframe=lessonframe,
  boxrule=0.5pt, arc=1.2mm, left=1.2mm, right=1.2mm, top=0.8mm, bottom=0.8mm,
  fonttitle=\bfseries\small, title={#1}, coltitle=white, colbacktitle=lessonframe,
  fontupper=\small}

\begin{document}

\title{\huge A Survey on Stacked Intelligent Metasurfaces:\\ Fundamentals, Recent Advances, and Challenges}

\author{Chandan~Kumar~Sheemar,~\IEEEmembership{Member,~IEEE,} Sourabh Solanki,~\IEEEmembership{Member,~IEEE,} Wali Ullah Khan,~\IEEEmembership{Member,~IEEE,} \\George C. Alexandropoulos,~\IEEEmembership{Senior Member,~IEEE,} Giovanni Iacovelli,~\IEEEmembership{Member,~IEEE,} \\and Symeon Chatzinotas,~\IEEEmembership{Fellow,~IEEE}
\thanks{C. K. Sheemar, W. U. Khan, G. Iacovelli and S. Chatzinotas are with SnT, University of Luxembourg, 1855 Luxembourg City, Luxembourg (e-mails: \{chandankumar.sheemar, waliullah.khan, giovanni.iacovelli, symeon.chatzinotas\}@uni.lu); S. Solanki is with the National Institute of Technology Warangal, TS 506004, India (e-mail: ssolanki@nitw.ac.in); G. C. Alexandropoulos is with the Department of Informatics and Telecommunications, National and Kapodistrian University of Athens, 16122 Athens, Greece, and with the Department of Electrical and Computer Engineering, University of Illinois Chicago, IL 60601, USA (e-mail: alexandg@di.uoa.gr).}
}

\markboth{Submitted to IEEE Communications Surveys \& Tutorials}{Sheemar \MakeLowercase{\textit{et al.}}: A Survey on Stacked Intelligent Metasurfaces}

\maketitle

\begin{abstract}
Stacked intelligent metasurfaces (SIMs) cascade several closely spaced programmable metasurface layers so that an electromagnetic wave undergoes repeated modulation--propagation interactions before leaving the stack. The stack thereby implements, at the speed of propagation and with few radio-frequency chains, high-dimensional analog transformations that would otherwise require digital precoding, combining, spatial filtering, focusing, or inference. Since 2023, SIM research has produced more than two hundred articles, a dedicated magazine special issue, and two survey-level overviews that are organized around applications and around wave-domain computing. This article complements them with a model-to-hardware perspective that the existing literature does not provide. First, we present the electromagnetic modeling hierarchy of SIMs, i.e., the widely used cascaded Rayleigh--Sommerfeld operator and the physically consistent multiport impedance ($Z$), scattering ($S$), and transfer-scattering ($T$) formulations, and we make explicit the assumptions under which they coincide, their validity regimes, and their computational complexity, together with a common notation for hardware non-idealities and power consumption. Second, we present a methodology-first taxonomy (SIM-aware channel acquisition, model-based optimization, statistical-CSI design, learning-based control, and hardware-aware robust design), compare the families in terms of CSI requirements, complexity, convergence, and hardware constraints, and reconcile conflicting findings reported across the literature. Third, we synthesize applications by the physical role that the SIM plays. Fourth, we compare experimentally demonstrated multilayer prototypes quantitatively (frequency, layers, meta-atoms, tuning technology, loss, latency, power, and measured performance), separate experimentally validated from conceptual architectures, and discuss thermal, tolerance, and calibration issues. Fifth, we provide a tutorial-style performance and trade-off analysis anchored in results reported in the literature, covering stacking depth, aperture, phase resolution, CSI acquisition, bandwidth, geometry sensitivity, energy efficiency, computational cost, and measured-versus-simulated performance. Finally, we map evidence-based challenges to research directions with measurable targets. Lessons learned are summarized at the end of every section to help readers avoid recurring pitfalls.
\end{abstract}


\IEEEpeerreviewmaketitle

\section{Introduction}
\label{sec:intro}

\IEEEPARstart{S}{tacked} intelligent metasurfaces (SIMs) are programmable multilayer electromagnetic (EM) structures in which several metasurface layers are placed in close proximity, so that an incident wave is modulated by one layer, diffracts to the next, is modulated again, and so on, until it leaves the stack~\cite{an2023stacked,an2024stacked,an2026stacked}. Each layer is a dense planar array of sub-wavelength meta-atoms whose transmission response (phase and, in some designs, amplitude, polarization, delay, or a nonlinear characteristic) is set by a controller. From a signal-processing viewpoint, the $l$-th layer applies a programmable operator to the wavefront, while each inter-layer region applies a fixed, dense mixing operator dictated by diffraction. The concatenation is an analog processor that implements precoding, combining, focusing, spatial filtering, interference suppression, and task-oriented feature extraction in the wave domain, at the speed of propagation and with far fewer radio-frequency (RF) chains than radiating elements~\cite{an2023stacked,an2025stacked,liu2025stacked}.

The interest in SIMs stems from a tension that dominates 6G transceiver design: extremely large apertures, holographic multiple-input multiple-output (MIMO), near-field links, integrated sensing and communication (ISAC), and non-terrestrial platforms require high-dimensional spatial processing, but implementing it digitally requires as many RF chains and data converters as radiating elements and baseband operations whose cost grows quadratically or cubically with the array size. SIMs move a substantial share of this processing into the physics of propagation. The price is paid elsewhere: in insertion loss that accumulates across layers, in a control plane that must set thousands of meta-atoms, in channel state information (CSI) that is entangled with the SIM configuration, and in models whose accuracy has rarely been confirmed against measurements. Whether, and when, the gain outweighs the price is the recurring question of this article.

\subsection{From Single-Layer Surfaces to SIMs: A Compact Lineage}
\label{sec:lineage}

SIMs descend from two lines of work. The first is base-station array processing: from the MIMO capacity results of the late 1990s~\cite{foschini_mimo_1998,telatar_capacity_1999} to massive MIMO~\cite{marzetta_noncooperative_2010,larsson_massive_mimo_2014,andrews_5g_2014}, its deployment~\cite{bjornson_massive_reality_2019}, holographic MIMO and large intelligent surfaces~\cite{huang_holographic_2020,dardari_large_2020}, and extremely large-scale near-field MIMO~\cite{wang_xlmimo_tutorial_2024}. The second is programmable metasurfaces and wave-domain computing: from metamaterials~\cite{smith_metamaterial_2000} and flat-optics metasurfaces~\cite{yu_flat_optics_2014} to coding and software-controlled metasurfaces~\cite{cui_coding_2014,liaskos_software_2018,pitilakis_sdm_2019}, reconfigurable intelligent surfaces (RISs)~\cite{huang2019reconfigurable,wu_irs_tcom_2019,AlexandropoulosRIS}, simultaneously transmitting and reflecting (STAR) RISs~\cite{xu_star_2021,liu_star_2021,mu_star_2022}, hybrid RISs performing tunable reflection and absorption simultaneously~\cite{10352433,10042447} or in orthogonal time slots~\cite{9053976,george2026hris_chapter}, fully-absorbing RISs as energy harvesters~\cite{10693440}, beyond-diagonal (BD) RISs~\cite{li_bdris_2023,nerini_bdris_2024}, active RISs realizing reflection amplification~\cite{long2021active,11045692}, and transmissive-RIS transceivers~\cite{tang_tris_2023,li_tris_transceiver_2024}, in parallel with the diffractive deep neural networks (D$^2$NNs) demonstrated at terahertz, optical, and microwave frequencies~\cite{lin2018all,qian2020performing,li2021spectrally,liu2022programmable}. The RIS literature is mature and has been surveyed extensively ~\cite{liu_ris_survey_2021,bjornson_ris_2021,basar2024reconfigurable,katwe2024overview}. What distinguishes an SIM is that a cascade of $L\geq2$ layers with inter-layer diffraction realizes an input--output map that is neither diagonal nor limited to a single scattering event and whose expressiveness grows with $L$: physically, an SIM is the microwave counterpart of a D$^2$NN; from the communication-theoretic side it is the multilayer generalization of a transmissive RIS in front of a small feed array~\cite{an_sim_icc_2023,an2023stacked}.

\subsection{Why Another SIM Survey? Three Gaps in the Existing Overviews}
\label{sec:gaps}

Since the SIM-aided holographic MIMO architecture of An~\emph{et al.}~\cite{an2023stacked}, research on SIMs has expanded quickly. By September 2026, our literature search identified more than 230 SIM-specific articles, a special issue of \emph{IEEE Wireless Communications} on SIM-empowered signal processing, and several overview-type contributions~\cite{an2024stacked,liu2025stacked,di2025state,an2025emerging,shi2025stacked,an2026sim_chapter,liu2026sim_electronics,an2026stacked,niu2026survey}. The two most complete are the tutorial-style overview of An~\emph{et al.} ~\cite{an2026stacked}, which presents SIMs as physical neural networks and organizes the material by function (communication, sensing, computing, and their fusion), and the comprehensive survey of Niu~\emph{et al.}~\cite{niu2026survey}, which is organized around signal-processing functions and application scenarios (beamforming, channel estimation, multiple access, sensing and ISAC, satellite, aerial, cell-free, security, OFDM, and machine-learning integration) and summarizes the three inter-layer propagation models qualitatively. Both adopt the cascaded diffraction model as the working description of the stack. Reading them against the primary literature reveals three gaps that define the scope of this article.

\emph{Gap 1 (modeling).} The cascaded Rayleigh--Sommerfeld operator underlying almost all SIM optimization papers is a first-order approximation. Physically consistent multiport formulations in the impedance~\cite{abrardo2025novel,pettanice2026modeling}, scattering~\cite{nerini2024physically}, and transfer-scattering~\cite{yahya2025t} domains show that inter-layer reflections, mutual coupling, loss, and amplitude--phase coupling change not only the numbers but sometimes the trends, e.g., whether adding layers helps~\cite{gonzalez2026lorentzian,dong2026geometry}. No overview places these models in a common notation, states when they coincide, quantifies their validity regimes, or compares their computational cost; the survey~\cite{niu2026survey} lists them side by side with qualitative limitations only.

\emph{Gap 2 (methodology).} Existing overviews are organized by application, yet the technical difficulties recur across applications: CSI must be acquired through the device being configured, the objective is a product of diagonal matrices interleaved with dense ones, instantaneous CSI is often unaffordable, learned controllers must respect EM constraints, and hardware non-idealities alter the feasible set. An application-first organization repeats these issues in every section and hides that different papers reach different conclusions under different assumptions.

\emph{Gap 3 (hardware and performance evidence).} Prototype descriptions in existing overviews are informative but not comparative: frequency, layers, meta-atoms, tuning mechanism, insertion loss, latency, power, and measured performance are not tabulated side by side, and validated designs are not separated from conceptual proposals. The quantitative trade-offs that the primary literature has established (saturation with depth, loss thresholds, sufficient quantization, training overhead) have not been consolidated, and the challenges listed in prior overviews therefore remain generic.

\subsection{Related Surveys, Tutorials, and Magazine Articles}
\label{sec:related}

Table~\ref{tab:survey_comparison} compares this article with the SIM-related overview literature along explicit attributes: article type, EM models covered, organizing principle, treatment of hardware, literature-anchored trade-off analysis, evidence-linked challenges, survey comparison, and literature window. It includes the architecture paper~\cite{an2023stacked}, which is the de-facto reference for the cascaded model; the magazine articles~\cite{an2024stacked,liu2025stacked,shi2025stacked,alwazani2025delay,abbas2025nonlinear,niu2026twolayer,liu2026inwave,jia2026calibration,wang2026learnable,cao2026element,zhang2026satellite,sun2026crossdomain,ma2026lain,zhang2026spherical,qi2026task}, each with one case study; the short state-of-the-art note~\cite{di2025state}; the invited overview~\cite{an2025emerging}, of which one section concerns SIMs; the book chapter~\cite{an2026sim_chapter}; the review~\cite{liu2026sim_electronics}; the edge-inference overview~\cite{stylianopoulos2026metasurfaces}; and the two survey-level works~\cite{an2026stacked,niu2026survey}. The comparison is meant to position rather than to rank: this article is complementary to~\cite{an2026stacked,niu2026survey}, which are organized by function and by signal-processing task, respectively, in the way that a modeling- and hardware-oriented survey complements function-oriented ones. Readers interested in wave-domain computing will find~\cite{an2026stacked} more detailed and readers who need a function-by-function review of signal-processing designs will find~\cite{niu2026survey} more detailed; readers who must decide which EM model to use, which methodological family fits their CSI and hardware budget, what prototypes can do, and which reported gains survive realistic assumptions are the audience of this article.

\begin{table*}[!t]
\centering
\caption{Positioning of this survey with respect to SIM-related overview, tutorial, magazine, and survey articles. EM models: C = cascaded diffraction operator; Z/S/T = multiport impedance / scattering / transfer-scattering models; V = validity regimes made explicit; X = computational complexity compared. Hardware: Q = qualitative; T = quantitative demonstrator table; G = theory--prototype gap; M = validated vs.\ conceptual architectures separated. \cmark\ covered; \pmark\ partial; \xmark\ not covered.}
\label{tab:survey_comparison}
\renewcommand{\arraystretch}{1.05}
\setlength{\tabcolsep}{2pt}
\scriptsize
\arrayrulecolor{SIMLine}
\begin{tabularx}{\textwidth}{|>{\raggedright\arraybackslash}p{1.55cm}|>{\raggedright\arraybackslash}p{1.6cm}|>{\raggedright\arraybackslash}p{1.75cm}|>{\raggedright\arraybackslash}p{1.95cm}|>{\centering\arraybackslash}p{1.25cm}|>{\centering\arraybackslash}p{1.05cm}|>{\centering\arraybackslash}p{1.0cm}|>{\centering\arraybackslash}p{0.95cm}|>{\raggedright\arraybackslash}X|}
\hline
\rowcolor{SIMHeader}
\textbf{Ref.} & \textbf{Type} & \textbf{EM models} & \textbf{Organizing principle} & \textbf{Hardware} & \textbf{Trade-off synthesis} & \textbf{Evidence-linked chal\-lenges} & \textbf{Survey compar\-ison} & \textbf{Scope / what is left to this article} \\
\hline
\cite{an2023stacked} & Architecture paper & C & Single architecture & \xmark & \pmark\ (own simulations) & \xmark & \xmark & Introduces SIM-aided holographic MIMO and the cascaded model; no literature review. \\
\hline
\cite{an2024stacked} & Magazine & C & Transceiver design & Q & \pmark\ (case study) & \pmark & \xmark & End-to-end SIM transceiver, wave-domain precoding/combining, open challenges. \\
\hline
\cite{abbas2024stacked} & Magazine & C & ISAC application & \xmark & \pmark & \pmark & \xmark & SIM-assisted MIMO ISAC; problem-specific. \\
\hline
\cite{di2025state} & State-of-the-art note & C (one paragraph on multiport) & Topic list & \xmark & \xmark & \xmark & \xmark & Annotated bibliography of early SIM work. \\
\hline
\cite{liu2025stacked} & Magazine & C & Applications & Q (3 hardware classes) & \pmark\ (estimator comparison) & \pmark & \xmark & Hardware classes, beamforming, semantic encoding, channel estimation. \\
\hline
\cite{an2025emerging} & Invited overview of RIS/SIM/FIM & C & Technology & Q & \pmark & \pmark & \xmark & One section on SIMs within a broader metasurface overview. \\
\hline
\cite{shi2025stacked} & Magazine & C & Network architectures & \xmark & \pmark & \pmark & \xmark & Centralized/hybrid/distributed SIM processing for cell-free networks. \\
\hline
\cite{an2026sim_chapter} & Book chapter & C & Modeling and optimization & Q & \pmark & \pmark & \xmark & Hardware, wave-domain beamforming, ISAC, channel estimation. \\
\hline
\cite{liu2026sim_electronics} & Review & C & Key technologies & Q & \pmark\ (algorithm table) & \pmark & \xmark & Architecture, prototypes, beamforming/estimation challenges, applications. \\
\hline
\cite{stylianopoulos2026metasurfaces} & Overview & C & Neural-network view & \xmark & \pmark & \pmark & \xmark & Metasurface-integrated neural networks for over-the-air edge inference. \\
\hline
\cite{alwazani2025delay,abbas2025nonlinear,niu2026twolayer,liu2026inwave,jia2026calibration,wang2026learnable,cao2026element,zhang2026satellite,sun2026crossdomain,ma2026lain,zhang2026spherical,qi2026task} & Magazine articles (2025--26, incl.\ a SIM special issue) & C (element-level in \cite{cao2026element}) & One architecture or scenario each & Q & \pmark\ (one case study each) & \pmark & \xmark & Delay-augmented, nonlinear, two-layer, in-wave computing, calibration, learnable, element-level, satellite, cross-domain, low-altitude, spherical, task-oriented SIMs. \\
\hline
\cite{an2026stacked} & Tutorial/overview & C (multiport mentioned) & Function: communication, sensing, computing, fusion & Q + prototype table (layers, size, frequency, task) & \pmark\ (own experiments) & \pmark & \xmark & SIMs as physical neural networks; training methods; experimental evidence of multifunctionality. \\
\hline
\cite{niu2026survey} & Survey & C; Z/T summarized qualitatively (three inter-layer models, no common notation, validity, or cost comparison) & Key issues (beamforming, estimation) $\rightarrow$ multiple access $\rightarrow$ sensing/ISAC $\rightarrow$ scenarios $\rightarrow$ ML & Q (six prototypes in a figure) & \pmark\ (per-work table of reported gains) & \pmark\ (four challenge areas with solution pathways) & \pmark\ (four articles) & Physical principles, wave-domain beamforming, channel estimation, SDMA/RSMA, DoA and ISAC, satellite/aerial/cell-free/security/OFDM scenarios, RL and semantic communications (concurrent with this work). \\
\hline
\rowcolor{SIMFirstCol}
\textbf{This article} & Survey, model-to-hardware & C, Z, S, T, V, X; non-ideality and power models in common notation & Methodology-first taxonomy (5 families) $\rightarrow$ application roles $\rightarrow$ hardware $\rightarrow$ trade-offs & Q + T + G + M; tuning-technology table; thermal, tolerance, calibration & \cmark\ (9 SIM-specific trade-offs anchored in published results) & \cmark\ (challenge $\rightarrow$ evidence $\rightarrow$ direction $\rightarrow$ metric) & \cmark\ (this table) & Complete Literature overview; $>$230 SIM-specific works; conflicting findings reconciled; lessons learned per section. \\
\hline
\end{tabularx}
\arrayrulecolor{black}
\end{table*}

\subsection{Contributions, Organization, and Reading Guide}
\label{sec:contributions}

The contributions map onto the three gaps, and Table~\ref{tab:structure} summarizes the structure of the article section by section. \emph{(Gap~1)} Section~\ref{sec:fund} develops the cascaded model, the multiport impedance model and its SIM specialization, and the scattering and transfer-scattering representations in one notation, states the four assumptions under which they collapse to the cascaded operator, compares their computational complexity, and introduces hardware non-ideality and power-consumption models used throughout. \emph{(Gap~2)} Section~\ref{sec:taxonomy} links architectures, EM models, methodological families, application roles, and hardware maturity in a unified taxonomy (Fig.~\ref{fig:unified_taxonomy}), classifies the literature into five methodological families, compares them (Table~\ref{tab:method_compare}), and reconciles conflicting findings (Table~\ref{tab:conflicts}); Section~\ref{sec:apps} synthesizes applications by the physical role of the SIM with the gains claimed and the assumptions behind them. \emph{(Gap~3)} Section~\ref{sec:hardware} tabulates experimentally demonstrated prototypes (Table~\ref{tab:prototypes}) and tuning technologies (Table~\ref{tab:tuning}), separates validated from conceptual architectures (Table~\ref{tab:maturity}), and discusses thermal management, tolerances, and calibration; Section~\ref{sec:performance} consolidates nine trade-offs established by published results (Table~\ref{tab:tradeoff_numbers}); Section~\ref{sec:challenges} maps each challenge to evidence, directions, and a progress metric; Section~\ref{sec:conclusion} provides the conclusions.

\begin{table*}[!t]
\centering
\caption{Structure of the survey. Each section ends with a ``lessons learned'' box; the last column names the gap of Section~\ref{sec:gaps} that the section addresses.}
\label{tab:structure}
\renewcommand{\arraystretch}{1.12}
\setlength{\tabcolsep}{3pt}
\scriptsize
\arrayrulecolor{SIMLine}
\begin{tabularx}{\textwidth}{|>{\raggedright\arraybackslash}p{2.35cm}|>{\raggedright\arraybackslash}X|>{\raggedright\arraybackslash}p{3.35cm}|>{\raggedright\arraybackslash}p{2.55cm}|}
\hline
\rowcolor{SIMHeader}\textbf{Section} & \textbf{Content (subsections)} & \textbf{Instruments} & \textbf{Gap addressed / outcome for the reader} \\
\hline
\cellcolor{SIMFirstCol}\textbf{I. Introduction} (Sec.~\ref{sec:intro}) & A. From single-layer surfaces to SIMs; B. Why another SIM survey? Three gaps in the existing overviews; C. Related surveys, tutorials, and magazine articles; D. Contributions, organization, and reading guide & Table~\ref{tab:survey_comparison} (positioning vs.\ 24 overview/survey articles); Table~\ref{tab:structure} (this table); Table~\ref{tab:notation} (notation) & Motivation and scope; where this article sits relative to~\cite{an2026stacked,niu2026survey} \\
\hline
\cellcolor{pastelpurple}\textbf{II. Electromagnetic foundations and modeling hierarchy} (Sec.~\ref{sec:fund}) & A. Physical architecture and wave-domain processing; B. Fundamentals of multiport network theory; C. Specialization of the multiport model to SIMs; D. Connection among $Z$, $S$, and $T$ representations; E. Modeling hierarchy and validity regimes; F. Computational complexity of the models; G. Hardware non-ideality and power-consumption models & Figs.~\ref{fig_SIM_base}--\ref{fig_ECO} (architecture, geometry, multiport network); Fig.~\ref{fig:model_hierarchy} (modeling hierarchy); Table~\ref{ZST_comparison} ($Z$/$S$/$T$); Table~\ref{tab:complexity} (complexity); Eqs.~\eqref{eq:Phi}--\eqref{eq:power} & \textbf{Gap 1 (modeling):} which model to trust in which regime, at what cost; common notation for loss, quantization, geometry errors, latency, and power \\
\hline
\cellcolor{pastelblue}\textbf{III. Unified taxonomy and methodological families} (Sec.~\ref{sec:taxonomy}) & A. SIM-aware channel acquisition and training; B. Model-based optimization; C. Statistical-CSI and low-overhead design; D. Learning-based and AI-driven control (incl.\ training overhead); E. Hardware-aware and robust optimization; F. Reconciling conflicting findings & Fig.~\ref{fig:unified_taxonomy} (five-coordinate taxonomy); Fig.~\ref{fig:method_taxonomy} (five families); Table~\ref{tab:method_compare} (CSI, complexity, convergence, adaptation, hardware constraints); Table~\ref{tab:conflicts} (six conflicting findings reconciled) & \textbf{Gap 2 (methodology):} a methodology-first map of the literature and the reasons why papers disagree \\
\hline
\cellcolor{pastelblue}\textbf{IV. Application synthesis by physical role} (Sec.~\ref{sec:apps}) & A. Wave-domain precoder and combiner (MU-MIMO, holographic MIMO, cell-free); B. Near-field focusing and pattern synthesis (incl.\ mixed-field and discrete control); C. Space--time processor for wideband and high-mobility links; D. Multifunctional aperture (ISAC, localization/imaging, security); E. Lightweight analog front end (energy efficiency/WPT/SWIPT, non-terrestrial and aerial, semantic/multiple access/full duplex) & Fig.~\ref{fig:applications} (application domains); Table~\ref{tab:SIM_application_unified} (function, tools, reported gains, assumptions per role) & \textbf{Gap 2:} which gains are claimed, under which assumptions, and which are fragile \\
\hline
\cellcolor{SIMBody}\textbf{V. Hardware demonstrators, tuning technologies, advanced architectures, theory--prototype gap} (Sec.~\ref{sec:hardware}) & A. Demonstrator classes and quantitative comparison; B. Tuning technologies for programmable layers; C. Advanced architectures and their maturity; D. Thermal and power management in active and dense stacks; E. Fabrication tolerances, misalignment, and calibration; F. Theory--prototype gap and a reporting checklist & Table~\ref{tab:prototypes} (13 measured prototypes: band, layers, elements, tuning, loss, latency, power, performance); Table~\ref{tab:tuning} (tuning technologies); Fig.~\ref{fig:Advanced_SIM} and Table~\ref{tab:maturity} (validated vs.\ conceptual architectures) & \textbf{Gap 3 (hardware):} what has been built and measured, what is conceptual, and what prototypes must report \\
\hline
\cellcolor{SIMBody}\textbf{VI. Performance and trade-off analysis} (Sec.~\ref{sec:performance}) & A. Stacking depth; B. Aperture versus depth; C. Phase resolution and discrete control; D. CSI acquisition, statistical CSI, and net rate; E. Bandwidth and beam squint; F. Geometry, alignment, and calibration sensitivity; G. Energy efficiency and the power model; H. Computational cost; I. Measured versus simulated performance & Figs.~\ref{fig:depth}--\ref{fig:measured} (results reproduced from the literature, each with trade-off, evidence, and design rule); Table~\ref{tab:tradeoff_numbers} (all numbers with their assumption level) & \textbf{Gap 3 (evidence):} tutorial-style design rules anchored in published results \\
\hline
\cellcolor{SIMFirstCol}\textbf{VII. Challenges and future research directions} (Sec.~\ref{sec:challenges}) & Eight strands: 1) physically consistent yet tractable modeling; 2) scalable joint optimization; 3) CSI acquisition and tracking; 4) wideband and doubly dispersive operation; 5) non-idealities, loss, and control overhead; 6) calibration and reproducibility; 7) network-level orchestration; 8) trustworthy AI control & challenge $\rightarrow$ evidence $\rightarrow$ directions $\rightarrow$ progress metric & \textbf{Gaps 1--3:} evidence-linked directions with measurable targets \\
\hline
\cellcolor{SIMFirstCol}\textbf{VIII. Conclusions} (Sec.~\ref{sec:conclusion}) & Five conclusions on modeling, methodology, applications, hardware, and trade-offs & -- & What decides the transition from concept to deployable front end \\
\hline
\end{tabularx}
\arrayrulecolor{black}
\end{table*}

\begin{table}[t]
\centering
\caption{Main notation.}
\label{tab:notation}
\renewcommand{\arraystretch}{1.02}
\setlength{\tabcolsep}{3pt}
\scriptsize
\begin{tabularx}{\linewidth}{|>{\raggedright\arraybackslash}p{1.8cm}|X|}
\hline
\rowcolor{SIMHeader}\textbf{Symbol} & \textbf{Meaning} \\
\hline
$L$, $M$, $S$, $K$ & Layers, meta-atoms per layer, RF chains (feeds), users \\
$\lambda$, $f_c$, $B$; $r_{e}$, $d$, $D$ & Wavelength, carrier, bandwidth; pitch, inter-layer spacing, stack thickness \\
$\theta_m^{(l)}$, $\phi_m^{(l)}$, $\bmm{\Phi}_l$ & Phase shift, transmission coefficient of meta-atom $m$ on layer $l$; diagonal layer matrix \\
$\bmm{W}_l$, $\bmm{P}$, $\bmm{H}$ & Inter-layer diffraction matrix ($\bmm{W}_1\in\bC^{M\times S}$ feed-to-layer-1); cascaded SIM response; channel to users \\
$\bmm{Z}_{TT},\bmm{Z}_{EE},\ldots$; $\bmm{Z}_E(\bm{\eta})$ & Blocks of the multiport impedance matrix (T transmitter, E ECO/SIM, R receiver); load network with parameters $\bm{\eta}$ \\
$\bmm{\Psi}(\bm{\eta})$, $\bmm{\Psi}_{i,j}$ & $(\bmm{Z}_{EE}+\bmm{Z}_E(\bm{\eta}))^{-1}$ and its $M\times M$ blocks \\
$\bmm{S}$, $\bmm{T}$, $Z_0$ & Scattering and transfer-scattering matrices; reference impedance \\
$\alpha_l$, $b$; $P_{\rm tot}$, $\eta_{\rm PA}$ & Per-layer amplitude (loss) factor, phase bits; total power, PA efficiency \\
$(\cdot)^{T}$, $(\cdot)^{H}$, $\diag$, $\blkdiag$, $\mathcal{O}(\cdot)$ & Transpose, Hermitian, (block-)diagonal, complexity order \\
\hline
\end{tabularx}
\end{table}

\section{Electromagnetic Foundations and the Modeling Hierarchy of SIMs}
\label{sec:fund}

This section is the technical backbone of SIM and this article. It introduces the SIM architecture and the cascaded diffraction model used by most of the literature (Section~\ref{sec:arch}), the multiport formulation that describes the stack as a physically consistent EM object (Sections~\ref{sec:multiport}--\ref{sec:sim_multiport}), and the scattering and transfer-scattering representations (Section~\ref{sec:ZST}). It then states when the models coincide and when they do not (Section~\ref{sec:validity}), compares their computational cost (Section~\ref{sec:complexity}), and introduces the non-ideality and power models needed to read the rest of the article critically (Section~\ref{sec:nonideal}).

\subsection{Physical Architecture and Wave-Domain Processing}
\label{sec:arch}

\begin{figure}[!t]
\centering
\includegraphics[width=0.9\linewidth,height=4.1cm]{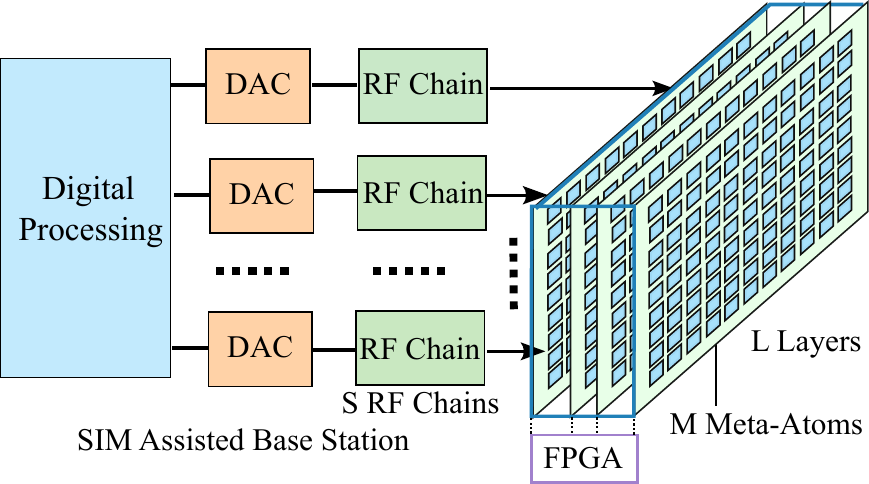}
\caption{An SIM-assisted base station: $S$ RF chains feed an $L$-layer stack of $M$-element programmable metasurfaces.}
\label{fig_SIM_base}
\end{figure}

Consider a base station (BS) whose transmit front end is an SIM with $L$ programmable metasurface layers aligned in parallel and separated by controlled distances (Fig.~\ref{fig_SIM_base}). Each layer carries a dense array of sub-wavelength meta-atoms whose resonant response is tuned by an embedded element (PIN diode, varactor, liquid crystal, micro-electro-mechanical element, or amplifier; Section~\ref{sec:tuning}) under the control of a field-programmable gate array (FPGA). A compact feed array driven by $S$ RF chains illuminates the first layer; the field undergoes successive modulation and diffraction and radiates from the last layer into the channel. The stack is a wave-domain precoder that synthesizes the aperture field without symbol-rate digital beamforming; in front of a receive array it is a wave-domain combiner.

Let each layer contain $M$ meta-atoms. The transmission coefficient of meta-atom $m$ on layer $l$ is $\phi_m^{(l)}=e^{j\theta_m^{(l)}}$, $\theta_m^{(l)}\in[0,2\pi)$, and, neglecting intra-layer coupling, the layer operation is
\begin{equation}
\bmm{\Phi}_l=\diag\!\big(\phi_1^{(l)},\ldots,\phi_M^{(l)}\big)\in\bC^{M\times M}.
\label{eq:Phi}
\end{equation}
This phase-only diagonal model is an abstraction: physically, tuning is implemented by adjustable loads, modeled through $\bmm{Z}_E(\bm{\eta})$ in Section~\ref{sec:multiport}, and practical meta-atoms exhibit loss and amplitude--phase coupling, modeled in Section~\ref{sec:nonideal}.

Each layer is a square lattice of $m_{\max}\times m_{\max}$ elements, $M=m_{\max}^2$ (Fig.~\ref{fig_SIM_conf_1}). With $m_y=\lceil m/m_{\max}\rceil$, $m_x=\mathrm{mod}(m-1,m_{\max})+1$, and pitch $r_e$, the in-plane separation of elements $m$ and $\tilde m$ is $r_{m,\tilde m}=r_{e}\sqrt{(m_y-\tilde m_y)^2+(m_x-\tilde m_x)^2}$; for a stack of thickness $D$ with uniform spacing $d=D/(L-1)$, the distance between element $\tilde m$ on layer $l-1$ and element $m$ on layer $l$ is $r^{(l)}_{m,\tilde m}=\sqrt{r_{m,\tilde m}^2+d^2}$ (Fig.~\ref{fig_SIM_conf}). Inter-layer coupling is governed by diffraction; the Rayleigh--Sommerfeld formula~\cite{wolf1964comparison,shen2006fast,freude2002rayleigh,gao2019rayleigh} gives
\begin{equation}
w_{m,\tilde m}^{(l)}=\frac{A_t\cos\chi_{m,\tilde m}^{(l)}}{r^{(l)}_{m,\tilde m}}\left(\frac{1}{2\pi r^{(l)}_{m,\tilde m}}-\frac{j}{\lambda}\right)e^{j\frac{2\pi}{\lambda}r^{(l)}_{m,\tilde m}},
\label{eq:RS}
\end{equation}
with $A_t$ the meta-atom area and $\chi_{m,\tilde m}^{(l)}$ the angle to the layer normal. With $[\bmm{W}_l]_{m,\tilde m}=w_{m,\tilde m}^{(l)}$ for $l\ge2$ and $\bmm{W}_1\in\bC^{M\times S}$ the feed-to-first-layer coupling, the end-to-end response is
\begin{equation}
\bmm{P}=\bmm{\Phi}_L\bmm{W}_L\cdots\bmm{\Phi}_2\bmm{W}_2\bmm{\Phi}_1\bmm{W}_1\in\bC^{M\times S},
\label{cascaded_model}
\end{equation}
and with a channel $\bmm{H}\in\bC^{K\times M}$ to $K$ users, $\bmm{y}=\bmm{H}\bmm{P}\bmm{x}+\bmm{n}$. Equation~\eqref{cascaded_model} is the model used by the overwhelming majority of SIM optimization, learning, and application papers, and the model with respect to which the results of Sections~\ref{sec:taxonomy}--\ref{sec:performance} must be read.

\begin{figure}[t]
\centering
\begin{subfigure}[t]{0.30\textwidth}
\centering
\resizebox{\linewidth}{!}{%
\begin{tikzpicture}[>=Latex, font=\Large]
\tikzset{atom/.style={draw=black, line width=1pt, fill=atomfill, minimum width=9mm, minimum height=9mm, inner sep=0pt}}
\node[anchor=west] at (0,6.2) {The $l$-th transmit metasurface};
\draw[fill=lightgreen, draw=none] (0.2,0.6) rectangle (6.2,5.6);
\node[atom] at (1.0,4.9) {$1$};
\node[atom] at (2.2,4.9) {$2$};
\node at (3.6,4.9) {$\cdots$};
\node[atom] (topright) at (5.4,4.9) {$-$};
\node[atom] at (1.0,3.6) {};
\node[atom] at (2.2,3.6) {};
\node at (3.6,3.6) {$\cdots$};
\node[atom] at (5.4,3.6) {};
\node at (1.0,2.6) {$\vdots$};
\node at (2.2,2.6) {$\vdots$};
\node at (5.4,2.6) {$\vdots$};
\node[atom] at (1.0,1.4) {};
\node[atom] at (2.2,1.4) {};
\node at (3.6,1.4) {$\cdots$};
\node[atom] at (5.4,1.4) {$M$};
\draw[densely dashed, ->, line width=1.0pt] (topright.east) -- ++(1.2,0) node[right] {$m_{\max}$};
\draw[densely dashed] (1.0,0.55) -- (1.0,0.05);
\draw[densely dashed] (2.2,0.55) -- (2.2,0.05);
\draw[<->, line width=1.0pt] (1.0,0.25) -- (2.2,0.25) node[midway, below] {$r_{e}$};
\end{tikzpicture}}
\caption{The $l$-th layer ($M=m_{\max}^2$ meta-atoms).}
\label{fig_SIM_conf_1}
\end{subfigure}
\hfill
\begin{subfigure}[t]{0.40\textwidth}
\centering
\resizebox{\linewidth}{!}{%
\begin{tikzpicture}[>=Latex, font=\Large]
\tikzset{layerline/.style={draw=black, line width=1.3pt}, port/.style={draw=black, line width=1.4pt, fill=white},
  atom/.style={draw=black, line width=1.3pt, fill=atomfill, minimum width=9mm, minimum height=9mm, inner sep=0pt},
  dashedbox/.style={draw=black, line width=1.2pt, dash pattern=on 7pt off 5pt}}
\node[font=\Large] at (7.2,6.7) {TX-SIM};
\draw[layerline] (0.7,5.6) -- (2.0,6.1);
\draw[layerline] (0.7,2.2) -- (2.0,1.7);
\draw[layerline] (2.0,6.1) -- (3.0,6.1);
\draw[layerline] (2.0,1.7) -- (3.0,1.7);
\foreach \y/\lab in {5.2/1,4.3/2,2.4/S}{ \draw[port] (1.0,\y) circle (0.16); \node[anchor=east] at (0.7,\y) {$\lab$}; }
\node at (0.85,3.35) {$\vdots$};
\draw[layerline] (3.0,1.7) rectangle (12.8,6.1);
\foreach \x in {4.6,6.2,7.8,9.4,11.0}{ \draw[layerline] (\x,1.7) -- (\x,6.1); }
\draw[layerline] (6.2,6.1) -- (6.2,4.7);
\draw[layerline] (7.8,6.1) -- (7.8,4.7);
\node[below] at (3.0,1.7) {$1$};
\node[below] at (6.2,1.7) {$l-1$};
\node[below] at (7.8,1.7) {$l$};
\node[below] at (12.8,1.7) {$L$};
\draw[dashedbox] (5.4,2.3) rectangle (8.6,5.5);
\node[atom] (mtil) at (6.2,4.9) {$\tilde m$};
\node[atom] (mL)   at (6.2,2.9) {$m$};
\node[atom] (mR)   at (7.6,2.9) {$m$};
\draw[densely dashed] (5.75,4.8) -- (5.75,3.0);
\draw[<->, line width=1.0pt] (5.75,4.7) -- (5.75,3.1);
\node[anchor=east] at (5.65,3.9) {$r_{m,\tilde m}$};
\draw[densely dashed] (mtil.south east) -- (mR.north west);
\node[anchor=west] at (6.95,3.95) {$r^{(l)}_{m,\tilde m}$};
\draw[<->, line width=1.0pt] ($(mL.south)+(0,-0.35)$) -- ($(mR.south)+(0,-0.35)$) node[midway,below] {$d$};
\node at (10.2,3.9) {$\cdots$};
\end{tikzpicture}}
\caption{Layered TX-SIM and inter-element distances.}
\label{fig_SIM_conf}
\end{subfigure}
\caption{Geometry of the cascaded model: (a) one layer and (b) the stack with the distances entering~\eqref{eq:RS}.}
\label{fig:updated_txsim}
\end{figure}

\subsection{Fundamentals of Multiport Network Theory}
\label{sec:multiport}

\begin{figure}[t]
\centering
\includegraphics[width=0.72\linewidth]{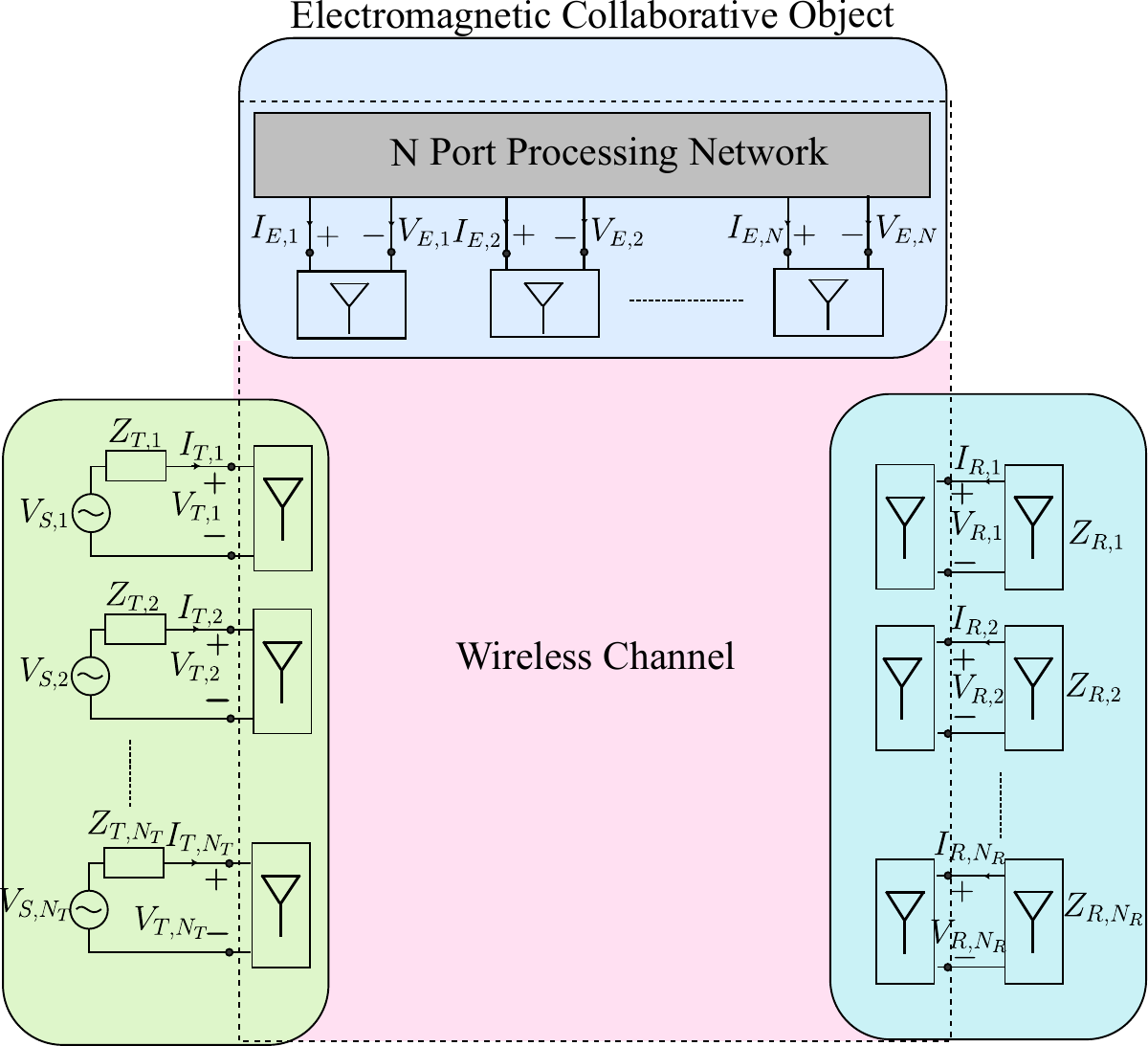}
\caption{A general multiport network: transmitter ($N_T$ ports), electromagnetic collaborative object (ECO, $N$ ports), and receiver ($N_R$ ports).}
\label{fig_ECO}
\end{figure}

The cascaded model assumes forward-only propagation, negligible mutual coupling, and uncoupled elements. A physically consistent description must account for radiation, mutual coupling, backward waves, and power conservation, which multiport network theory provides without far-field or uncoupled-element assumptions~\cite{abrardo2024design,zheng2024mutual,abrardo2025novel,pettanice2026modeling}. Consider three radiatively coupled subsystems (Fig.~\ref{fig_ECO}): a transmitter with $N_T$ ports, a receiver with $N_R$ ports, and an electromagnetic collaborative object (ECO) with $N$ ports, which abstracts reflective RISs, STAR-RISs, and SIMs depending on its internal impedance network. The impedance ($Z$-parameter) formulation relates port voltages and currents and captures near-field interactions, radiation loss, and matching. With $\bmm{V}_T,\bmm{I}_T\in\bC^{N_T}$, $\bmm{V}_E,\bmm{I}_E\in\bC^{N}$, $\bmm{V}_R,\bmm{I}_R\in\bC^{N_R}$,
\begin{equation}
\begin{bmatrix}\bmm{V}_T\\ \bmm{V}_E\\ \bmm{V}_R\end{bmatrix}=
\begin{bmatrix}\bmm{Z}_{TT}&\bmm{Z}_{TE}&\bmm{Z}_{TR}\\ \bmm{Z}_{ET}&\bmm{Z}_{EE}&\bmm{Z}_{ER}\\ \bmm{Z}_{RT}&\bmm{Z}_{RE}&\bmm{Z}_{RR}\end{bmatrix}
\begin{bmatrix}\bmm{I}_T\\ \bmm{I}_E\\ \bmm{I}_R\end{bmatrix},
\label{eq:Zfull}
\end{equation}
where diagonal blocks are self-impedances and off-diagonal blocks describe coupling through the medium. Programmability comes from the load network terminating the ECO ports: $\bmm{V}_E=-\bmm{Z}_E\bmm{I}_E$ with $\bmm{Z}_E\in\bC^{N\times N}$~\cite{abrardo2025novel}. Eliminating the ECO variables gives $\bmm{V}_R=\bmm{H}_Z\bmm{V}_T$ with~\cite{abrardo2025novel}
\begin{equation}
\bmm{H}_Z=\frac{1}{4Z_0}\Big[\bmm{Z}_{RT}-\bmm{Z}_{RE}\big(\bmm{Z}_{EE}+\bmm{Z}_E\big)^{-1}\bmm{Z}_{ET}\Big].
\label{eq:hz_general}
\end{equation}
With $P$ controllable parameters $\bm{\eta}\in\bC^{P}$ in $\bmm{Z}_E(\bm{\eta})$, $\bmm{\Psi}(\bm{\eta})\triangleq(\bmm{Z}_{EE}+\bmm{Z}_E(\bm{\eta}))^{-1}$, a matrix $\bmm{A}\in\bC^{N_R\times N}$ that includes $\bmm{Z}_{RE}$ and any linear receive processing, and $\bmm{b}\in\bC^{N}$ the excitation induced at the ECO ports, the ECO-induced transfer function is the bilinear form~\cite{abrardo2025novel}
\begin{equation}
\bmm{h}_Z(\bm{\eta})=\bmm{A}\,\bmm{\Psi}(\bm{\eta})\,\bmm{b},
\label{eq:generic_transfer}
\end{equation}
which unifies the optimization of any ECO, and in particular of an SIM, across applications.

\subsection{Specialization of the Multiport Model to SIMs}
\label{sec:sim_multiport}

Associating one port with each meta-atom, each layer is an $M$-port subnetwork and $N=LM$. For even $L$, the layers are grouped into $Q=L/2$ facing pairs; each pair $(2q-1,2q)$ is a transmissive RIS modeled as a $2M$-port network~\cite{abrardo2025novel}. The first layer faces the feeds, the last faces the environment, and inner layers are alternately coupled through internal load networks (within a pair) and through free-space propagation (between pairs). Because each layer interacts only with its neighbors, $\bmm{Z}_{ET}\in\bC^{LM\times N_T}$ is nonzero only in its first $M$ rows and $\bmm{Z}_{RE}\in\bC^{N_R\times LM}$ only in its last $M$ columns. Partitioning $\bmm{\Psi}(\bm{\eta})$ into $L\times L$ blocks $\bmm{\Psi}_{i,j}\in\bC^{M\times M}$,
\begin{equation}
\bmm{\Psi}(\bm{\eta})=
\begin{bmatrix}\bmm{\Psi}_{1,1}&\cdots&\bmm{\Psi}_{1,L}\\ \vdots&\ddots&\vdots\\ \bmm{\Psi}_{L,1}&\cdots&\bmm{\Psi}_{L,L}\end{bmatrix},
\label{eq:t_block_structure}
\end{equation}
and with $\bmm{Z}_{ET}^{(0)}$ ($\bmm{Z}_{RE}^{(0)}$) the nonzero rows (columns) of $\bmm{Z}_{ET}$ ($\bmm{Z}_{RE}$),
\begin{equation}
\bmm{H}_Z=\frac{1}{4Z_0}\Big[\bmm{Z}_{RT}-\bmm{Z}_{RE}^{(0)}\,\bmm{\Psi}_{L,1}(\bm{\eta})\,\bmm{Z}_{ET}^{(0)}\Big],
\label{eq:HZ_SIM}
\end{equation}
so that only the corner block $\bmm{\Psi}_{L,1}$ matters and~\eqref{eq:generic_transfer} becomes $\bmm{h}_Z(\bm{\eta})=\bmm{A}\,\bmm{\Psi}_{L,1}(\bm{\eta})\,\bmm{b}$~\cite{abrardo2025novel}. The internal matrices are banded. $\bmm{Z}_{EE}$ models the coupling between adjacent layers through free space,
\begin{equation}
\bmm{Z}_{EE}=
\begin{bmatrix}
\bmm{Z}_{2,2}^{(0)}&\bmm{0}&\bmm{0}&\bmm{0}&\cdots&\bmm{0}\\
\bmm{0}&\bmm{Z}_{1,1}^{(1)}&\bmm{Z}_{1,2}^{(1)}&\bmm{0}&\cdots&\bmm{0}\\
\bmm{0}&\bmm{Z}_{2,1}^{(1)}&\bmm{Z}_{2,2}^{(1)}&\bmm{0}&\cdots&\bmm{0}\\
\bmm{0}&\bmm{0}&\bmm{0}&\bmm{Z}_{1,1}^{(2)}&\cdots&\bmm{0}\\
\vdots&\vdots&\vdots&\vdots&\ddots&\vdots\\
\bmm{0}&\bmm{0}&\bmm{0}&\bmm{0}&\cdots&\bmm{Z}_{1,1}^{(Q)}
\end{bmatrix},
\label{eq:zee_band}
\end{equation}
where each $2\times2$ block $\{\bmm{Z}^{(q)}_{i,j}\}$ captures the bidirectional coupling between layers $2q$ and $2q+1$, $\bmm{Z}^{(0)}_{2,2}$ and $\bmm{Z}^{(Q)}_{1,1}$ are boundary self-impedances, and reciprocity gives $\bmm{Z}^{(q)}_{2,1}=(\bmm{Z}^{(q)}_{1,2})^{T}$. The load network is block-diagonal,
\begin{equation}
\begin{aligned}
\bmm{Z}_E(\bm{\eta})&=\blkdiag\big(\bmm{Z}_E^{(1)}(\bm{\eta}_1),\ldots,\bmm{Z}_E^{(Q)}(\bm{\eta}_Q)\big),\\
\bmm{Z}_E^{(q)}(\bm{\eta}_q)&=\begin{bmatrix}\bmm{X}^{(q)}_{1,1}(\bm{\eta}_q)&\bmm{X}^{(q)}_{1,2}(\bm{\eta}_q)\\ \bmm{X}^{(q)}_{2,1}(\bm{\eta}_q)&\bmm{X}^{(q)}_{2,2}(\bm{\eta}_q)\end{bmatrix}\in\bC^{2M\times 2M},
\end{aligned}
\label{eq:ze_band}
\end{equation}
with $\bm{\eta}=[\bm{\eta}_1^T,\ldots,\bm{\eta}_Q^T]^T$. Within a pair, $\bmm{X}^{(q)}_{1,2}$ is diagonal for element-wise through-connections and full for BD-type interconnections~\cite{nerini2024physically,kim2026bdsim}. When the inter-pair blocks $\bmm{Z}^{(q)}_{1,2}$ are approximated by the Rayleigh--Sommerfeld coefficients~\eqref{eq:RS}, the model reproduces the geometry of Section~\ref{sec:arch}; when they are computed by full-wave simulation or measured, it captures the actual coupling, which is how~\cite{pettanice2026modeling} validates the multiport model against full-wave results for communication and sensing.

\subsection{Connection Among $Z$, $S$, and $T$ Representations}
\label{sec:ZST}

Equivalent descriptions exist in terms of scattering ($S$) parameters~\cite{nerini2024physically} and transfer-scattering ($T$) parameters~\cite{yahya2025t}; they are related by invertible transformations for a fixed reference impedance $Z_0$ and differ in what is natural to measure, cascade, and optimize. For a block with impedance matrix $\bmm{Z}$, $\bmm{S}=(\bmm{Z}+Z_0\bmm{I})^{-1}(\bmm{Z}-Z_0\bmm{I})$. In~\cite{nerini2024physically}, each layer and inter-layer medium is described by a local $S$-matrix and the overall response is obtained by cascading multiport blocks with elimination of internal ports; the formulation is measurement-native and exposes reflection and transmission at each interface. The $T$-matrix relates input-side to output-side waves, $[\bmm{b}_{\rm in};\bmm{a}_{\rm in}]=\bmm{T}[\bmm{a}_{\rm out};\bmm{b}_{\rm out}]$, and follows from the block partition of $\bmm{S}$ as~\cite{yahya2025t}
\begin{equation}
\begin{bmatrix}\bmm{T}_{11}&\bmm{T}_{12}\\ \bmm{T}_{21}&\bmm{T}_{22}\end{bmatrix}=
\begin{bmatrix}\bmm{S}_{12}-\bmm{S}_{11}\bmm{S}_{21}^{-1}\bmm{S}_{22}&\bmm{S}_{11}\bmm{S}_{21}^{-1}\\ -\bmm{S}_{21}^{-1}\bmm{S}_{22}&\bmm{S}_{21}^{-1}\end{bmatrix};
\label{eq:S_to_T}
\end{equation}
cascading is then a plain product, $\bmm{T}=\bmm{T}_{L}\bmm{T}_{d}\bmm{T}_{L-1}\cdots\bmm{T}_{d}\bmm{T}_{1}$ (alternating layer and inter-layer blocks), followed by one terminal inversion. This keeps the multiplicative structure of~\eqref{cascaded_model} while retaining backward waves and reflections.

\begin{table*}[t]
\centering
\caption{Comparison of the $Z$-, $S$-, and $T$-parameter approaches to SIM modeling (~\cite{abrardo2025novel,pettanice2026modeling}, \cite{nerini2024physically}, \cite{yahya2025t}).}
\label{ZST_comparison}
\renewcommand{\arraystretch}{1.05}
\setlength{\tabcolsep}{2pt}
\scriptsize
\begin{tabularx}{\textwidth}{|p{2.0cm}|X|X|X|}
\hline
\rowcolor{SIMHeader}\textbf{Aspect} & \textbf{$Z$-parameters (impedance)} & \textbf{$S$-parameters (scattering)} & \textbf{$T$-parameters (transfer scattering)} \\
\hline
Primary viewpoint & Circuit-complete multiport description preserving self/mutual coupling and inter-layer feedback & Wave-based reflection/transmission description aligned with RF characterization & Wave-based transfer description tailored to cascaded stacks \\
\hline
Key strength & Highest fidelity; validated against full-wave simulations~\cite{pettanice2026modeling}; avoids assumptions that cause model mismatch in optimization & Measurement-native; convenient for interfaces and per-layer reflection/transmission & Cascade-friendly: stack response is a product of per-layer and inter-layer blocks \\
\hline
Cascading across layers & Not inherently multiplicative; global model assembled from banded blocks~\eqref{eq:zee_band}--\eqref{eq:ze_band} & Repeated multiport cascading with internal-port elimination; nested operations & Linear matrix products; no nested operators \\
\hline
Computational bottleneck & $\mathcal{O}(L^3M^3)$ if structure is ignored; $\mathcal{O}(LM^3)$ with the block-banded structure (Section~\ref{sec:complexity}) & Number of $M\times M$ inversions grows with $L$ through nesting & Multiplication-dominated, $\mathcal{O}(L(2M)^3)$, plus one terminal inversion \\
\hline
Optimization tractability & Adjoint gradient of the corner block $\bmm{\Psi}_{L,1}$; structure-exploiting iterations & Variables entangled inside nested inverses & Well suited to gradient-based optimization over many layers \\
\hline
Physical constraints & Enforced by the complete network model & Lossless/reciprocal conditions on per-layer $\bmm{S}$ & Explicit lossless/reciprocal constraints in the transfer domain \\
\hline
Main limitation & Requires the coupling blocks (Rayleigh--Sommerfeld, full-wave, or measurement) & Poor scalability with $L$ & Best suited to balanced multiports \\
\hline
\end{tabularx}
\end{table*}

Table~\ref{ZST_comparison} summarizes the trade-offs. The cascaded response~\eqref{cascaded_model} is recovered from any of the three physically consistent models under four assumptions: (i) unilateral propagation between transmitter and first layer and between last layer and receiver (no backward waves); (ii) perfect matching of transmitter and receiver to $Z_0$; (iii) negligible mutual coupling among feeds, among receive antennas, and among first-layer elements; and (iv) diagonal modulation matrices and forward-only inter-layer propagation. Under (i)--(iv), the corner block $\bmm{\Psi}_{L,1}$ factorizes into layer-wise load terms and inter-layer coupling terms, which is~\eqref{cascaded_model}.

\subsection{Modeling Hierarchy and Validity Regimes}
\label{sec:validity}

\begin{figure}[t]
\centering
\resizebox{0.9\columnwidth}{!}{%
\begin{tikzpicture}[font=\scriptsize, >=Latex,
  lvl/.style={draw=black!60, rounded corners=1mm, fill=#1, text width=4.5cm, align=center, inner sep=2.5pt, minimum height=7mm},
  asm/.style={draw=phframe!70, dashed, rounded corners=1mm, fill=white, text width=3.15cm, align=left, inner sep=2pt, font=\tiny}]
\node[lvl=pastelpurple] (fw) {\textbf{Full-wave EM} (MoM/FDTD/FEM)\\ $N_{\rm unk}\gg LM$ unknowns; offline only};
\node[lvl=pastelblue, below=8.5mm of fw] (z) {\textbf{Multiport $Z$ / $S$ / $T$ models}~\cite{abrardo2025novel,nerini2024physically,yahya2025t}\\ coupling, reflections, loss, matching retained};
\node[lvl=SIMBody, below=8.5mm of z] (casc) {\textbf{Cascaded operator}~\eqref{cascaded_model}~\cite{an2023stacked}\\ forward-only diffraction, diagonal layers};
\node[lvl=SIMFirstCol, below=8.5mm of casc] (ideal) {\textbf{Ideal phase-only model}\\ $|\phi_m^{(l)}|=1$, continuous $\theta$, lossless, calibrated};
\draw[->, thick] (fw) -- (z) node[midway, right=1mm, asm] {coupling blocks from Rayleigh--Sommerfeld~\eqref{eq:RS}, full-wave, or measurement};
\draw[->, thick] (z) -- (casc) node[midway, right=1mm, asm] {assumptions (i)--(iv): no backward waves, matched ports, weak coupling, diagonal layers};
\draw[->, thick] (casc) -- (ideal) node[midway, right=1mm, asm] {no insertion loss, no amplitude--phase coupling, infinite resolution, perfect geometry};
\node[left=1.5mm of fw, align=right, font=\tiny, text width=1.55cm] {fidelity $\uparrow$\\ cost $\uparrow$};
\node[left=1.5mm of ideal, align=right, font=\tiny, text width=1.55cm] {tractability $\uparrow$\\ optimism $\uparrow$};
\end{tikzpicture}}
\caption{Modeling hierarchy of SIMs. Each downward step drops physical effects (right) in exchange for tractability; results obtained at a lower level should be read with the dropped effects in mind.}
\label{fig:model_hierarchy}
\end{figure}

Fig.~\ref{fig:model_hierarchy} organizes the models into a hierarchy in which each downward step removes physical effects and gains tractability. The cascaded operator is accurate when inter-layer propagation is predominantly forward, ports are matched, intra-layer coupling is weak or absorbed into effective coefficients, and each layer is well approximated by a diagonal matrix, i.e., roughly for spacings of the order of a wavelength or more, sub-wavelength pitch, and weakly resonant cells. It fails, in ways that change trends and not only numbers, in four documented regimes: \emph{closely spaced stacks} ($d\lesssim\lambda/2$), where backward waves and multiple reflections matter, as quantified by the multiport model and its full-wave validation~\cite{abrardo2025novel,pettanice2026modeling}, by the SNR loss incurred when coupling is ignored in active stacks~\cite{mandalika2026active}, and by the 13-dB measured dependence of received power on spacing between $\lambda/4$ and $\lambda$ in a three-layer 5.8-GHz prototype~\cite{wang2024multi}; \emph{strongly resonant or lossy cells}, where a Lorentzian element model extended to multiple layers through generalized sheet transition conditions shows that the ideal model overestimates the sum rate of deep stacks by up to 7.5~bit/s/Hz and reverses the depth-versus-aperture preference~\cite{gonzalez2026lorentzian}; \emph{near-field geometry}, where coupling-aware hop matrices make the sum rate non-monotonic in $L$ and in thickness~\cite{dong2026geometry} and plane-wave-spectrum models control the phase error of the scalar approximation~\eqref{eq:RS}~\cite{zhang2026eit}; and \emph{wideband, nonlinear, or active operation}, where layer responses become frequency dependent (Section~\ref{sec:perf_wideband}), nonlinear terminations require a fixed-point solution~\cite{abrardo2026nonlinear}, and unilateral active cells lead to a feed-forward recursive cascade that is again cheap to differentiate~\cite{abrardo2026cascade}. The rule applied when reading the literature in Sections~\ref{sec:taxonomy}--\ref{sec:performance} is that the cascaded model is adequate for algorithmic exploration and for comparisons under identical assumptions, whereas hardware-accurate prediction requires a multiport, full-wave, or measurement-calibrated model. Many reported gains were obtained at the bottom of Fig.~\ref{fig:model_hierarchy}; they identify potential, not deliverable performance.

\subsection{Computational Complexity of the Models}
\label{sec:complexity}

A frequent objection to physically consistent models is their cost. Table~\ref{tab:complexity} compares one forward evaluation and one gradient with respect to all $LM$ parameters. The cascaded model costs $\mathcal{O}(LM^2S)$ (each stage multiplies an $M\times M$ dense matrix by an $M\times S$ matrix); its gradient by reverse-mode differentiation costs about twice as much. The impedance model needs the corner block $\bmm{\Psi}_{L,1}$ of the inverse of the $LM\times LM$ matrix $\bmm{Z}_{EE}+\bmm{Z}_E$: $\mathcal{O}(L^3M^3)$ if structure is ignored, but $\mathcal{O}(LM^3)$ with a block forward-elimination (block Thomas) recursion over the block-tridiagonal structure of~\eqref{eq:zee_band}--\eqref{eq:ze_band}, with the gradient at the same order from the adjoint recursion~\cite{abrardo2025novel,abrardo2026nonlinear}. The $S$-parameter cascade eliminates internal ports one interface at a time with an $M\times M$ inversion each, but nested formulations multiply the inversions with $L$ and entangle the variables inside nested inverses~\cite{yahya2025t}. The $T$-parameter cascade replaces nesting with products of $2M\times2M$ blocks, $\mathcal{O}(L(2M)^3)$, plus one inversion, with a gradient of the same structure as that of the cascaded model. Full-wave solvers act on $N_{\rm unk}\gg LM$ unknowns and are used offline to obtain or validate the coupling blocks~\cite{pettanice2026modeling}.

\begin{table}[t]
\centering
\caption{Complexity of one forward evaluation and one gradient (all $LM$ parameters); numerical example for $M=100$, $S=4$ (complex multiplications).}
\label{tab:complexity}
\renewcommand{\arraystretch}{1.05}
\setlength{\tabcolsep}{2.5pt}
\scriptsize
\begin{tabularx}{\linewidth}{|>{\raggedright\arraybackslash}p{1.75cm}|>{\centering\arraybackslash}p{1.75cm}|>{\centering\arraybackslash}p{1.05cm}|>{\centering\arraybackslash}X|>{\centering\arraybackslash}X|}
\hline
\rowcolor{SIMHeader}\textbf{Model} & \textbf{Forward} & \textbf{Gradient} & $\mathbf{L=4}$ & $\mathbf{L=7}$ \\
\hline
Cascaded~\eqref{cascaded_model} & $\mathcal{O}(LM^2S)$ & $\approx 2\times$ & $1.6\!\times\!10^5$ & $2.8\!\times\!10^5$ \\
\hline
$Z$, dense inversion & $\mathcal{O}(L^3M^3)$ & $\approx 2\times$ & $6.4\!\times\!10^7$ & $3.4\!\times\!10^8$ \\
\hline
$Z$, block-banded & $\mathcal{O}(LM^3)$ & same order (adjoint) & $4\!\times\!10^6$ & $7\!\times\!10^6$ \\
\hline
$S$, nested cascade & $\mathcal{O}(LM^3)$ to $\mathcal{O}(L^2M^3)$ & entangled & $4\!\times\!10^6$--$1.6\!\times\!10^7$ & $7\!\times\!10^6$--$4.9\!\times\!10^7$ \\
\hline
$T$, product cascade & $\mathcal{O}(L(2M)^3)$ & $\approx 2\times$ & $3.2\!\times\!10^7$ & $5.6\!\times\!10^7$ \\
\hline
Full-wave (MoM) & $\mathcal{O}(N_{\rm unk}^{2\text{--}3})$, $N_{\rm unk}\!\gg\! LM$ & n/a & offline & offline \\
\hline
\end{tabularx}
\end{table}

Two observations follow. The structured $Z$ and $T$ models are one to two orders of magnitude more expensive per evaluation than the cascaded model but scale linearly in $L$, so the gap does not widen with depth. And since every SIM optimizer (Section~\ref{sec:model_based}) evaluates the response and its gradient hundreds of times, the structured $Z$ model costs about $10^7$ multiplications per iteration at $M=100$, $L=7$: affordable for design, not for per-coherence-block reconfiguration on embedded controllers. This motivates the hybrid strategy of recent works: calibrate an effective cascaded model (or an effective coupling matrix) offline with the physically consistent model, then optimize the cheap model online~\cite{li2026fullywave,jia2026calibration}. The complexity of the optimizers is discussed in Section~\ref{sec:perf_complexity}.

\subsection{Hardware Non-Ideality and Power-Consumption Models}
\label{sec:nonideal}

To read the literature critically one needs a common vocabulary for the effects that separate the ideal model from hardware; the measured values behind the parameters are given in Section~\ref{sec:hardware}.

\emph{Phase resolution.} Programmable meta-atoms support $\theta_m^{(l)}\in\{0,2\pi/2^b,\ldots\}$ with $b$ bits: typically 1 bit for PIN-diode cells, 2--3 bits or quasi-continuous for multi-diode or varactor cells, quasi-continuous but slow for liquid crystal. Quantization applied after continuous optimization is not benign; inside the optimizer most of the loss is recovered (Section~\ref{sec:perf_quant}).

\emph{Insertion loss and amplitude--phase coupling.} A practical layer multiplies the field by $\alpha_l\bmm{\Phi}_l$ with $\alpha_l<1$ (or by a state-dependent $\diag(\alpha_1^{(l)}e^{j\theta_1^{(l)}},\ldots)$). Measured 1-bit transmissive PIN cells lose 1.2--1.9~dB from X- to Ka-band~\cite{rana2021experimental,huang2023wideband,zhang2023onebit}, so three programmable layers lose 3.6--5.7~dB in the cells alone, before diffraction spillover; static dielectric layers can stay below 1~dB~\cite{shi2026sliding}, and 3D-printed terahertz layers lose about 3~dB each~\cite{lin2018all}. State-dependent amplitude is captured by the Lorentzian response $\phi=(j+e^{j\theta})/2$ and its multilayer extension~\cite{gonzalez2026lorentzian,cao2026element}. Because loss compounds multiplicatively, the end-to-end amplitude scales as $\prod_l\alpha_l$, which bounds the useful depth (Section~\ref{sec:perf_depth}).

\emph{Geometry and calibration errors.} Spacing irregularity, height offsets, and planar rotation perturb~\eqref{eq:RS}. The calibration study of~\cite{jia2026calibration} models them as uniform errors of $\pm0.1\%\lambda$, $\pm1\%\lambda$, and $\pm0.01^\circ$ and reduces the inter-layer NMSE by 9--12~dB with multi-stage pilot-based calibration; in diffractive networks, lateral shifts of a few wavelengths reduce classification accuracy from 98\% to below 40\% unless training includes misalignment augmentation~\cite{mengu2020misalignment}. Millimeter-wave PCB stacks are manufactured to $\pm0.025$--$0.1$~mm~\cite{gao2024superresolution,yang2025metatraining,yu2026movable}.

\emph{Reconfiguration latency.} The reconfiguration time sets the minimum update interval: nanoseconds to microseconds for PIN diodes and RF switches, sub-microsecond for varactors, 15--80~ms for liquid crystal~\cite{neuder2024lcris,abantoleon2024liquiris}, hundreds of milliseconds for panel-level liquid-crystal control~\cite{gao2024superresolution}, minutes for mechanical or manual reconfiguration~\cite{yu2026movable}. A slow layer forces a two-timescale design (Section~\ref{sec:statcsi}).

\emph{Active layers.} Amplifying meta-atoms replace $\alpha_l\le1$ by a gain $g_l$ ($-22$ to $+13$~dB per element in~\cite{liu2022programmable}) at the price of amplifier noise, several hundred milliwatts per element (about 429~mW per nonlinear unit in~\cite{ning2025multilayer}), heat, and stability margins (Section~\ref{sec:thermal}).

\emph{Power-consumption model.} The energy-efficiency literature uses~\cite{shi2025energy,perovic2025energy,wang2024power}
\begin{equation}
\begin{aligned}
P_{\rm tot}=\;&P_{\rm static}+\frac{1}{\eta_{\rm PA}}\sum_{k}\|\bmm{p}_k\|^2+N_{\rm RF}P_{\rm RF}\\
&+LM\,P_{\rm meta}+P_{\rm ctrl}+L_aM\,P_{\rm amp},
\end{aligned}
\label{eq:power}
\end{equation}
where $\bmm{p}_k$ is the digital precoder of user $k$, $N_{\rm RF}$ the number of active RF chains, $P_{\rm meta}$ the per-element bias/control power, $P_{\rm ctrl}$ the controller power, and $L_a$ the number of active layers with per-element amplifier power $P_{\rm amp}$. Representative values are $P_{\rm RF}\approx0.4$~W, $P_{\rm meta}\approx10$~mW, $P_{\rm ctrl}\approx0.3$~W in~\cite{shi2025energy}; measured values for programmable surfaces are 12--12.6~mW per ON PIN diode, 4.8~W per FPGA board, and 15.7~W (all off) to 103~W (all on) for a $60\times60$ dual-polarized 35-GHz panel~\cite{wang2024power}, versus about 21.5~nW per liquid-crystal phase shifter~\cite{abantoleon2024liquiris}. Since the wave-domain gain saturates with $L$ (Section~\ref{sec:perf_depth}) while $LM\,P_{\rm meta}$ and the loss-induced transmit power grow at least linearly, energy efficiency is not monotonic in $L$.

\begin{lessons}
\begin{itemize}[leftmargin=*, nosep]
\item The cascaded operator~\eqref{cascaded_model} is a first-order approximation of the multiport models, valid under assumptions (i)--(iv); adequate for algorithm design and for comparisons under identical assumptions, not for hardware-accurate prediction.
\item Closely spaced, strongly resonant, near-field, wideband, or active/nonlinear stacks require a $Z$/$S$/$T$ or measurement-calibrated model, because these regimes change trends, not only numbers.
\item Structured $Z$ and $T$ models cost $\mathcal{O}(LM^3)$ per evaluation, linear in depth; use them offline to calibrate a cheap effective model for online configuration.
\item Report $b$, $\alpha_l$, the geometry-error model, the latency, and the terms of~\eqref{eq:power} included with every result; otherwise gains cannot be compared across papers.
\end{itemize}
\end{lessons}

\section{A Unified Taxonomy and the Methodological Families of SIM Research}
\label{sec:taxonomy}

\begin{figure*}[t]
\centering
\includegraphics[width=0.84\linewidth]{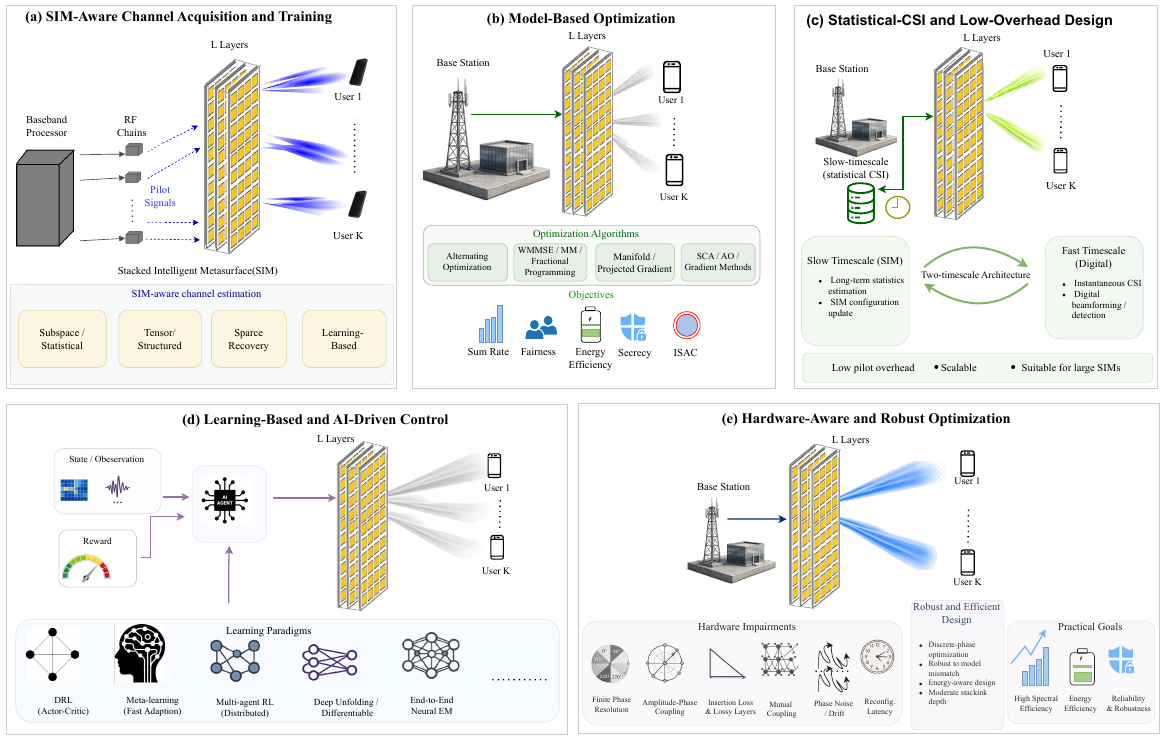}
\caption{The five methodological families: (a) SIM-aware channel acquisition and training; (b) model-based optimization; (c) statistical-CSI and low-overhead design; (d) learning-based and AI-driven control; (e) hardware-aware and robust optimization.}
\label{fig:method_taxonomy}
\end{figure*}

This section overviews two instruments: a unified taxonomy (Fig.~\ref{fig:unified_taxonomy}) relating the five dimensions along which SIM works differ (architecture, EM model, methodological family, physical role, hardware maturity), and a methodology-first classification into five families (Fig.~\ref{fig:method_taxonomy}): SIM-aware channel acquisition and training (Section~\ref{sec:csi}), model-based optimization (Section~\ref{sec:model_based}), statistical-CSI and low-overhead design (Section~\ref{sec:statcsi}), learning-based control (Section~\ref{sec:learning}), and hardware-aware robust design (Section~\ref{sec:hwaware}). The families overlap (statistical designs use model-based solvers, unfolded networks derive from iterative algorithms, robust designs can be either), but each is defined by a dominant assumption, a set of tools, and a characteristic limitation, compared in Table~\ref{tab:method_compare}. Section~\ref{sec:conflicts} reconciles findings that contradict each other across the literature.

\begin{figure*}[t]
\centering
\resizebox{\textwidth}{!}{%
\begin{tikzpicture}[font=\scriptsize, >=Latex,
  band/.style={draw=black!50, rounded corners=1.5mm, fill=#1, minimum height=11mm, inner sep=2pt},
  item/.style={draw=black!55, rounded corners=0.8mm, fill=white, inner sep=2pt, minimum height=5mm, font=\scriptsize, align=center},
  lbl/.style={font=\bfseries\scriptsize, text width=2.1cm, align=right, anchor=east},
  arr/.style={->, thick, black!65}]
\node[lbl] at (-0.2,0) {\textbf{Architecture}\\ (Sec.~\ref{sec:hardware})};
\node[band=pastelpurple, minimum width=17.4cm, anchor=west] (b1) at (0,0) {};
\node[item] (a1) at (1.2,0) {Passive phase-only\\ (PIN/varactor/LC)};
\node[item] (a2) at (3.55,0) {Active /\\ amplifying};
\node[item] (a3) at (5.6,0) {Beyond-diagonal\\ layers};
\node[item] (a4) at (7.7,0) {Two-layer: meta-fiber,\\ FILM};
\node[item] (a5) at (9.9,0) {Delay-augmented /\\ dual-polarized};
\node[item] (a6) at (12.1,0) {Nonlinear /\\ neural};
\node[item] (a7) at (14.1,0) {Flexible, fluid,\\ spherical, STAR};
\node[item] (a8) at (16.2,0) {Static (D$^2$NN)\\ 3D-printed};
\node[lbl] at (-0.2,-1.35) {\textbf{EM model}\\ (Sec.~\ref{sec:fund})};
\node[band=pastelblue, minimum width=17.4cm, anchor=west] (b2) at (0,-1.35) {};
\node[item] (m1) at (2.0,-1.35) {Ideal phase-only};
\node[item] (m2) at (5.3,-1.35) {Cascaded diffraction~\eqref{cascaded_model}\\ + loss / quantization / Lorentzian};
\node[item] (m3) at (9.3,-1.35) {Multiport $Z$ / $S$ / $T$\\ (coupling, reflections, matching)};
\node[item] (m4) at (12.7,-1.35) {Fixed-point / recursive\\ (nonlinear, unilateral active)};
\node[item] (m5) at (15.8,-1.35) {Full-wave /\\ measurement-calibrated};
\node[lbl] at (-0.2,-2.7) {\textbf{Methodology}\\ (Sec.~\ref{sec:taxonomy})};
\node[band=SIMBody, minimum width=17.4cm, anchor=west] (b3) at (0,-2.7) {};
\node[item] (f1) at (1.7,-2.7) {SIM-aware CSI\\ acquisition \& training};
\node[item] (f2) at (5.1,-2.7) {Model-based \\optimization (AO, PGA, \\WMMSE, MM, FP, manifold, QUBO)};
\node[item] (f3) at (8.7,-2.7) {Statistical-CSI /\\ two-timescale};
\node[item] (f4) at (11.9,-2.7) {Learning-based control\\ (RL, unfolding, EM-NN)};
\node[item] (f5) at (15.3,-2.7) {Hardware-aware /\\ robust design};
\node[lbl] at (-0.2,-4.05) {\textbf{Physical role}\\ (Sec.~\ref{sec:apps})};
\node[band=SIMFirstCol, minimum width=17.4cm, anchor=west] (b4) at (0,-4.05) {};
\node[item] (r1) at (2.1,-4.05) {Wave-domain precoder /\\ combiner (MU-MIMO, HMIMO, CF)};
\node[item] (r2) at (5.5,-4.05) {Near-field focusing /\\ pattern synthesis};
\node[item] (r3) at (8.6,-4.05) {Space--time / wideband\\ processor (OFDM, DD)};
\node[item] (r4) at (11.9,-4.05) {Multifunctional aperture\\ (ISAC, DoA, imaging, PLS)};
\node[item] (r5) at (15.3,-4.05) {Lightweight analog front end\\ (NTN, UAV, semantic, FD)};
\node[lbl] at (-0.2,-5.4) {\textbf{Hardware maturity}\\ (Table~\ref{tab:maturity})};
\node[band=pastelgray, minimum width=17.4cm, anchor=west] (b5) at (0,-5.4) {};
\node[item, fill=green!12] (h1) at (3.5,-5.4) {Experimentally validated: static D$^2$NN stacks; 1-bit programmable \\passive (5.8 GHz); programmable-active (5.4 GHz); \\nonlinear ReLU layers (5.8 GHz)};
\node[item, fill=yellow!15] (h2) at (9.6,-5.4) {Full-wave / multiport validated:\\ $Z$-model stacks, unilateral active cascades,\\ unit-cell designs};
\node[item, fill=red!8] (h3) at (14.9,-5.4) {Conceptual (simulation only): delay-augmented,\\ dual-pol, meta-fiber, FILM, \\fluid, spherical, STAR-SIM, BD-SIM};
\draw[arr] (b1.south) ++(-6.5,0) -- ++(0,-0.2) node[right, font=\scriptsize] {is described by};
\draw[arr] (b2.south) ++(-6.5,0) -- ++(0,-0.2) node[right, font=\scriptsize] {is configured through};
\draw[arr] (b3.south) ++(-6.5,0) -- ++(0,-0.2) node[right, font=\scriptsize] {serves as};
\draw[arr] (b4.south) ++(-6.5,0) -- ++(0,-0.2) node[right, font=\scriptsize] {rests on hardware of maturity};
\end{tikzpicture}}
\caption{Unified taxonomy of SIM research. Any SIM paper is located by one item in each band; reading a claim requires all five coordinates, and Section~\ref{sec:conflicts} shows that most apparent contradictions in the literature arise from works that differ in the second or fifth one.}
\label{fig:unified_taxonomy}
\end{figure*}

\subsection{SIM-Aware Channel Acquisition and Training}
\label{sec:csi}

CSI acquisition in SIM systems differs from array training in one respect: the effective channel $\bmm{H}\bmm{P}$ is partly created by the device being trained. Pilot observations depend on $\bmm{W}_1$, on the $\bmm{W}_l$, on the training states $\{\bmm{\Phi}_l\}$, and on $\bmm{H}$; $S$ is small; and the layers reconfigure at the latency of their tuning technology. Five directions have emerged: \emph{1) Subspace and statistical estimation:} pilots collected across several SIM training configurations are processed by a subspace estimator whose training phases are optimized to minimize the MSE rather than drawn at random~\cite{yao2024channel}, which establishes the SIM as an active part of the sensing mechanism; MMSE estimators exploit a deterministic LoS component under Rician fading~\cite{papazafeiropoulos2025channel}. \emph{2) Learning-assisted recovery:} FlatCE-Net refines coarse LS estimates with dilated convolutions that exploit the correlation induced by multilayer coupling~\cite{dong2026deep}, and a one-dimensional convolutional estimator maps pilots to the effective channel in RF-chain-limited holographic MIMO~\cite{lawal2025channel}; both depend on the representativeness of the training set. \emph{3) Structured hybrid and tensor estimation:} hybrid wave-domain/digital estimation uses optimized SIM states followed by baseband refinement~\cite{an2024hybrid}, nested tensor decompositions exploit the multiway structure across pilots, layers, users, and receive dimensions~\cite{ginige2025nested}, and TenSIM~\cite{dealmeida2026tensim} shows that training only the central layers turns the cascade into a PARAFAC (odd $L$) or known-core Tucker (even $L$) model, so that alternating least squares needs about $N$ pilot blocks instead of the $N^2$ of an unstructured aggregate LS estimator (Section~\ref{sec:perf_csi}). \emph{4) Sparse and near-field estimation:} angular-domain compressed sensing~\cite{yao2024sparse} is replaced, for large apertures where users lie in the Fresnel region, by polar-domain sparse Bayesian learning capturing angle and range sparsity~\cite{yao2025sparse,yao2026sparse}, whose low-complexity variant is four times faster at equal accuracy~\cite{yao2026sparse}; near-field beam training by matched-filter search over distance and angles~\cite{wei2026nearfield}, hierarchical codebooks whose layers cover the full angular range to avoid beam aliasing~\cite{yu2026hierarchical}, code-aided iterative estimation for SIM and BD-SIM holographic MIMO~\cite{porto2026code}, and multiport-consistent projections built on coarse location priors~\cite{abrardo2026localization} reduce overhead when explicit estimation is unaffordable. \emph{5) Dynamics and overhead accounting:} channel aging degrades SIM-assisted cell-free systems under mobility, with some robustness from LoS-dominated wave-domain control~\cite{ding2025channel,wang2026aging,wang2026uplink}, and the training--throughput analysis of~\cite{bahanshal2026training} shows that the number of training sub-phases must be optimized for net rate, with an optimum that depends on spatial correlation.

\textbf{Critical synthesis:} Accuracy improves when the multilayer structure is exploited through subspace, tensor, sparse, statistical, or learned priors, and the strongest recent results obtain quadratic pilot reductions from the structure of~\eqref{cascaded_model} rather than from more pilots. Almost all works, however, assume that the SIM response in each training state is exactly known and stable; the calibration results of~\cite{jia2026calibration} (9--12~dB NMSE improvement of the inter-layer coefficients) show that the device response must be estimated too. A mature CSI framework will estimate the channel and the device jointly and report net rate after training rather than NMSE.

\subsection{Model-Based Optimization}
\label{sec:model_based}

Model-based optimization is the backbone of SIM research: digital variables (precoders, powers, covariances, subcarrier or antenna assignments) are coupled with $LM$ phases entering through the product~\eqref{cascaded_model}, so the problem is non-convex before unit-modulus, discrete-phase, power, sensing, fairness, or secrecy constraints are added. Four patterns dominate it: \emph{1) Alternating optimization (AO) with projected gradient} updates digital and wave-domain blocks in turn, with per-layer projected-gradient ascent onto the unit circle, as in the statistical-CSI rate design~\cite{papazafeiropoulos2024achievable}, the joint transmit--receive design~\cite{papazafeiropoulos2024achievable_2}, the refined AO for strongly coupled regimes~\cite{bahingayi2025refined}, and the ISAC, secrecy, energy-efficiency, and cell-free designs~\cite{niu2024stacked,niu2024enhancing,shi2025energy,shi2024harnessing}; closed-form per-layer updates replace the gradient in~\cite{park2026efficient}, reducing runtime by more than 99\% at the same sum rate. \emph{2) Surrogate-based methods:} weighted MMSE (WMMSE) for sum-rate and max--min objectives~\cite{bahingayi2025scaling,ginige2025max}; majorization--minimization (MM), fractional programming, and quadratic transforms for ratio objectives~\cite{shi2025energy,perovic2025energy,yu2025energy,sheemar2026ee}; Riemannian optimization on the product of unit circles~\cite{bahingayi2025scaling,li2025stacked_NF,li2024stacked_analog}; layer-wise Takagi-decomposition designs realizing regularized zero forcing with BD layers using fewer layers than diagonal stacks~\cite{kim2026bdsim}; and a unified AO for fully connected, group-connected, and diagonal layers in ISAC~\cite{jiao2026bdsim}. \emph{3) Discrete and combinatorial solvers:} projection schedules~\cite{an2025stacked,nassirpour2025sum,jiao2026efficient}, quantization-aware projected gradients~\cite{hassan2026robust}, and quadratic unconstrained binary formulations solved by quantum-inspired simulated bifurcation~\cite{ji2026qubo,ji2026hmimo}, which reach near-continuous performance with 3-bit phases. \emph{4) Analytical characterization:} closed-form or asymptotic mutual information, outage, and ergodic capacity~\cite{papazafeiropoulos2025ergodic_outage_3,papazafeiropoulos2025ergodic_HMIMO,papazafeiropoulos2025performance,perovic2024mutual,papazafeiropoulos2026fas} give scaling laws in $L$, $M$, and the fading statistics at the price of simplified channel and hardware models.

\textbf{Critical synthesis:} The literature has converged on AO with per-layer gradient or closed-form updates and on WMMSE/MM/fractional-programming surrogates. All methods reach stationary points, are initialization sensitive (random restarts are standard), and mostly report utilities under the ideal model. Three practices distinguish the more useful studies: reporting per-iteration complexity and wall-clock time~\cite{an2025stacked,park2026efficient,wang2026unfolding}; putting discreteness and loss inside the optimizer~\cite{an2025stacked,rostamikafaki2026loss,hassan2026robust}; and comparing against fully digital and hybrid baselines with equal RF chains and power model~\cite{shi2025energy,ibrahim2026wideband}. Whether a deeper stack is better cannot be answered without the last two (Section~\ref{sec:conflicts}).

\subsection{Statistical-CSI and Low-Overhead Design}
\label{sec:statcsi}

Instantaneous CSI becomes unaffordable as $M$ and $L$ grow, and slow tuning technologies cannot follow small-scale fading anyway. Statistical designs configure the SIM on the timescale of the large-scale statistics and leave short-term adaptation to the digital part: deterministic equivalents from random matrix theory make ergodic-rate objectives tractable~\cite{xia2025statistical}; two-timescale control saves fronthaul in cell-free systems~\cite{shi2024harnessing,shi2024uplink} and is unavoidable in satellite systems~\cite{lin2024stacked,zhang2026satellite}; TD3-based statistical-CSI control stays competitive with instantaneous-CSI optimization for large SIMs at a fraction of the overhead~\cite{hu2026td3}; and closed-form outage analyses give the objective for stacks with fluid-antenna receivers~\cite{papazafeiropoulos2026fas}. The limitation is stationarity: mobility, blockage, near-field range variation, and frequency selectivity violate it, and a configuration optimized on average may fail for cell-edge users or under sudden LoS blockage.

\textbf{Critical synthesis:} Statistical designs are necessary for scalability and match slow tuning technologies, but do not replace instantaneous adaptation. The literature is converging on a hierarchical architecture: a slowly updated SIM shaped by statistics and geometry, fast digital updates based on local CSI, and occasional SIM refinement when the statistics drift, which is also the natural entry point for learned controllers.

\subsection{Learning-Based and AI-Driven Control}
\label{sec:learning}

Learning addresses fast inference-time configuration in dynamic environments and task-oriented objectives that are hard to express analytically. The methods developed as: \emph{1) Reinforcement learning (RL):} deep RL for multi-user downlink~\cite{liu2024drl}, TD3 for continuous phase and power control~\cite{yang2024joint,hu2026td3}, meta-learning for fast adaptation~\cite{mohammadzadeh2024meta,yang2025low}, and multi-agent RL for distributed cell-free orchestration~\cite{zhu2025joint}. \emph{2) Deep unfolding} unrolls projected-gradient iterations with learned step sizes, for wideband OFDM with index modulation~\cite{li2025stacked_OFDM_IM}, multiband downlink with one configuration for all subcarriers~\cite{ibrahim2026unfolding}, joint antenna selection, phase design, and power allocation~\cite{wang2026unfolding}, and holographic MIMO~\cite{xue2026unfolded}; unfolded networks match high-iteration benchmarks at a fraction of the runtime and are reported robust to model mismatch and imperfect CSI~\cite{wang2026unfolding}. \emph{3) Electromagnetic neural networks (EM-NNs)} treat the coefficients as trainable complex parameters~\cite{zayat2025deep} and train stacks end to end for OFDM~\cite{zhang2025stacked_E2E_OFDM}, over-the-air inference~\cite{stylianopoulos2025over,stylianopoulos2026over,stylianopoulos2025integrating,stylianopoulos2026metasurfaces}, extreme learning machines with a fixed nonlinear front layer~\cite{stylianopoulos2026over}, DoA estimation from amplitude-only measurements~\cite{lin2026emnn}, functional coding with unfolded detection~\cite{chen2026functional}, and all-analog semantic mapping~\cite{hou2026semantic,choi2026jscc}; the learnable-SIM paradigm is reviewed in~\cite{wang2026learnable}.
\begin{table*}[t]
\centering
\caption{Comparison of the methodological families ($I$ iterations, $N_{\rm p}$ pilot blocks; complexity per configuration of an $L$-layer, $M$-element SIM serving $K$ users under the cascaded model).}
\label{tab:method_compare}
\renewcommand{\arraystretch}{1.05}
\setlength{\tabcolsep}{2pt}
\scriptsize
\arrayrulecolor{SIMLine}
\begin{tabularx}{\textwidth}{|>{\columncolor{pastelpurple}\raggedright\arraybackslash}p{2.0cm}|>{\raggedright\arraybackslash}p{2.3cm}|>{\raggedright\arraybackslash}p{2.1cm}|>{\raggedright\arraybackslash}p{2.0cm}|>{\raggedright\arraybackslash}p{1.9cm}|>{\raggedright\arraybackslash}X|>{\raggedright\arraybackslash}X|}
\hline
\rowcolor{pastelblue}\textbf{Family} & \textbf{CSI requirement} & \textbf{Online complexity} & \textbf{Convergence / guarantee} & \textbf{Adaptation speed} & \textbf{Hardware constraints handled} & \textbf{Characteristic limitation} \\
\hline
\textbf{SIM-aware CSI acquisition} \cite{yao2024channel,an2024hybrid,ginige2025nested,dealmeida2026tensim,yao2026sparse,bahanshal2026training} & Produces CSI; $N_{\rm p}\ge N$ (tensor) vs.\ $N^2$ (aggregate LS) pilot blocks & $\mathcal{O}(I\,N_{\rm p}M^2)$ (ALS/SBL); CNN inference $\mathcal{O}(M^2)$ & ALS: monotone decrease; SBL: local; CNN: none & Per coherence block; limited by layer reconfiguration latency & Known training states assumed; calibration errors rarely modeled & Assumes the SIM response in each training state is exactly known and stable \\
\hline
\textbf{Model-based optimization} \cite{papazafeiropoulos2024achievable,bahingayi2025refined,an2025stacked,park2026efficient,ji2026qubo,kim2026bdsim} & Instantaneous (or statistical) CSI & $\mathcal{O}(I\,LM^2K)$ (per-layer PGA); closed-form per-layer updates~\cite{park2026efficient}; QUBO solvers heavier & Stationary point; initialization dependent & Tens to hundreds of iterations per block; $\gg$ coherence time for large $M$ & Unit modulus, discrete phases, power, loss (when included) & Results under the ideal model; complexity and runtime often unreported \\
\hline
\textbf{Statistical-CSI / two-timescale} \cite{papazafeiropoulos2024achievable,xia2025statistical,shi2024harnessing,lin2024stacked,hu2026td3} & Long-term statistics only & Same solvers, once per large-scale coherence interval & Stationary point of the deterministic equivalent & Slow (seconds to minutes); matches LC/mechanical tuning & Slow reconfiguration; fronthaul limits & Fails under non-stationarity, blockage, near-field range variation, frequency selectivity \\
\hline
\textbf{Learning-based control} \cite{liu2024drl,yang2024joint,zhu2025joint,li2025stacked_OFDM_IM,ibrahim2026unfolding,wang2026unfolding,zayat2025deep,stylianopoulos2026over} & Instantaneous CSI (unfolding), state observations (RL), or raw data (EM-NN) & Fixed: one forward pass, $\mathcal{O}(LM^2K)$ per unfolded layer; RL policy inference & None in general; unfolding inherits the base algorithm's structure & Milliseconds or less at inference; hours of offline training & Quantization by projection layers; passivity only if built in & Training cost unreported; simulation-to-hardware transfer; unsafe exploration \\
\hline
\textbf{Hardware-aware / robust} \cite{an2025stacked,hassan2026robust,rostamikafaki2026loss,fang2026hwi,quran2026robust,darsena2025design,shen2026active,dong2026geometry} & Instantaneous or imperfect CSI with error model & As model-based; sample-average robust variants $\times$ samples & Stationary point of the robust surrogate & As model-based & $b$, $\alpha_l$, amplitude--phase coupling, phase noise, DAC resolution, active-layer noise/power, geometry & Requires impairment models and calibration data that prototypes have not standardized \\
\hline
\end{tabularx}
\arrayrulecolor{black}
\end{table*}
\emph{Training overhead:} A fair treatment must include the cost of training, which SIM-for-communication papers rarely report. Diffractive-network prototypes give the order of magnitude: 55\,000 images and about 8~h on one GPU for a five-layer terahertz classifier~\cite{lin2018all}, about 80~h for a spectrally encoded three-layer network~\cite{li2021spectrally}, $10^4$ samples per task over 100 epochs for a millimeter-wave DoA network~\cite{gao2024superresolution}, and minutes of in-situ fine-tuning with 2\% of the data after hours of pre-training for a reconfigurable optical processor~\cite{zhou2021reconfigurable_dpu}. SIM-for-communication works report 250 episodes for multi-agent RL~\cite{zhu2025joint} and tens of thousands of images for semantic tasks~\cite{huang2024stacked}, but no wall-clock comparison with iterative optimizers; the unfolding papers state that inference has fixed complexity~\cite{ibrahim2026unfolding,wang2026unfolding}, which is the relevant advantage once training is amortized. For slow hardware, reconfiguration (20--70~min of manual work in~\cite{yu2026movable}) rather than retraining (12--15~s) dominates.

\textbf{Critical synthesis:} Learned controllers trade an offline cost that is rarely reported for a small, predictable online cost. Their open problems are simulation-to-hardware transfer (a policy trained on the ideal model inherits its optimism) and constraint satisfaction (passivity, reciprocity, quantization, actuator latency). Physics-informed learning, with the structure of~\eqref{cascaded_model} or of the multiport model built into the architecture and measurement-calibrated data for fine-tuning, is the most promising direction; the in-wave computation case study of~\cite{liu2026inwave} (Section~\ref{sec:perf_measured}) shows the accuracy still to be recovered.

\subsection{Hardware-Aware and Robust Optimization}
\label{sec:hwaware}

Hardware-aware optimization puts the non-idealities of Section~\ref{sec:nonideal} inside the design, as discussed in the following. \emph{1) Discrete control:} low-resolution phases approach continuous performance when the discrete set is part of the optimizer (loss below 0.5~bit/s/Hz at 2 bits~\cite{an2025stacked}; near-continuous 3-bit designs~\cite{ji2026hmimo}); statistical-CSI discrete designs~\cite{jiao2026efficient} and quantized control~\cite{nassirpour2025sum}. \emph{2) Impairments:} receive combining under phase noise and amplitude distortion~\cite{rezvani2025uplink}, near-field design under phase errors~\cite{li2025stacked_NF}, transceiver impairments (finite-resolution DACs, amplifier nonlinearity, I/Q imbalance)~\cite{fang2026hwi}, robust max--min fairness~\cite{quran2026robust}, sample-driven robust WMMSE with quantization-aware projection under imperfect CSI~\cite{hassan2026robust}, and loss-aware finite-blocklength beamforming~\cite{rostamikafaki2026loss} show that impairments saturate performance at high SNR and shift the optimal depth. \emph{3) Active and hybrid stacks} compensate loss and enlarge the amplitude--phase response~\cite{darsena2025design,darsena2024downlink,shen2026active,jiang2026doa,mandalika2026active,abrardo2026cascade} at the cost of noise, power, heat, and stability (Section~\ref{sec:thermal}). \emph{4) Geometry:} optimized inter-layer spacings beat uniform spacing in the near field~\cite{dong2026geometry}, and fluid-element layers optimize positions jointly with phases~\cite{wei2026fsim}.

\textbf{Critical synthesis:} Hardware-aware design should be the default. A study should state $b$, $\alpha_l$, the amplitude--phase model, the geometry-error model, the latency, and the terms of~\eqref{eq:power} it includes, and test robustness to phase errors, amplitude mismatch, coupling deviations, and imperfect calibration. Studies that do so find a moderately deep optimum (Section~\ref{sec:perf_depth}); studies that do not find that deeper is always better.

\subsection{Reconciling Conflicting Findings}
\label{sec:conflicts}

Different SIM papers reach different conclusions on the same question. Table~\ref{tab:conflicts} collects six such questions, the positions found in the literature, and their reconciliation; in every case the disagreement dissolves once the EM model, the hardware assumptions, or the baseline are aligned, which is the practical value of the taxonomy of Fig.~\ref{fig:unified_taxonomy}.

\begin{table*}[t]
\centering
\caption{Conflicting findings in the SIM literature and their reconciliation.}
\label{tab:conflicts}
\renewcommand{\arraystretch}{1.05}
\setlength{\tabcolsep}{2.5pt}
\scriptsize
\arrayrulecolor{SIMLine}
\begin{tabularx}{\textwidth}{|>{\columncolor{pastelpurple}\raggedright\arraybackslash}p{2.2cm}|>{\raggedright\arraybackslash}X|>{\raggedright\arraybackslash}X|>{\raggedright\arraybackslash}X|}
\hline
\rowcolor{pastelblue}\textbf{Question} & \textbf{Position A (assumptions)} & \textbf{Position B (assumptions)} & \textbf{Reconciliation} \\
\hline
Does adding layers always help? & Capacity and sum rate increase and saturate around $L\approx5$--7 under the lossless cascaded model~\cite{an2023stacked,an2025stacked}. & Sum rate is non-monotonic in $L$ and thickness with coupling-aware near-field hop matrices~\cite{dong2026geometry}; with 17\% power loss per layer a 7-layer stack falls below a 4-layer one and at 26\% below conventional MIMO~\cite{niu2026twolayer}; six lossy layers lose $\approx$17\% of the lossless rate~\cite{rostamikafaki2026loss}; energy efficiency peaks at 2--5 layers~\cite{shi2025energy}; the ideal model overestimates deep stacks by up to 7.5~bit/s/Hz~\cite{gonzalez2026lorentzian}. & A holds only at the ideal level of Fig.~\ref{fig:model_hierarchy}. With $\alpha_l<1$, amplitude--phase coupling, or~\eqref{eq:power}, the optimum depth is moderate (Section~\ref{sec:perf_depth}); two-layer designs with richer inter-layer connectivity~\cite{niu2026twolayer,kim2026bdsim} obtain the expressiveness of deep diagonal stacks with less loss. \\
\hline
Is an SIM better than a single-layer RIS transceiver? & Multilayer stacks give about twice the capacity gain of a single layer with the same total meta-atoms~\cite{an2023stacked} and 130--200\% sum-rate gains over single-layer or zero-forcing baselines~\cite{an2025stacked}. & With a coupling-aware wideband circuit model, a single-layer RIS suffices in narrowband or low-SNR settings; the SIM wins in wideband multi-user settings; partially reconfigurable SIMs are near-optimal~\cite{ibrahim2026wideband}. & The advantage of depth is interference-limited, wideband multi-user separability, not SNR. Compare under equal aperture, loss, and RF-chain count, and state the SNR regime. \\
\hline
Is the cascaded model adequate? & Almost all optimization and learning papers (Sections~\ref{sec:model_based}--\ref{sec:learning}) use~\eqref{cascaded_model} without validation. & Multiport models show mismatch for closely spaced or strongly coupled stacks~\cite{abrardo2025novel,pettanice2026modeling}; ignoring coupling costs SNR in active stacks~\cite{mandalika2026active}; measured received power varies by 13~dB with spacing~\cite{wang2024multi}. & Adequate for $d\gtrsim\lambda/2$, weakly resonant cells, and comparisons under identical assumptions; inadequate for absolute predictions and for dense, active, or near-field stacks (Section~\ref{sec:validity}). \\
\hline
Is statistical CSI sufficient? & Statistical designs attain most of the instantaneous-CSI rate for large SIMs~\cite{papazafeiropoulos2024achievable,xia2025statistical,hu2026td3}. & Aging and mobility degrade cell-free SIM systems~\cite{ding2025channel,wang2026aging}; near-field range variation and blockage violate stationarity. & Sufficient for the slow layer of a two-timescale design in LoS-dominated, low-mobility, far-field settings; instantaneous or tracked CSI is needed for the fast digital layer and for near-field and mobile users. \\
\hline
Is phase quantization benign? & 2-bit phases lose $<0.5$~bit/s/Hz and 3--4 bits are near-continuous when quantization is inside the optimizer~\cite{an2025stacked,ji2026hmimo}. & Null depth for secrecy, covertness, and clutter suppression is limited by resolution; robust designs need quantization-aware projections~\cite{hassan2026robust,niu2024enhancing}. & Benign for rate objectives with many degrees of freedom; not for null-type objectives, where the required resolution grows with the target null depth (Section~\ref{sec:perf_quant}). \\
\hline
Is learning faster than optimization? & Unfolded networks match high-iteration benchmarks at a fraction of the runtime~\cite{wang2026unfolding,ibrahim2026unfolding}; closed-form AO cuts runtime by $>99\%$~\cite{park2026efficient}. & Training needs $10^4$--$10^5$ samples and hours of GPU time~\cite{lin2018all,li2021spectrally,gao2024superresolution}; no like-for-like wall-clock comparison exists (Section~\ref{sec:learning}). & Faster \emph{online} once training is amortized; the comparison must include training cost, retraining under drift, and the reconfiguration latency of the hardware (Section~\ref{sec:perf_complexity}). \\
\hline
\end{tabularx}
\arrayrulecolor{black}
\end{table*}

\begin{lessons}
\begin{itemize}[leftmargin=*, nosep]
\item Locate every SIM paper in the five coordinates of Fig.~\ref{fig:unified_taxonomy} before comparing it with another; most contradictions are differences in the EM model or the hardware assumptions, not in the algorithms.
\item CSI acquisition and configuration are one problem: exploit the tensor/subspace structure of the cascade (pilots $N$ instead of $N^2$), optimize training length for net rate, and calibrate the device response, not only the channel.
\item AO with per-layer projected-gradient or closed-form updates is the workhorse; report complexity, runtime, and initialization sensitivity, with discreteness and loss inside the optimizer.
\item Use statistical CSI for the slow layer of a two-timescale design, learning for fast online inference (reporting training cost), and hardware-aware design by default.
\end{itemize}
\end{lessons}

\section{Application Synthesis\\ by the Physical Role of SIM}
\label{sec:apps}

\begin{figure*}[t]
\centering
\includegraphics[width=0.84\linewidth]{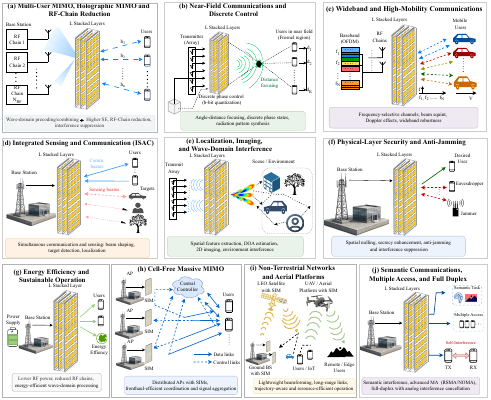}
\caption{SIM application domains grouped by the physical role of the stack: (a)--(b) wave-domain precoding/combining and near-field focusing; (c) wideband and high-mobility processing; (d)--(f) multifunctional apertures for ISAC, localization/imaging, and security; (g)--(j) lightweight analog front ends for energy-efficient, cell-free, non-terrestrial, and semantic/multiple-access/full-duplex systems.}
\label{fig:applications}
\end{figure*}

Section~\ref{sec:taxonomy} classified the literature by how the SIM is configured; this section examines what it is used for, organized by the physical role of the stack, because the role determines the modeling assumptions, the relevant hardware limits, and the meaningful baselines. Five roles account for essentially all works (Fig.~\ref{fig:applications}): (i) wave-domain precoder or combiner; (ii) near-field focusing and pattern-synthesis structure; (iii) space--time processor for wideband and mobile links; (iv) multifunctional aperture for sensing, localization, imaging, and security; and (v) lightweight analog front end for energy-, payload-, or complexity-constrained platforms. Table~\ref{tab:SIM_application_unified} summarizes representative works, tools, the gains as reported, and the assumptions behind them; the gains should be read with Section~\ref{sec:performance}.

\subsection{Role I: Wave-Domain Precoder and Combiner}
\label{sec:role_precoder}

\emph{Multi-user and holographic MIMO:} In SIM-aided holographic MIMO the transmit and receive stacks make the end-to-end channel approximately diagonal, enabling parallel streams without digital precoding, with about 150\% capacity improvement over MIMO and RIS baselines and about twice the gain of a single layer with the same total meta-atoms~\cite{an2023stacked} (Section~\ref{sec:perf_depth}). In multi-user downlink, wave-domain beamforming with $S=K$ RF chains reaches a sum rate that zero forcing needs almost twice as many antennas to match, with a 130\% gain of a 7-layer stack over a single layer under 2-bit phases~\cite{an2025stacked}. Statistical-CSI designs retain most of this gain~\cite{papazafeiropoulos2024achievable}; joint transmit--receive designs~\cite{papazafeiropoulos2024achievable_2}, analytical benchmarks~\cite{papazafeiropoulos2025ergodic_outage_3,papazafeiropoulos2025ergodic_HMIMO,papazafeiropoulos2025performance,perovic2024mutual}, refined AO and Riemannian/WMMSE frameworks for strongly coupled regimes~\cite{bahingayi2025refined,bahingayi2025scaling}, interference-free precoding with about 20~dB of suppression at 12~dB SNR~\cite{wang2026interferencefree}, and BD layers achieving regularized zero forcing with fewer layers~\cite{kim2026bdsim} extend it. Hybrid passive--active stacks compensate the cascade loss~\cite{darsena2025design,darsena2024downlink}; finite-blocklength designs serve latency-constrained IoT uplinks~\cite{zhang2025stacked} and, with loss included, report up to 140\% gain over no SIM but 17\% loss-induced degradation at six layers~\cite{rostamikafaki2026loss}; fairness designs protect the worst user~\cite{ginige2025max,fang2025stacked,quran2026robust}.

\emph{Cell-free massive MIMO:} SIM-equipped access points (APs) aggregate and suppress interference locally, reducing baseband dimension and fronthaul: uplink statistical-CSI combining~\cite{shi2024harnessing,shi2024uplink,shi2025uplink}, holographic AP processing~\cite{li2024stacked}, downlink interference suppression~\cite{hu2024joint}, layers substituting for APs~\cite{shi2025joint}, fronthaul-constrained hybrid beamforming~\cite{park2025sim}, and closed-form per-AP updates with more than 99\% runtime reduction~\cite{park2026efficient}. Coordination is the bottleneck: multi-agent RL~\cite{zhu2025joint}, rate splitting against channel-aging interference in high-speed railways~\cite{ding2026hsr}, aging and impairment analyses~\cite{ding2025channel,wang2026aging,wang2026uplink}, cell-free SWIPT~\cite{hua2025swipt}, and UAV-mounted SIM-aided cell-free systems~\cite{wang2026spatialwave} converge, with~\cite{shi2025stacked}, on a two-timescale design.

\textbf{Synthesis.} Role~I has the strongest evidence and the clearest baselines. The gains are spatial degrees of freedom per RF chain, largest in interference-limited multi-user settings and vanishing in noise-limited ones, where a single layer or a conventional array performs as well~\cite{ibrahim2026wideband}. Comparisons should fix RF-chain count, aperture, loss, and power model.

\subsection{Role II: Near-Field Focusing and Pattern Synthesis}
\label{sec:role_nearfield}

Large apertures at high frequencies place users in the Fresnel region, where the multilayer operator can synthesize focal spots rather than far-field beams: spherical-wave designs~\cite{papazafeiropoulos2024near}, multi-user near-field transceivers separating users by range and angle~\cite{li2025stacked_NF,jia2024stacked}, geometry-aware designs optimizing the spacings~\cite{dong2026geometry}, covert near-field transmission with 120--225\% sum-covert-rate gains over projected-gradient designs~\cite{benaya2026covert}, orbital-angular-momentum synthesis in which aperture limits the mode order and depth controls crosstalk~\cite{torcolacci2026oam}, and EM-domain pattern synthesis~\cite{hassan2024efficient}. The usable near-field distance is bounded: lateral resolution degrades linearly and axial resolution quadratically with distance, and more layers extend the usable range only until loss saturates the gain~\cite{zhao2026nearfield}.

\emph{Mixed-field operation:} The same stack must typically serve near-field and far-field users. The mixed-field problem, studied for single-layer surfaces~\cite{li2025near,droulias2024near}, is only beginning to be addressed for SIMs: hierarchical codebooks covering the full angular range~\cite{yu2026hierarchical} and distance-and-angle beam training~\cite{wei2026nearfield} provide CSI in both regimes, and geometry design~\cite{dong2026geometry} shows that the spacing optimal for far-field beamforming is not optimal for focusing. A mixed-field design therefore needs a model exact in both regimes (the plane-wave-spectrum formulation of~\cite{zhang2026eit} or the multiport model) and an allocation of layers between angular and range selectivity. \emph{Discrete control:} low-resolution phases approach continuous performance when discreteness is inside the optimizer~\cite{an2025stacked,jiao2026efficient,nassirpour2025sum,ji2026hmimo}, but focusing relies on coherent superposition across a large aperture and is sensitive to phase errors~\cite{li2025stacked_NF} and geometry errors (Section~\ref{sec:tolerance}).

\textbf{Synthesis.} Focusing gains are the most model- and calibration-dependent gains in the SIM literature; they require range-aware estimation (Section~\ref{sec:csi}), a Fresnel-valid model, and a tolerance analysis, as the measured 13-dB dependence on inter-layer spacing~\cite{wang2024multi} shows.

\subsection{Role III: Space--Time Processor for Wideband and High-Mobility Links}
\label{sec:role_wideband}

One configuration acts on all frequencies while each meta-atom and inter-layer path is frequency dependent, so wideband design trades spatial resolution against frequency robustness (beam squint). Fully analog wideband beamforming~\cite{li2024stacked_analog}, MIMO-OFDM with stacks at both ends~\cite{li2025stacked_OFDM}, a coupling-aware circuit model showing that the SIM beats a single-layer RIS in wideband multi-user but not narrowband or low-SNR settings~\cite{ibrahim2026wideband}, a subcarrier-reuse trade-off with an optimum near 60\%~\cite{li2025fundamental}, unfolding for one configuration across subcarriers~\cite{ibrahim2026unfolding}, baseband-free OFDM transmitters implementing symbol loading, precoding, inverse DFT, and cyclic prefix in the wave domain with a second stack fitted offline to the multiport OFDM operator~\cite{li2026basebandfree,li2026fullywave}, randomized space--time coding~\cite{darsena2025randomized}, delay--Doppler modeling~\cite{ranasinghe2025doubly}, and delay-augmented stacks engineering group delay~\cite{alwazani2025delay} define the space.

\textbf{Synthesis.} Wideband SIMs require space--frequency co-design of spacing, dispersion, subcarrier objectives, and equalization; the depth optimal for narrowband focusing is not optimal for wideband operation, and the measured dispersion of real cells (Section~\ref{sec:tuning}) is the limiting factor that simulations omit.

\subsection{Role IV: Multifunctional Aperture for Sensing, Localization, Imaging, and Security}
\label{sec:role_multifunction}

\emph{ISAC and sensing:} ISAC designs balance rate against beampattern matching and Cram\'er--Rao bounds~\cite{niu2024stacked,ebrahimi2025stacked}; cognitive wideband ISAC nulls primary users~\cite{fadakar2025stacked}; parametrized stacks decouple sensing and communication wavefronts~\cite{ranasinghe2025parametrized}; satellite and discrete-phase variants~\cite{jiang2025stacked,zhang2025joint}, passive transmit stacks~\cite{li2024transmit}, RIS-assisted position-aware designs with about 1.5 times the sum rate under a sensing constraint~\cite{wang2026positionaware}, STAR-SIMs~\cite{wang2026starsim}, BD-SIM ISAC~\cite{jiao2026bdsim}, age-of-information-aware allocation with about 230\% energy-efficiency gain~\cite{ataeebojd2026isac}, fluid-antenna combinations~\cite{wu2026fas}, and joint edge-computing/ISAC with SIM backhaul~\cite{chen2026mec} extend the role. The only experimental SIM-ISAC validation is the 5.8-GHz three-layer testbed of~\cite{wang2024multi}, with the DoA error falling from 3.6$^\circ$ to 1.4$^\circ$ from one to three layers.

\emph{Localization, DoA, and imaging:} Wave-domain DFT approximations~\cite{an2024stacked_DFT}, directional scanning~\cite{an2024two}, EM-NNs from amplitude-only measurements~\cite{lin2026emnn}, active stacks countering multiplicative fading~\cite{jiang2026doa}, multiport-consistent projections with few RF chains~\cite{abrardo2026localization,abrardo2026nonlinear}, location signatures for indoor positioning~\cite{javed2025sim}, onboard terrain classification~\cite{liu2025onboard}, and in-wave inference~\cite{liu2026inwave}; measured angular resolutions are 3$^\circ$/1$^\circ$ with three/four millimeter-wave layers~\cite{gao2024superresolution} and 0.5$^\circ$ with meta-trained layers~\cite{yang2025metatraining}.

\emph{Security and anti-jamming:} Wiretap designs null eavesdroppers~\cite{niu2024enhancing,niu2024efficient}; multi-user secrecy designs~\cite{kavianinia2025secrecy,hoang2025secure}, covert near-field~\cite{benaya2026covert} and aerial designs with up to 189\% energy-efficiency gain under covertness~\cite{wu2026adsim}, and receive-side jammer suppression by wave-domain phase inversion~\cite{pei2024stacked} extend the objective.

\textbf{Synthesis.} The opportunity is multifunctionality; the limitation is objective conflict, since the configuration that maximizes rate rarely gives the best beampattern, null depth, or localization accuracy, and nulls need more resolution than rate (Table~\ref{tab:conflicts}). Pareto fronts and metrics evaluated under the same hardware and calibration assumptions are the appropriate standard.

\subsection{Role V: Lightweight Analog Front End}
\label{sec:role_frontend}

\emph{Energy efficiency, wireless power transfer, SWIPT:} With the full power model~\eqref{eq:power}, 2--5 layers give the best compromise between spectral and energy efficiency and hybrid precoding improves energy efficiency by about 80\% over digital precoding without SIM~\cite{shi2025energy}; broadcast MIMO~\cite{perovic2025energy}, energy-efficient UAV designs~\cite{sheemar2026ee}, active layers with their power included~\cite{shen2026active,yeganeh2025enhancing}, power minimization under QoS~\cite{niu2025transmit}, and semantic energy efficiency~\cite{sun2025towards} broaden the objective. Wireless power transfer uses the focusing of Role~II: SIM-assisted SWIPT enlarges the information--energy region~\cite{amiri2025stacked}, cell-free SWIPT improves harvested energy~\cite{hua2025swipt}, and multilayer RISs are proposed for edge power transfer~\cite{zhang2025multilayerris}.

\emph{Non-terrestrial and aerial platforms:} Payload, power, RF-chain, and CSI constraints motivate lightweight beam shaping for LEO downlinks, beam hopping, NOMA, and solar-panel-mounted active stacks~\cite{lin2024stacked,khan2025stacked,yu2025energy,yeganeh2025enhancing}; a 5-layer stack yields about 54\% more throughput than one layer in~\cite{zhang2026satellite}; HAPS unicast/multicast is analyzed in~\cite{le2026haps}. UAV-mounted stacks are optimized with trajectory and power~\cite{zarini2025orchestration,fan2025joint,sheemar2026uav,sheemar2026ee}, digital twins~\cite{xiong2025digital}, generative AI~\cite{sun2026Generative}, adaptive deep stacks for covert links~\cite{wu2026adsim}, network-calculus delay guarantees for eVTOL~\cite{chen2025stacked_vtol}, offshore low-altitude networks~\cite{pan2026offshore}, and unified channel-model-driven optimization~\cite{ma2026lain}.

\emph{Semantic communications, multiple access, full duplex:} The stack acts as a physical feature encoder or semantic equalizer: multimodal transmission~\cite{huang2025stacked}, over-the-air semantic alignment~\cite{pandolfo2025over}, task-oriented recognition~\cite{huang2024stacked}, all-analog semantic mapping matching deep joint source--channel coding at low SNR~\cite{hou2026semantic}, residual EM-NNs for images~\cite{choi2026jscc}, and task-oriented wave processing~\cite{qi2026task}. Rate-splitting (RSMA) designs combine signal-domain interference management with wave-domain shaping~\cite{quran2025max,hashida2025precoding,huai2025stacked,quran2026rsma}, with up to 700\% higher minimum rate under hardware impairment~\cite{quran2026robust}, and hybrid RIS--SIM deployments address blockage~\cite{liu2025sum}. Full-duplex systems use the stack as an analog self-interference pre-filter~\cite{zhang2025sim} and combine it with RSMA~\cite{liang2025stacked}.

\textbf{Synthesis.} Role~V is where SIMs are most compelling, because RF chains, payload, and power are the binding constraints, and where the evidence is thinnest: results are simulation-only, power models are borrowed from RIS measurements, and thermal, Doppler, and mechanical constraints are seldom included.

\begin{table*}[!t]
\centering
\caption{Application synthesis by physical role: representative works, SIM function, tools, reported gains (as stated by the cited works), and the assumptions behind them.}
\label{tab:SIM_application_unified}
\renewcommand{\arraystretch}{1.05}
\setlength{\tabcolsep}{1.5pt}
\scriptsize
\arrayrulecolor{SIMLine}
\begin{tabularx}{\textwidth}{|>{\columncolor{pastelpurple}\raggedright\arraybackslash}p{1.5cm}|>{\raggedright\arraybackslash}p{2.0cm}|>{\raggedright\arraybackslash}p{2.0cm}|>{\raggedright\arraybackslash}p{2.3cm}|>{\raggedright\arraybackslash}X|>{\raggedright\arraybackslash}X|}
\hline
\rowcolor{pastelblue}\textbf{Role / domain} & \textbf{Representative works} & \textbf{SIM function} & \textbf{Tools} & \textbf{Reported gains (as stated)} & \textbf{Assumptions / open issues} \\
\hline
\textbf{I: MU-MIMO, HMIMO} & \cite{an2023stacked,an2025stacked,papazafeiropoulos2024achievable,papazafeiropoulos2024achievable_2,bahingayi2025refined,bahingayi2025scaling,kim2026bdsim,wang2026interferencefree,rostamikafaki2026loss,darsena2025design,ginige2025max} & Wave-domain precoding/combining; channel diagonalization; fairness & AO, PGA, Riemannian, WMMSE, Takagi/BD layers, capacity/outage analysis & $\approx$150\% capacity vs.\ MIMO/RIS~\cite{an2023stacked}; 130\% vs.\ single layer, 200\% vs.\ ZF~\cite{an2025stacked}; $\approx$20~dB interference suppression~\cite{wang2026interferencefree}; up to 140\% vs.\ no SIM, $-17$\% with loss at $L=6$~\cite{rostamikafaki2026loss} & Ideal or 2-bit phases, lossless (except~\cite{rostamikafaki2026loss}), perfect/statistical CSI, cascaded model, no measurements \\
\hline
\textbf{I: Cell-free} & \cite{shi2024harnessing,shi2024uplink,shi2025uplink,li2024stacked,hu2024joint,shi2025joint,park2025sim,park2026efficient,zhu2025joint,ding2026hsr,ding2025channel,hua2025swipt} & AP-side aggregation; interference suppression; fronthaul/RF-chain reduction & Two-timescale statistical CSI, closed-form AO, MARL, rate splitting, aging analysis & $>$99\% runtime reduction at equal rate~\cite{park2026efficient}; layers substitute for APs~\cite{shi2025joint}; aging mitigated by rate splitting~\cite{ding2026hsr} & Coordination, synchronization, fronthaul signaling, distributed calibration unaddressed \\
\hline
\textbf{II: Near-field, mixed-field} & \cite{papazafeiropoulos2024near,li2025stacked_NF,jia2024stacked,dong2026geometry,benaya2026covert,torcolacci2026oam,zhao2026nearfield,hassan2024efficient,yu2026hierarchical,wei2026nearfield} & Distance--angle focusing; pattern synthesis; OAM modes & Spherical-wave / plane-wave-spectrum models, geometry design, codebooks, beam training & Range-aware user separation; 120--225\% covert-rate gain~\cite{benaya2026covert}; geometry-aware spacing beats uniform~\cite{dong2026geometry} & Highly sensitive to geometry, calibration, phase errors; mixed-field designs only beginning \\
\hline
\textbf{III: Wideband, mobility} & \cite{li2024stacked_analog,li2025stacked_OFDM,ibrahim2026wideband,li2025fundamental,ibrahim2026unfolding,li2026basebandfree,li2026fullywave,darsena2025randomized,ranasinghe2025doubly,alwazani2025delay} & Frequency-robust beamforming; wave-domain OFDM; space--time filtering & Multi-subcarrier AO, circuit models, unfolding, offline fitting to multiport operator, DD modeling & SIM $>$ RIS in wideband multi-user, not narrowband/low SNR~\cite{ibrahim2026wideband}; $\approx$60\% reuse optimum~\cite{li2025fundamental}; baseband-free OFDM near digital benchmark~\cite{li2026basebandfree} & Dispersion and frequency-dependent loss rarely measured; beam squint at large $B/f_c$ \\
\hline
\textbf{IV: ISAC, sensing} & \cite{niu2024stacked,ebrahimi2025stacked,fadakar2025stacked,ranasinghe2025parametrized,jiang2025stacked,zhang2025joint,li2024transmit,wang2026positionaware,wang2026starsim,jiao2026bdsim,ataeebojd2026isac,wang2024multi} & Joint beams, illumination, nulls; CRB-constrained rate & Multi-objective AO, CRB design, BD/STAR layers, two-timescale allocation, testbed & $\approx$1.5$\times$ rate under sensing constraint~\cite{wang2026positionaware}; $\approx$230\% EE~\cite{ataeebojd2026isac}; measured DoA error 3.6$^\circ\!\rightarrow$1.4$^\circ$ (1$\rightarrow$3 layers)~\cite{wang2024multi} & Pareto fronts rarely reported; one testbed; clutter/mobility models missing \\
\hline
\textbf{IV: Localization, imaging, inference} & \cite{an2024stacked_DFT,an2024two,lin2026emnn,jiang2026doa,abrardo2026localization,javed2025sim,liu2025onboard,liu2026inwave,gao2024superresolution,yang2025metatraining} & Spatial spectrum analysis; location signatures; physical feature extraction & Wave-domain DFT, EM-NN, multiport projections, diffractive training & Measured resolution 3$^\circ$/1$^\circ$ (3/4 layers)~\cite{gao2024superresolution}, 0.5$^\circ$~\cite{yang2025metatraining}; in-wave 60--67\% vs.\ $>$79\% digital at ns vs.\ $\mu$s--ms latency~\cite{liu2026inwave} & Task-specific overfitting; accuracy gap of in-wave inference \\
\hline
\textbf{IV: Security, covertness, anti-jamming} & \cite{niu2024enhancing,niu2024efficient,kavianinia2025secrecy,hoang2025secure,benaya2026covert,wu2026adsim,pei2024stacked} & Eavesdropper nulling; covert focusing; jammer suppression & AO, WMMSE/SCA, convex approximation, ADMM, deep stacks & Up to 189\% EE under covertness~\cite{wu2026adsim}; secrecy gains over digital beamforming~\cite{niu2024enhancing} & Null depth limited by $b$ and calibration; eavesdropper/jammer uncertainty seldom modeled \\
\hline
\textbf{V: EE, WPT, SWIPT} & \cite{shi2025energy,perovic2025energy,sheemar2026ee,shen2026active,yeganeh2025enhancing,niu2025transmit,sun2025towards,amiri2025stacked,hua2025swipt,zhang2025multilayerris} & Low-power spatial processing; information--energy shaping & Fractional programming, Dinkelbach, quadratic transform, BC--MAC duality, Bayesian optimization & $\approx$80\% EE gain vs.\ digital precoding, best at 2--5 layers~\cite{shi2025energy}; active layers improve EE when power is included~\cite{shen2026active} & Power models borrowed from RIS measurements; control, calibration, thermal terms omitted \\
\hline
\textbf{V: NTN, UAV, HAPS} & \cite{lin2024stacked,khan2025stacked,yu2025energy,zhang2026satellite,le2026haps,zarini2025orchestration,fan2025joint,sheemar2026uav,xiong2025digital,sun2026Generative,chen2025stacked_vtol,pan2026offshore,ma2026lain} & Lightweight beam shaping; trajectory-aware control & Statistical-CSI AO, NOMA/RSMA, MARL, digital twins, network calculus, 3D placement & $\approx$54\% throughput gain (5 vs.\ 1 layer)~\cite{zhang2026satellite}; gains saturate in $L$, grow with $M$~\cite{sheemar2026uav} & Doppler, CSI delay, payload/thermal limits, mechanical integration rarely modeled \\
\hline
\textbf{V: Semantic, MA, FD} & \cite{huang2025stacked,pandolfo2025over,huang2024stacked,hou2026semantic,choi2026jscc,qi2026task,quran2025max,hashida2025precoding,huai2025stacked,quran2026rsma,quran2026robust,liu2025sum,zhang2025sim,liang2025stacked} & Physical feature encoding; RSMA interference shaping; self-interference pre-filtering & EM-NN training, cGAN reconstruction, SCA/AO, manifold optimization & Matches Deep-JSCC at low SNR~\cite{hou2026semantic}; up to 700\% minimum-rate gain~\cite{quran2026robust} & Task accuracy, latency, energy, reliability rarely reported together; strong digital baselines needed \\
\hline
\end{tabularx}
\arrayrulecolor{black}
\end{table*}

\begin{lessons}
\begin{itemize}[leftmargin=*, nosep]
\item Gains are largest where the task is wave-domain and interference-limited and where RF chains, payload, or power are the binding constraints; they shrink in noise-limited settings.
\item Gains of Roles~II and IV (focusing, nulling, sensing) are the most sensitive to model accuracy, resolution, and calibration; report tolerance analyses and Pareto fronts.
\item Reported percentages are comparable across papers only under the same baseline, RF-chain count, aperture, loss, and power model, which is why Table~\ref{tab:SIM_application_unified} lists the assumptions.
\item Mixed-field operation, wireless power transfer, and thermal/Doppler-aware non-terrestrial designs are the least developed topics.
\end{itemize}
\end{lessons}

 \section{Hardware Demonstrators, Tuning Technologies, Advanced Architectures, and the Theory--Prototype Gap}
\label{sec:hardware}

This section addresses what the modeling and algorithm literature postpones: \emph{what has been built, what has been measured, and how far the measured numbers are from the assumptions of Sections~\ref{sec:taxonomy}--\ref{sec:apps}}. It compares demonstrators (Section~\ref{sec:demos}, Table~\ref{tab:prototypes}) and tuning technologies (Section~\ref{sec:tuning}, Table~\ref{tab:tuning}), separates validated from conceptual architectures (Section~\ref{sec:advanced}, Table~\ref{tab:maturity}), and discusses thermal and power management (Section~\ref{sec:thermal}), tolerances and calibration (Section~\ref{sec:tolerance}), and the resulting gap together with a reporting checklist (Section~\ref{sec:gap}).

\subsection{Demonstrator Classes and Quantitative Comparison}
\label{sec:demos}

Three classes have emerged~\cite{liu2025stacked}: \emph{1) static} stacks with fixed phase profiles (3D-printed or machined dielectric, etched silicon, engraved PCB), which perform one task at zero operating power; \emph{2) programmable-passive} stacks, in which FPGA-controlled PIN diodes, varactors, or liquid crystal set the phase while the layer remains passive; and \emph{3) programmable-active} stacks, in which per-element amplifiers set gain and phase and can emulate nonlinear activations. A fourth class, \emph{mechanically reconfigurable} stacks (sliding layers, detachable meta-atoms), has recently appeared. Table~\ref{tab:prototypes} compares the multilayer demonstrators that report measurements; optical processors are included for the latency and energy figures they establish.

Three facts stand out. Only one demonstrator is an SIM in the communication-theoretic sense (a programmable transmissive stack in front of a feed evaluated with wireless metrics): the 5.8-GHz, three-layer, 1-bit PIN-diode prototype of~\cite{wang2024multi}, which measured received power and DoA error for one to three layers and five spacings. The only programmable-active multilayer demonstrators are the 5.4-GHz PDNN~\cite{liu2022programmable} and the 5.8-GHz nonlinear network~\cite{ning2025multilayer}; the latter is the only one reporting power (about 429~mW per active unit, about 252~W in total) and per-layer latency (17.7~ns). And no multilayer prototype reports per-layer insertion loss, array reconfiguration time, or thermal behavior explicitly; loss appears indirectly (about 3~dB per 3D-printed terahertz layer~\cite{lin2018all}, 0.048\% end-to-end efficiency to a single-pixel detector~\cite{li2021spectrally}, a 2-dB gap between simulated and measured focal efficiency~\cite{gu2024classification}, below 1~dB for a passive sliding bilayer~\cite{shi2026sliding}). These omissions are the hardware side of Gap~3.

\begin{table*}[!t]
\centering
\caption{Experimentally demonstrated multilayer metasurface / diffractive-network stacks with measured results. SP = static-passive; PP = programmable-passive; PA = programmable-active; MR = mechanically reconfigurable; n.r. = not reported. Apertures in wavelengths are derived from reported dimensions.}
\label{tab:prototypes}
\renewcommand{\arraystretch}{1.05}
\setlength{\tabcolsep}{1.5pt}
\scriptsize
\arrayrulecolor{SIMLine}
\begin{tabularx}{\textwidth}{|>{\columncolor{pastelpurple}\raggedright\arraybackslash}p{1.5cm}|>{\centering\arraybackslash}p{0.95cm}|>{\centering\arraybackslash}p{0.55cm}|>{\raggedright\arraybackslash}p{1.65cm}|>{\centering\arraybackslash}p{0.85cm}|>{\raggedright\arraybackslash}p{1.75cm}|>{\raggedright\arraybackslash}p{1.3cm}|>{\raggedright\arraybackslash}p{1.6cm}|>{\raggedright\arraybackslash}p{1.3cm}|>{\raggedright\arraybackslash}p{1.25cm}|>{\raggedright\arraybackslash}X|}
\hline
\rowcolor{pastelblue}\textbf{Ref.\ (year)} & \textbf{Band} & \textbf{$L$} & \textbf{Elements / layer (aperture)} & \textbf{Spacing} & \textbf{Tuning / class} & \textbf{Resolution} & \textbf{Loss / efficiency} & \textbf{Latency} & \textbf{Power} & \textbf{Task and measured performance} \\
\hline
Lin \emph{et al.}~\cite{lin2018all} (2018) & 0.4 THz & 5 & $200\times200$, 8~cm ($\approx$107$\lambda$) & 3~cm (40$\lambda$) & 3D-printed thickness / SP & continuous & $\approx$51\% power attenuation per layer ($\approx$3~dB) & static & 0 & MNIST 91.75\% numerical, 88\% experimental agreement (50 digits); Fashion-MNIST 81--86\% numerical \\
\hline
Qian \emph{et al.}~\cite{qian2020performing} (2020) & 17 GHz & 2 (3 sim.) & $30\times42$, pitch 0.57$\lambda$ & 17$\lambda$ & machined F4B / SP & continuous & transmission close to unity & static & 0 & NOT/OR/AND logic; contrast $>$9.6~dB \\
\hline
Li \emph{et al.}~\cite{li2021spectrally} (2021) & 0.2--0.3 THz & 3 & 0.5-mm features & n.r. & 3D-printed / SP & 12 thickness levels & 0.048\% end-to-end power efficiency & static & 0 & Spectrally encoded single-pixel MNIST: 95\% numerical, 88\% experimental agreement \\
\hline
Liu \emph{et al.}~\cite{liu2022programmable} (2022) & 5.4 GHz & 5 & $8\times8$, 56~cm ($\approx$10$\lambda$) & 10~cm & 2 amplifier chips per cell, FPGA / PA & gain $-22$ to $+13$~dB; 4--9-bit (sources disagree) & n.r. (active gain) & n.r. & n.r. (640 amplifier chips) & Shape classification 100\%; 4-user coded link with Wi-Fi interference, error rate 49\%$\rightarrow$0.5\%; multi-beam leakage $<$10\% \\
\hline
Gu \emph{et al.}~\cite{gu2024classification} (2024) & 11 GHz & 3 & 1024 ($32\times32$) & n.r. & engraved input, static layers / SP & n.r. & focal efficiency 18.7\% sim.\ vs.\ 11.7\% meas.\ ($\approx$2~dB) & static & 0 & Metal handwritten digits: 90\% numerical; 18/18 samples correct \\
\hline
Wang \emph{et al.}~\cite{wang2024multi} (2024) & 5.8 GHz & 1--3 & $16\times16$, 31~cm ($\approx$6$\lambda$) & $\lambda/4$--$2\lambda$ & 2 PIN diodes per cell, FPGA / PP & 1-bit & n.r. & n.r. (PIN) & n.r. (5-V logic boards) & Received power $-53$~dBm (1 layer) $\rightarrow$ $-46$~dBm (3 layers, $\lambda/4$), $-59$~dBm (3 layers, $\lambda$); DoA error 3.57$^\circ$/1.49$^\circ$/1.39$^\circ$ (1/2/3 layers) \\
\hline
Gao \emph{et al.}~\cite{gao2024superresolution} (2024) & 25--30 GHz & 3--4 (+LC-RIS) & $32\times32$, pitch $\lambda/2$, 16$\lambda$ & 5$\lambda$ & CNC-machined / SP; LC-RIS / PP & $\approx$7-bit; LC 5-bit & n.r. & LC $<$500~ms; snapshot 1.67~$\mu$s / 67~ns & 0 (passive); $\approx$0.5~W per LC panel & DoA resolution 3$^\circ$ (3 layers), 1$^\circ$ (4 layers); 6.94 TOPS; ISAC +16.1~dB detected amplitude \\
\hline
Ning \emph{et al.}~\cite{ning2025multilayer} (2025) & 5.8 GHz & 3 lin.\ + 3 nonlin. & $22\times22$, 19-mm cells ($\approx$8$\lambda$) & 10~cm & static linear layers; amplifier + detector + adder ReLU cells / PA & continuous; 2 params.\ per nonlinear layer & nonlinear cell $<-10$~dB below threshold to stable gain & 17.7~ns per nonlinear layer & $\approx$429~mW per unit; $\approx$252~W total & MNIST 92.6\% measured (92.8\% simulated); Fashion-MNIST 77.4\% (nonlinear) vs.\ 72.5\% (linear) \\
\hline
Yang \emph{et al.}~\cite{yang2025metatraining} (2025) & 27--31 GHz & 3 & $50\times50$, 200~mm ($\approx$18$\lambda$) & 100~mm & meta-trained dual-pol.\ cells / SP & n.r. & n.r. & static & 0 & DoA resolution 0.5$^\circ$ ($\approx$7$\times$ Rayleigh limit); MAE 0.028$^\circ$; FoV $\pm$11.5$^\circ$ \\
\hline
Yu \emph{et al.}~\cite{yu2026movable} (2026) & 14 GHz & 3 & $20\times20$ & 120~mm (5.6$\lambda$) & detachable movable-type meta-atoms / MR & 3-bit; amplitude $>$0.9 & per-atom amplitude $>$0.9 & manual: 20--70~min & 0 & 4-class EMNIST digits/letters: 92.5\% measured (95--97\% simulated); retraining 12--15~s \\
\hline
Shi \emph{et al.}~\cite{shi2026sliding} (2026) & 26.5--29.5 GHz & 2 & $40\times40$, 140~mm ($\approx$13$\lambda$) & 1~mm & lateral sliding / MR & continuous & $<$1~dB in band; $>$20~dB out of band & mechanical & 0 & Continuous $\pm57^\circ$ steering, sidelobes $<-10$~dB \\
\hline
Zhou \emph{et al.}~\cite{zhou2021reconfigurable_dpu} (2021, optical) & 698 nm & 3 & millions of neurons (SLM) & -- & DMD + phase SLM / PP & 8-bit & n.r. & 56--70 frames/s & 2.9--5.9 TOPS/J & MNIST 96.0--96.8\%; video recognition 94--97\% \\
\hline
Chen \emph{et al.}~\cite{chen2023accel} (2023, optical) & 532 nm & 2 + analog electronic & $400\times400$ & -- & static phase masks + CMOS / SP & 8-level & n.r. & 72~ns per frame & 4.4~nJ per frame & MNIST 90.9\%; Fashion-MNIST 80.9\%; 3-class ImageNet 82\% \\
\hline
\end{tabularx}
\arrayrulecolor{black}
\end{table*}

\subsection{Tuning Technologies for Programmable Layers}
\label{sec:tuning}

The tuning mechanism determines phase resolution, insertion loss, reconfiguration latency, and power, and therefore which methodological family can be used with a layer. Table~\ref{tab:tuning} compiles measured values from transmissive and reflective programmable-surface cells and arrays. Two conclusions follow. Fast technologies (PIN diodes, RF switches) are 1-bit and lossy (1.2--1.9~dB per transmissive cell), whereas low-loss or high-resolution technologies (liquid crystal, varactors at low frequency, mechanical) are slow, lossy at millimeter wave (6.6~dB for liquid-crystal delay lines), or both. And the bias power of diode-based arrays is not negligible when multiplied by $LM$: at 12~mW per ON diode, three $16\times16$ 1-bit layers draw up to about 9~W in the diodes plus about 5~W per FPGA board~\cite{wang2024power}, the same order as the RF-chain power the SIM is meant to save. Liquid crystal draws nanowatts per cell but reconfigures in tens of milliseconds and loses phase range as temperature rises (Section~\ref{sec:thermal}).

\begin{table*}[t]
\centering
\caption{Tuning technologies for programmable metasurface layers: measured unit-cell and array data relevant to SIM design.}
\label{tab:tuning}
\renewcommand{\arraystretch}{1.05}
\setlength{\tabcolsep}{2pt}
\scriptsize
\arrayrulecolor{SIMLine}
\begin{tabularx}{\textwidth}{|>{\columncolor{pastelpurple}\raggedright\arraybackslash}p{2.2cm}|>{\raggedright\arraybackslash}p{1.6cm}|>{\raggedright\arraybackslash}p{2.6cm}|>{\raggedright\arraybackslash}p{2.0cm}|>{\raggedright\arraybackslash}p{3.0cm}|>{\raggedright\arraybackslash}X|}
\hline
\rowcolor{pastelblue}\textbf{Technology} & \textbf{Frequency} & \textbf{Insertion loss (measured)} & \textbf{Switching time} & \textbf{Power} & \textbf{Notes for SIMs} \\
\hline
PIN diode, 1-bit transmissive & 7.5--35 GHz & 1.2--1.9~dB per cell~\cite{rana2021experimental,huang2023wideband,zhang2023onebit}; $16\times16$ transmitarray 12.1\% aperture efficiency~\cite{zhang2023onebit} & ns~\cite{etsi2025ris002} & 12--12.6~mW per ON diode; 4.8~W per FPGA; $60\times60$ dual-pol.\ panel 15.7--103~W~\cite{wang2024power} & Used in the only programmable SIM testbed~\cite{wang2024multi}; loss compounds with $L$; bias power $\propto LM$ \\
\hline
PIN + amplifier, 2-bit active transmissive & 2.6 GHz & below 4~dB per element (before gain)~\cite{song2024active} & ns & PA up to 120~mA at 5.5~V per element~\cite{song2024active} & +11.9~dB over passive transmissive RIS; adds noise and heat \\
\hline
Varactor & $<$20 GHz (24.5 GHz design) & 1.5--4.15~dB (2-bit, 270$^\circ$)~\cite{dasilva2023varactor} & ns--$\mu$s & diode current negligible; 430~mW per DAC/op-amp channel~\cite{wang2024power} & Quasi-continuous phase; loss and bias complexity grow at mmWave \\
\hline
RF switch & -- & -- & ns & $\approx$495~$\mu$W per cell~\cite{wang2024power} & Low power; 1-bit \\
\hline
Liquid crystal & 25--62 GHz & 6.6~dB (delay-line cells)~\cite{neuder2024lcris,abantoleon2024liquiris} & 15--80~ms~\cite{neuder2024lcris,abantoleon2024liquiris}; $<$500~ms panel~\cite{gao2024superresolution} & $\approx$21.5~nW per phase shifter~\cite{abantoleon2024liquiris}; $\approx$0.5~W per $20\times20$ panel~\cite{gao2024superresolution} & Quasi-continuous, ultra-low power, slow; phase range shrinks with temperature~\cite{gholian2025temperature} \\
\hline
RF-MEMS & $<$40 GHz & low & $\mu$s--ms~\cite{etsi2025ris002} & low & Reliability and packaging of $LM$ moving parts \\
\hline
Sub-THz switches (RF-SOI, BiCMOS, GaN, PCM, memristor) & up to 480 GHz & technology dependent & ns (PCM: 1--500~nJ per switch) & 0.05--50~mW DC per switch~\cite{matos2024thz} & Candidates for THz SIMs; no multilayer prototype yet \\
\hline
Static (3D-printed, machined, etched) & 11 GHz--0.6 THz & $<$1~dB per passive bilayer~\cite{shi2026sliding}; $\approx$3~dB per 3D-printed THz layer~\cite{lin2018all} & none & 0 & Lowest loss and power; task-specific; tolerance $\pm$0.025--0.1~mm at mmWave~\cite{gao2024superresolution,yang2025metatraining} \\
\hline
Mechanical / manual & 14--29 GHz & $<$1~dB~\cite{shi2026sliding} & seconds to 70~min~\cite{yu2026movable} & 0 & Continuous tuning by displacement; not for per-block reconfiguration \\
\hline
\end{tabularx}
\arrayrulecolor{black}
\end{table*}

\subsection{Advanced Architectures and Their Maturity}
\label{sec:advanced}

\begin{figure*}[t]
\centering
\includegraphics[width=0.6\linewidth]{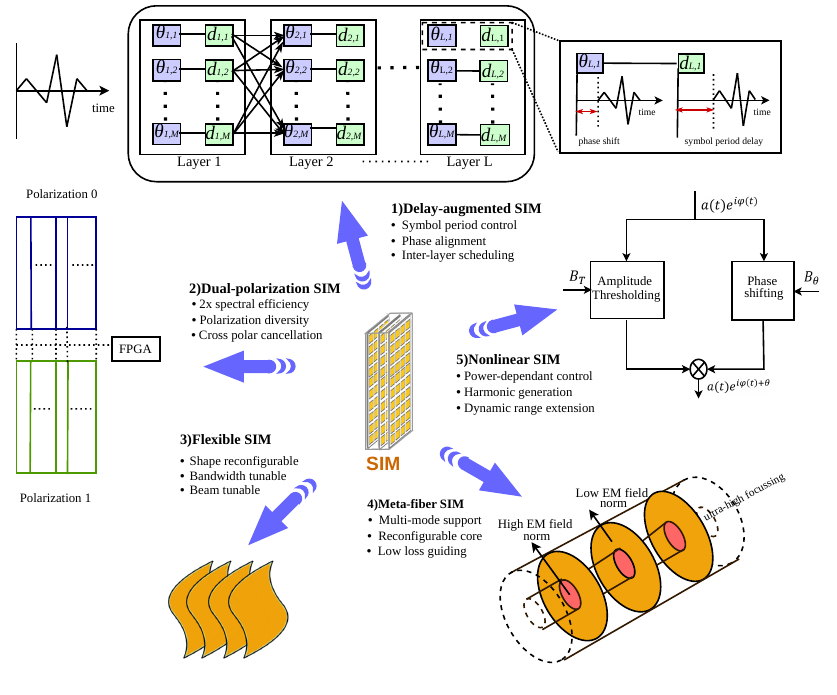}
\caption{Advanced SIM architectures: (1) delay-augmented, (2) dual-polarization, (3) flexible, (4) meta-fiber-assisted, and (5) nonlinear SIMs. Table~\ref{tab:maturity} states which have been demonstrated, validated by full-wave simulation, or proposed conceptually.}
\label{fig:Advanced_SIM}
\end{figure*}

Beyond stacked passive phase-modulating layers, several architectures embed additional functionality in the stack (Fig.~\ref{fig:Advanced_SIM}). \emph{Delay-augmented} SIMs engineer dispersive layers and spacings to provide controllable group delay, making the stack an analog space--time processor~\cite{alwazani2025delay}. \emph{Dual-polarization} SIMs process the two polarization components independently with anisotropic meta-atoms, doubling the processing dimension at fixed aperture~\cite{zhang2025dual}. \emph{Flexible} SIMs and flexible intelligent layered metasurfaces (FILMs) use compliant substrates for conformal and mechanically tuned stacks~\cite{magbool2025stacked,niu2026twolayer}. \emph{Meta-fiber-assisted} SIMs route energy between layers through guided conduits so that two layers emulate the mixing of deeper free-space cascades with less loss~\cite{niu2025introducing,niu2026twolayer}. \emph{Nonlinear} SIMs embed diodes, transistors, or saturating amplifiers to realize amplitude-dependent transformations analogous to activation functions~\cite{abbas2025nonlinear,abrardo2026nonlinear}. Further proposals with limited studies also include \emph{spherical} SIMs~\cite{zhang2026spherical}, \emph{fluid-element} SIMs~\cite{wei2026fsim}, \emph{STAR-SIMs}~\cite{wang2026starsim}, \emph{BD-SIMs}~\cite{nerini2024physically,kim2026bdsim,jiao2026bdsim}, \emph{active} SIMs~\cite{darsena2025design,shen2026active,abrardo2026cascade}, and reliability-aware unit designs~\cite{cao2026element}.

These architectures differ widely in maturity, and the distinction matters because performance claims attached to conceptual architectures rest on models not yet checked against hardware. Table~\ref{tab:maturity} classifies each by the strongest available evidence and lists the barriers that explain why the conceptual designs have not been built: independently tunable dispersion with acceptable loss (delay-augmented), cross-polarization coupling in tunable anisotropic cells (dual-polarization), layer alignment under mechanical stress (flexible), low-loss transitions between free space and guided conduits with reconfigurability (meta-fiber), biasing, stability, and heat (nonlinear and active), and curved multilayer fabrication (spherical).

\begin{table*}[t]
\centering
\caption{Maturity of SIM architectures: strongest available evidence and main implementation barriers.}
\label{tab:maturity}
\renewcommand{\arraystretch}{1.05}
\setlength{\tabcolsep}{2.5pt}
\scriptsize
\arrayrulecolor{SIMLine}
\begin{tabularx}{\textwidth}{|>{\columncolor{pastelpurple}\raggedright\arraybackslash}p{2.5cm}|>{\centering\arraybackslash}p{2.1cm}|>{\raggedright\arraybackslash}p{4.6cm}|>{\raggedright\arraybackslash}X|}
\hline
\rowcolor{pastelblue}\textbf{Architecture} & \textbf{Maturity} & \textbf{Strongest evidence} & \textbf{Main implementation barriers} \\
\hline
Static multilayer (D$^2$NN) & Experimental & THz, microwave, mmWave, optical prototypes~\cite{lin2018all,qian2020performing,li2021spectrally,gu2024classification,gao2024superresolution,yang2025metatraining} & Task-specific; no reconfiguration; tolerance $\pm$0.025--0.1~mm at mmWave \\
\hline
Programmable-passive (1-bit PIN) & Experimental & 5.8-GHz three-layer testbed with wireless metrics~\cite{wang2024multi} & 1-bit resolution; 1.2--1.9~dB per cell; bias power $\propto LM$; no reported latency, loss, or thermal data \\
\hline
Programmable-active (amplifying) & Experimental & 5.4-GHz PDNN~\cite{liu2022programmable}; 5.8-GHz nonlinear ReLU layers~\cite{ning2025multilayer} & $\approx$0.4~W per active unit, $\approx$250~W per network; noise, stability, heat; no thermal characterization \\
\hline
Nonlinear SIM & Experimental (ReLU layers) / conceptual (wireless use) & Nonlinear layers measured~\cite{ning2025multilayer}; wireless designs simulated~\cite{abbas2025nonlinear,abrardo2026nonlinear,stylianopoulos2026over} & Biasing networks, oscillation/stability, thermal drift, harmonics \\
\hline
Mechanically reconfigurable & Experimental & Sliding bilayer~\cite{shi2026sliding}; movable-type meta-atoms~\cite{yu2026movable} & Seconds to minutes reconfiguration; not for per-block adaptation \\
\hline
Multiport-validated stacks & Full-wave validated & $Z$-model vs.\ full-wave for communication and sensing~\cite{pettanice2026modeling}; unilateral active cascades~\cite{abrardo2026cascade} & Coupling blocks must be obtained per design (full-wave or measurement) \\
\hline
Two-layer meta-fiber SIM & Conceptual (cascaded-model simulation) & $>$11~dB transmit-power reduction vs.\ 7-layer SIM at BER $10^{-5}$~\cite{niu2026twolayer,niu2025introducing} & Low-loss free-space-to-conduit transitions; reconfigurable guided coupling \\
\hline
Two-layer FILM / flexible SIM & Conceptual & $>$7~dB transmit-power reduction vs.\ 7-layer SIM~\cite{niu2026twolayer}; conformal designs~\cite{magbool2025stacked} & Alignment and EM stability under deformation; actuation \\
\hline
Delay-augmented SIM & Conceptual & Space--time processing designs~\cite{alwazani2025delay} & Dispersion vs.\ loss vs.\ bandwidth uniformity; no dispersive tunable cell demonstrated \\
\hline
Dual-polarization SIM & Conceptual & Vector-valued transformations~\cite{zhang2025dual} & Independent tunability per polarization; cross-polarization coupling \\
\hline
BD-SIM & Conceptual (physically consistent model) & $S$-parameter model~\cite{nerini2024physically}; RZF with fewer layers~\cite{kim2026bdsim}; ISAC~\cite{jiao2026bdsim} & Inter-element interconnection circuits within a transmissive layer \\
\hline
Spherical, fluid-element, STAR-SIM & Conceptual & \cite{zhang2026spherical,wei2026fsim,wang2026starsim} & Curved multilayer fabrication; fluidic actuation; per-layer transmission/reflection control \\
\hline
\end{tabularx}
\arrayrulecolor{black}
\end{table*}

\subsection{Thermal and Power Management in Active and Dense Stacks}
\label{sec:thermal}

Programmable-active SIMs are advocated for joint amplitude and phase control and for compensating loss, but their thermal and power budget is largely unaddressed. The available numbers make the issue concrete: about 429~mW per active unit (150~mW amplifier, 53~mW detector, 224~mW adder) and about 252~W for the nonlinear network of~\cite{ning2025multilayer}; 640 amplifier chips of unreported power in the PDNN~\cite{liu2022programmable}; a power amplifier rated at 120~mA and 5.5~V per element in a 2-bit active transmissive cell~\cite{song2024active}. Extrapolating the per-unit figure of~\cite{ning2025multilayer} to $L=4$ active layers of $M=100$ elements gives about 170~W dissipated within a few hundred square centimeters of dielectric, poorly ventilated multilayer, a heat flux comparable to a power-amplifier module. Passive diode stacks are not exempt: a $60\times60$ dual-polarized 35-GHz PIN panel draws up to 103~W with all diodes on~\cite{wang2024power}. Heat has three effects the ideal model ignores: it shifts the resonance of the cells (and thus their phase and amplitude states); it reduces the phase range of liquid crystal, modeled in~\cite{gholian2025temperature} as $\theta_{\max}(T)=2\pi((T_c-T)/(T_c-T_r))^{\alpha}$ with $\alpha\approx0.2$--$0.25$; and it drives amplifier gain and noise figure. Active layers also add noise that propagates through subsequent layers and require stability margins against oscillation through inter-layer feedback, precisely the effect the forward-only model neglects. Energy-efficiency models of active SIMs~\cite{shen2026active,yeganeh2025enhancing} include amplifier efficiency but no thermal limit, and the unilateral-amplifier multiport model~\cite{abrardo2026cascade} is ideal with respect to stability and heat. Until active prototypes report temperature rise, gain drift, and noise figure, the amplitude--phase advantage of active SIMs must be weighed against a power and thermal budget at least as large as that of the RF chains they replace.

\subsection{Fabrication Tolerances, Misalignment, and Calibration}
\label{sec:tolerance}

Focusing and nulling rely on coherent superposition across a large aperture and across layers, so they are sensitive to spacing, alignment, meta-atom positions, and phase states. The evidence is sparse but consistent. In the 5.8-GHz testbed~\cite{wang2024multi}, the received power of the three-layer stack varies by 13~dB between $\lambda/4$ and $\lambda$ spacing although the configuration was re-optimized for each spacing. For a five-layer terahertz diffractive network, random lateral shifts up to $\pm2.12\lambda$ reduce the blind-testing accuracy from 97.8\% to 38.4\%, axial shifts of $19.2\lambda$ leave 49.8\%, and misalignment-augmented training recovers 94.4\%; experimentally the standard design failed in 23 cases versus 2 for the misalignment-aware one~\cite{mengu2020misalignment}. In a two-layer optical chip, overlay errors within $\pm2$~$\mu$m have minimal effect while a $\pm50$-$\mu$m substrate-thickness error costs about 2\% accuracy~\cite{dong2024eternal}. The measured focal efficiency of a three-layer 11-GHz network is about 2~dB below simulation~\cite{gu2024classification}, and millimeter-wave stacks are machined to $\pm0.025$--$0.1$~mm, about 7-bit effective phase accuracy at 27.5~GHz~\cite{gao2024superresolution,yang2025metatraining}. At the model level, \cite{jia2026calibration} considers spacing irregularities of $\pm0.1\%\lambda$, height offsets of $\pm1\%\lambda$, and rotations of $\pm0.01^\circ$ in a four-layer 28-GHz stack and shows that multi-stage pilot-based calibration reduces the NMSE of the inter-layer matrices by 12.07~dB (layers 1--2; 9.25~dB on average) and of the user-to-SIM channel by 8.48~dB; element failure and misalignment reliability is analyzed in~\cite{cao2026element}. Three consequences follow for near-field SIM beamforming: the tolerance analysis belongs in the design (misalignment-aware training or robust optimization over a geometry-error distribution costs little and prevents the collapse observed in~\cite{mengu2020misalignment}); calibration of the inter-layer coefficients is a prerequisite for focusing, and since exhaustive measurement of $2^{bLM}$ states is impossible, pilot-based, compressed, or digital-twin-assisted calibration is needed; and prototypes must report the alignment method and residual error, because the 2-dB and 13-dB discrepancies above are the scale of the gains many simulation papers claim.

\subsection{Theory--Prototype Gap and a Reporting Checklist}
\label{sec:gap}

Comparing Tables~\ref{tab:prototypes}--\ref{tab:maturity} with the assumptions of Sections~\ref{sec:taxonomy}--\ref{sec:apps} quantifies the gap. Most optimization papers assume continuous or 2--4-bit phases; the only programmable SIM testbed is 1-bit. Most assume lossless layers; measured cells lose 1.2--1.9~dB each and measured focal efficiencies are 2~dB below simulation. Most assume known propagation matrices; calibration improves them by about 10~dB, so the uncalibrated error is of the order of the signal. Most assume instantaneous reconfiguration; fast technologies are 1-bit and slow ones take tens of milliseconds to minutes. Most compare against digital baselines without bias, controller, and amplifier power; measured values are 12~mW per diode, about 5~W per FPGA, and about 0.4~W per active unit. And the only wireless measurements (received power and DoA error for one to three layers~\cite{wang2024multi}) used configurations found by exhaustive search rather than by the algorithms proposed in the same paper, itself a symptom of the calibration problem.

To close the gap, prototype papers should report: frequency and bandwidth; $L$, $M$, aperture, and spacing; tuning mechanism, phase and amplitude resolution, and the measured state-dependent amplitude of the cells; insertion loss per layer and total radiation efficiency; control and bias power including the controller; reconfiguration latency; alignment method, residual geometry error, and calibration procedure; measured beampatterns or channel responses; and end-to-end communication or sensing performance against a digital baseline with equal RF chains and power budget. Active stacks should add noise figure, gain range, linearity, stability margin, and temperature rise; wideband stacks the measured frequency dependence of the cells. Algorithm papers should state which of these quantities their model includes, so that a reader can place the result in Fig.~\ref{fig:model_hierarchy}.

\begin{lessons}
\begin{itemize}[leftmargin=*, nosep]
\item Only one programmable SIM with wireless metrics has been measured (three 1-bit layers at 5.8~GHz) and only one multilayer prototype reports power and latency; all other performance claims are model-based.
\item Fast tuning is 1-bit and lossy; low-loss or high-resolution tuning is slow. Match the methodological family to the tuning latency.
\item Active layers cost about 0.4~W per element and add noise and heat; their amplitude--phase advantage must be evaluated against a thermal budget no prototype has reported.
\item Geometry errors of a fraction of a wavelength move measured performance by 2--13~dB; misalignment-aware design and calibration of the inter-layer coefficients are prerequisites for focusing and nulling.
\item Delay-augmented, dual-polarization, flexible, meta-fiber, spherical, fluid, STAR, and BD architectures are conceptual; their reported gains are cascaded-model results awaiting hardware validation.
\end{itemize}
\end{lessons}

\section{Performance and Trade-off Analysis}
\label{sec:performance}

This section consolidates the quantitative trade-offs that the SIM literature has established. Each subsection states the trade-off, shows the published result that establishes it (figures are reproduced from the cited works and identified by their original figure numbers), states the parameters, and derives a design rule. Nothing is re-simulated; the results are those of the cited works, whose assumption level in Fig.~\ref{fig:model_hierarchy} is stated. Table~\ref{tab:tradeoff_numbers} collects the numbers.

\subsection{Stacking Depth: Expressiveness, Saturation, and Loss}
\label{sec:perf_depth}

\textbf{Trade-off:} Each layer adds $M$ degrees of freedom and one diffraction stage, hence transformation richness; it also adds one insertion loss, one set of geometry errors, and $M$ control lines. Under the lossless cascaded model performance increases with $L$ and saturates; with loss it peaks and then decreases.

 \begin{figure}[t]
\centering

\begin{subfigure}[t]{0.48\columnwidth}
\centering
\includegraphics[width=\linewidth,height=3.4cm,keepaspectratio]{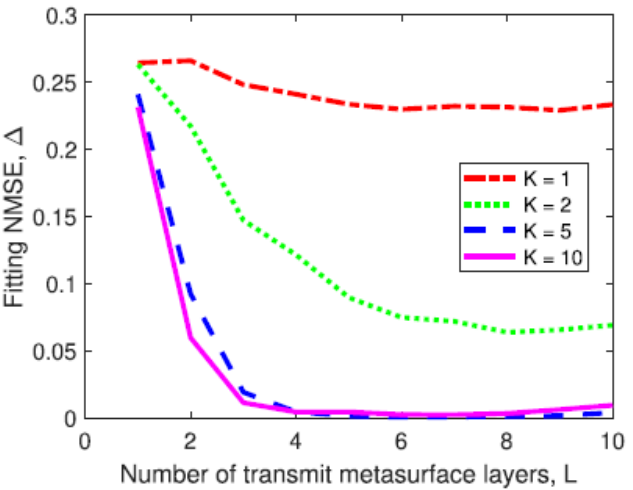}
\caption{NMSE vs.\ $L$~\cite{an2023stacked}.}
\label{fig:depth_a}
\end{subfigure}\hfill
\begin{subfigure}[t]{0.48\columnwidth}
\centering
\includegraphics[width=\linewidth,height=3.4cm,keepaspectratio]{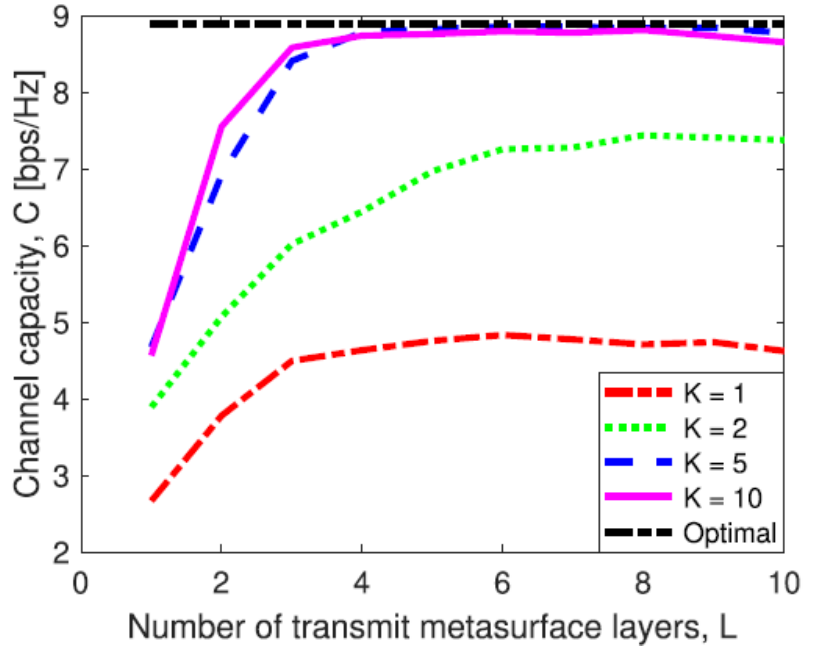}
\caption{Capacity vs.\ $L$~\cite{an2023stacked}.}
\label{fig:depth_b}
\end{subfigure}

\vspace{1mm}

\begin{subfigure}[t]{0.48\columnwidth}
\centering
\includegraphics[width=\linewidth,height=3.4cm,keepaspectratio]{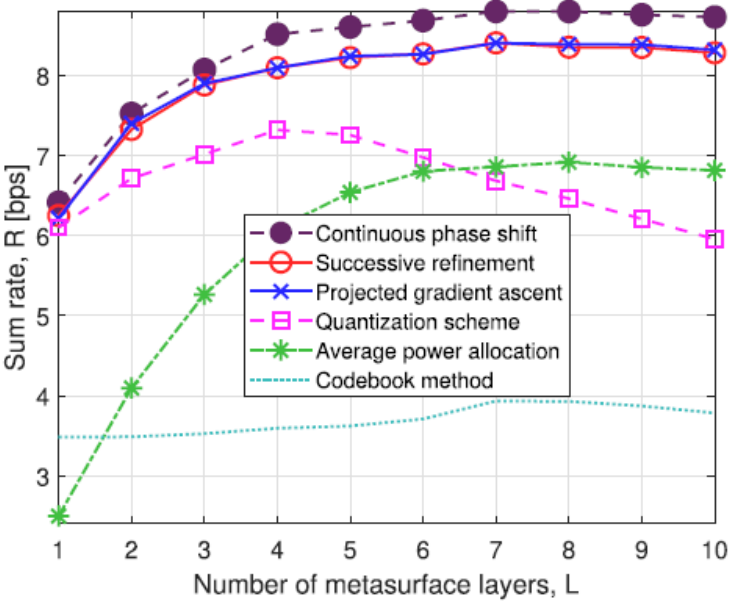}
\caption{Multi-user sum rate vs.\ $L$~\cite{an2025stacked}.}
\label{fig:depth_c}
\end{subfigure}\hfill
\begin{subfigure}[t]{0.48\columnwidth}
\centering
\includegraphics[width=\linewidth,height=3.4cm,keepaspectratio]{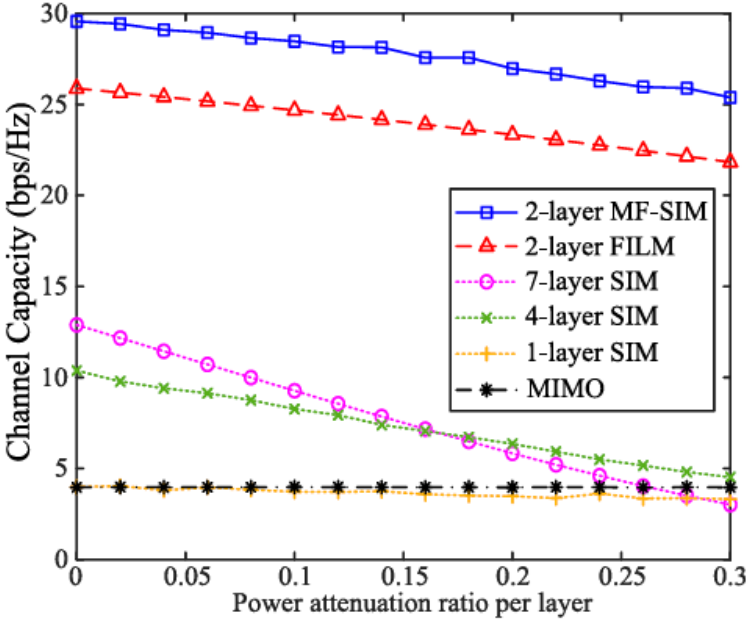}
\caption{Capacity vs.\ per-layer loss~\cite{niu2026twolayer}.}
\label{fig:depth_d}
\end{subfigure}

\caption{Depth trade-off. (a)--(b) Lossless cascaded model at 28~GHz, $M=100$ meta-atoms per layer at $\lambda/2$ pitch, thickness $5\lambda$, $S=4$, $P_t=20$~dBm~\cite{an2023stacked}. (c) Multi-user downlink with 2-bit phases~\cite{an2025stacked}. (d) Per-layer power attenuation~\cite{niu2026twolayer}: the 7-layer stack drops below the 4-layer one at 17\% loss per layer and below MIMO at 26\%.}
\label{fig:depth}
\end{figure}

\textbf{Evidence:} Under the lossless cascaded model (Fig.~\ref{fig:depth}(a)--(c)), the fitting NMSE saturates for $L\geq5$ and the capacity follows, with about 150\% improvement over MIMO and RIS baselines and twice the gain of a single layer with the same total meta-atoms~\cite{an2023stacked}, and the multi-user sum rate peaks at $L=7$, 130\% above a single layer under 2-bit phases~\cite{an2025stacked}. With a per-layer attenuation (Fig.~\ref{fig:depth}(d)), the 7-layer stack loses its advantage over the 4-layer stack at 17\% per layer (0.8~dB) and over MIMO at 26\% (1.3~dB)~\cite{niu2026twolayer}, below the 1.2--1.9~dB measured per 1-bit transmissive cell (Table~\ref{tab:tuning}). Consistently, \cite{rostamikafaki2026loss} reports 17\% loss-induced degradation at six layers, \cite{dong2026geometry} finds the near-field sum rate non-monotonic in $L$ and thickness, \cite{shi2025energy} finds the best compromise at 2--5 layers, and \cite{sheemar2026uav} finds gains that saturate in $L$ while growing in $M$.

\textbf{Design rule:} Under realistic loss the useful depth is 2--5 layers for diode-based layers; deeper stacks pay off only with low-loss (static or mechanically tuned) layers or with active compensation and its power and thermal costs (Section~\ref{sec:thermal}). Two-layer designs with richer inter-layer connectivity (meta-fiber, FILM, BD layers) obtain the expressiveness of deep diagonal stacks with fewer loss stages.

\subsection{Aperture Versus Depth}
\label{sec:perf_aperture}

\textbf{Trade-off:} Performance can be bought with meta-atoms per layer ($M$: aperture, array gain, angular resolution, focusing range) or with layers ($L$: transformation richness at fixed aperture); the two are not interchangeable.

 \begin{figure}[t]
\centering
\begin{subfigure}[t]{0.48\columnwidth}
\centering
\includegraphics[
    width=\linewidth,
    height=3.4cm,
    keepaspectratio
]{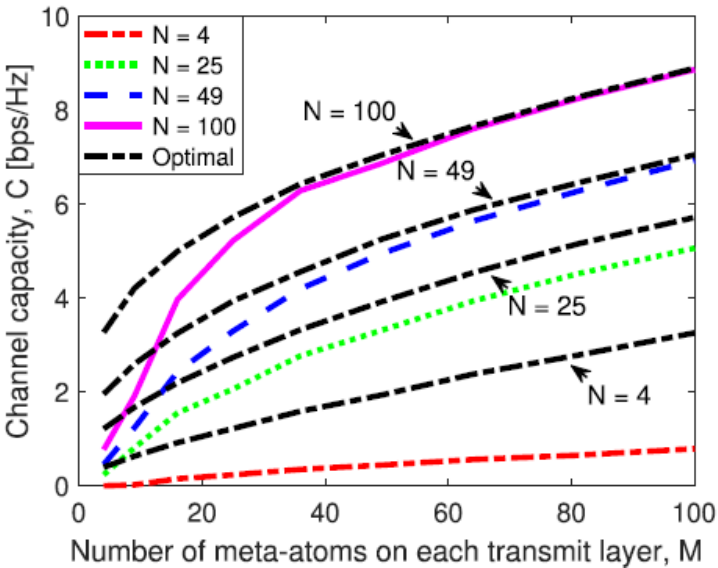}
\caption{Capacity vs.\ number of meta-atoms $M$ per layer~\cite{an2023stacked}.}
\label{fig:aperture_a}
\end{subfigure}
\hfill
\begin{subfigure}[t]{0.48\columnwidth}
\centering
\includegraphics[
    width=\linewidth,
    height=3.4cm,
    keepaspectratio
]{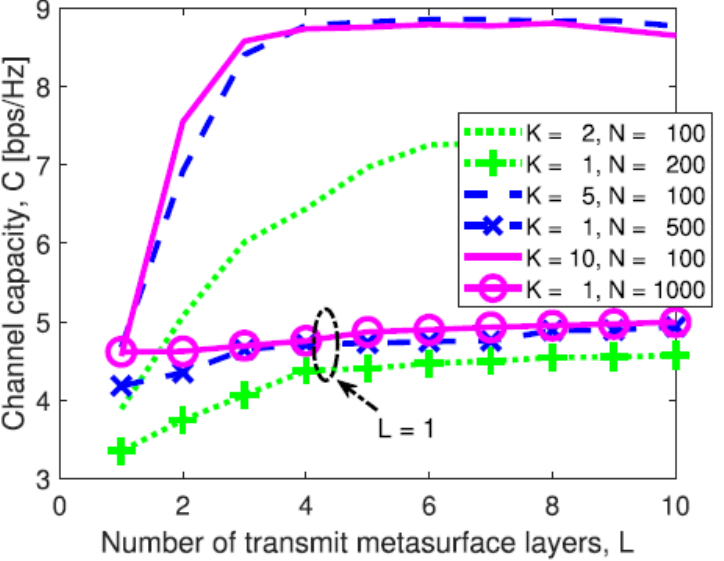}
\caption{Multilayer vs.\ single-layer SIM with equal total
meta-atoms~\cite{an2023stacked}.}
\label{fig:aperture_b}
\end{subfigure}
\caption{Aperture versus depth under the lossless cascaded
model~\cite{an2023stacked} (parameters as in Fig.~\ref{fig:depth}).
(a) Channel capacity versus the number of meta-atoms $M$ on each
transmit metasurface layer for different values of $N$, together with
the corresponding optimal benchmarks.
(b) Channel capacity versus the number of transmit metasurface layers
$L$ for different $(K,N)$ configurations, illustrating the trade-off
between increasing the aperture of each layer and distributing a fixed
number of meta-atoms across multiple layers. With a Lorentzian element
model, \cite{gonzalez2026lorentzian} finds that the ideal model
overestimates deep stacks by up to 7.5~bit/s/Hz and that aperture should
take priority over depth.  }
\label{fig:aperture}
\end{figure}

\textbf{Evidence:} Quadrupling $M$ from 25 to 100 adds about 8~bit/s/Hz (Fig.~\ref{fig:aperture}(a)), a quadratic array-gain law, and for the same total meta-atoms a multilayer stack achieves about twice the improvement of a single layer (Fig.~\ref{fig:aperture}(b))~\cite{an2023stacked}. With state-dependent amplitude, deep stacks lose more than shallow ones, the ideal model overestimates the sum rate by up to 7.5~bit/s/Hz, and aperture becomes preferable~\cite{gonzalez2026lorentzian}; the UAV results of~\cite{sheemar2026uav} and the near-field range analysis of~\cite{zhao2026nearfield} agree.

\textbf{Design rule:} Spend the hardware budget first on aperture, then on a small number of layers; the marginal value of a layer decreases with $L$ and with cell loss, that of aperture does not until near-field and control-overhead limits are reached.

\subsection{Phase Resolution and Discrete Control}
\label{sec:perf_quant}

\textbf{Trade-off:} Fewer bits mean cheaper, faster, lower-loss cells but a coarser feasible set; the loss depends on whether quantization is applied after continuous optimization or inside it, and on the objective.

 \begin{figure}[t]
\centering
\begin{subfigure}[t]{0.48\columnwidth}
\centering
\includegraphics[
    width=\linewidth,
    height=3.4cm,
    keepaspectratio
]{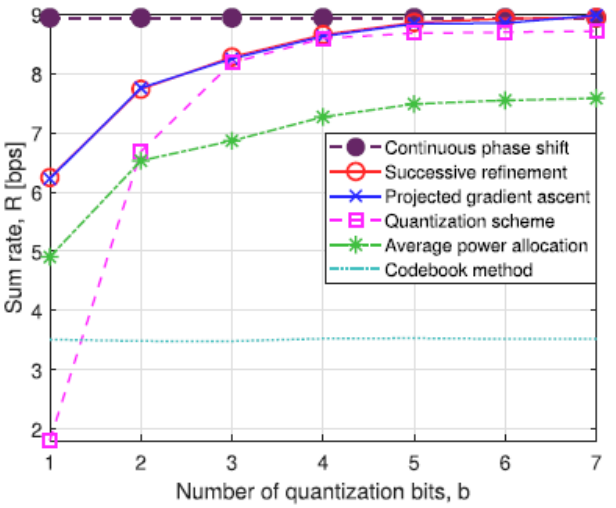}
\caption{Sum rate vs.\ quantization bits $b$ for $L=7$, $M=49$,
and $K=4$, including the continuous-phase benchmark
~\cite{an2025stacked}.}
\label{fig:quant}
\end{subfigure}
\hfill
\begin{subfigure}[t]{0.48\columnwidth}
\centering
\includegraphics[
    width=\linewidth,
    height=3.4cm,
    keepaspectratio
]{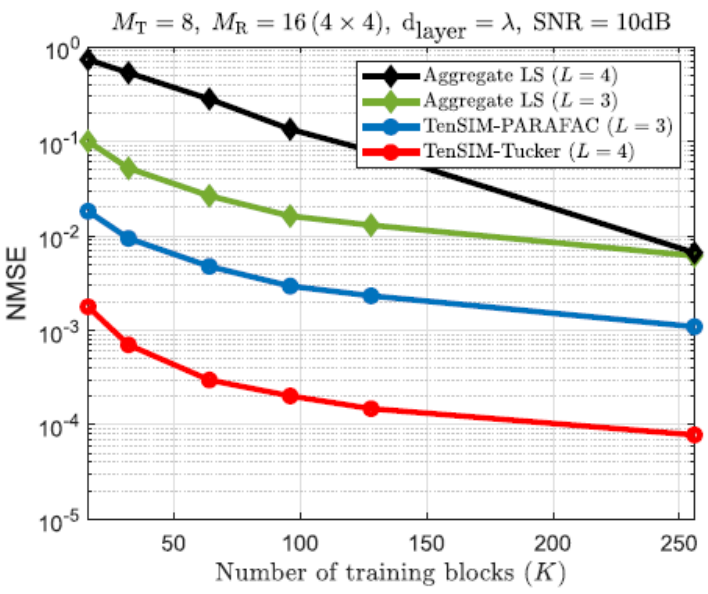}
\caption{Cascaded-channel NMSE vs.\ number of pilot blocks,
comparing TenSIM-PARAFAC/Tucker with aggregate LS
~\cite{dealmeida2026tensim}.}
\label{fig:csi}
\end{subfigure}
\caption{
Practical control and channel-acquisition trade-offs.
(a) Phase-resolution trade-off~\cite{an2025stacked}: for a
7-layer SIM with $M=49$ meta-atoms per layer and $K=4$ users,
quantization within the optimizer rapidly approaches continuous-phase
performance; 2-bit control incurs only about 0.3~bit/s/Hz loss,
4 bits closely approach the continuous benchmark, and $b\geq6$
is essentially indistinguishable from it.
(b) Pilot-overhead trade-off~\cite{dealmeida2026tensim}:
cascaded-channel NMSE versus the number of training blocks $K$ for
$M_T=8$, $M_R=16$ ($4\times4$), $d_{\rm layer}=\lambda$, and
SNR $=10$~dB. Tensor-structured TenSIM-PARAFAC and TenSIM-Tucker
substantially outperform unstructured aggregate LS and remain effective
with approximately $N$ pilot blocks, whereas aggregate LS requires
on the order of $N^2$. 
}
\label{fig:quant_csi}
\end{figure}

\textbf{Evidence:} With quantization inside the optimizer (Fig.~\ref{fig:quant}), 2-bit phases cost about 0.3~bit/s/Hz (below 0.5 across the depth range), 4 bits approach the continuous benchmark, and 6 bits are indistinguishable~\cite{an2025stacked}; quantum-inspired discrete designs reach near-continuous performance with 3 bits~\cite{ji2026qubo,ji2026hmimo}, and robust quantization-aware projection preserves the gains under imperfect CSI~\cite{hassan2026robust}. These are rate objectives with $LM\gg K$ degrees of freedom; for null-type objectives the achievable null depth is bounded by phase and amplitude resolution and the required resolution grows with the target depth~\cite{niu2024enhancing,hassan2026robust}.

\textbf{Design rule:} For rate objectives, 2--3 bits inside the optimizer suffice, so multi-diode or low-resolution varactor cells are adequate; 1-bit cells (the only ones demonstrated in an SIM testbed) cost more and must be evaluated explicitly. For nulling, choose the resolution from the required null depth and consider amplitude control.

\subsection{CSI Acquisition: Pilot Overhead, Statistical CSI, and Net Rate}
\label{sec:perf_csi}

\textbf{Trade-off:} The effective channel has $KM$ (or $KLM$) unknowns observed through $S$ RF chains; more pilots and training states improve the estimate but consume the coherence interval; statistical designs avoid the overhead but lose adaptivity.

\textbf{Evidence:} Training only the central layers makes the cascaded channel a PARAFAC or Tucker tensor, so alternating least squares needs about $N$ pilot blocks instead of $N^2$ (Fig.~\ref{fig:csi})~\cite{dealmeida2026tensim}. The training sub-phases of a hybrid digital/EM estimator should be optimized for net rate, with an optimum depending on spatial correlation and significant gains over fixed training lengths~\cite{bahanshal2026training}. Statistical-CSI design~\cite{papazafeiropoulos2024achievable} and TD3 control~\cite{hu2026td3} stay competitive with instantaneous-CSI optimization for large SIMs at a fraction of the overhead, and the polar-domain estimator of~\cite{yao2026sparse} is four times faster than standard sparse Bayesian learning at equal accuracy.

\textbf{Design rule:} Use structured (tensor, subspace, sparse) estimators exploiting the cascade, optimize training length for net rate rather than NMSE, configure the slow SIM layer from statistical CSI, reserve instantaneous estimation for the digital layer and for near-field or mobile users, and calibrate the device with the channel (Section~\ref{sec:perf_geometry}).

 \subsection{Bandwidth: Beam Squint and When Depth Helps}
\label{sec:perf_wideband}

\textbf{Trade-off:} One configuration serves all subcarriers while each cell and inter-layer path is frequency dependent; depth adds frequency-selective phase accumulation, so the configuration optimal at the center frequency squints at the band edges.

 \begin{figure}[t]
\centering
\begin{subfigure}[t]{0.48\columnwidth}
\centering
\includegraphics[width=\linewidth,height=3.4cm,keepaspectratio]{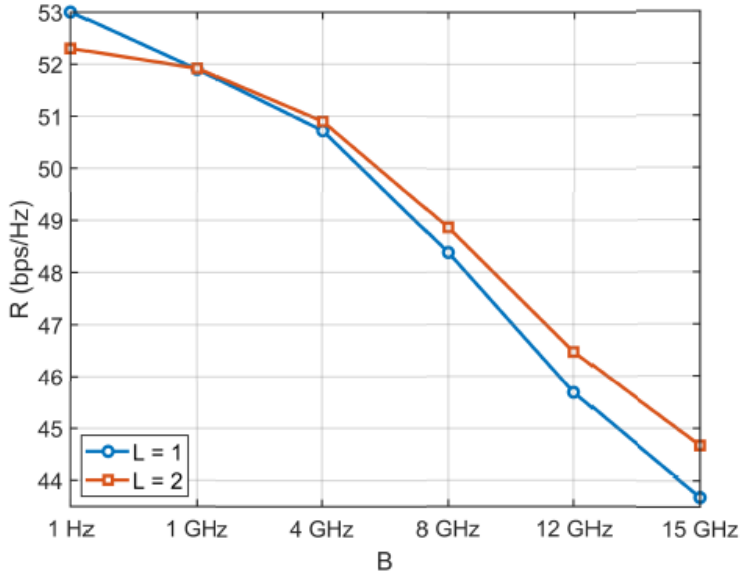}
\caption{Spectral efficiency vs.\ bandwidth~\cite{ibrahim2026wideband}.}
\label{fig:wideband_a}
\end{subfigure}\hfill
\begin{subfigure}[t]{0.48\columnwidth}
\centering
\includegraphics[width=\linewidth,height=3.4cm,keepaspectratio]{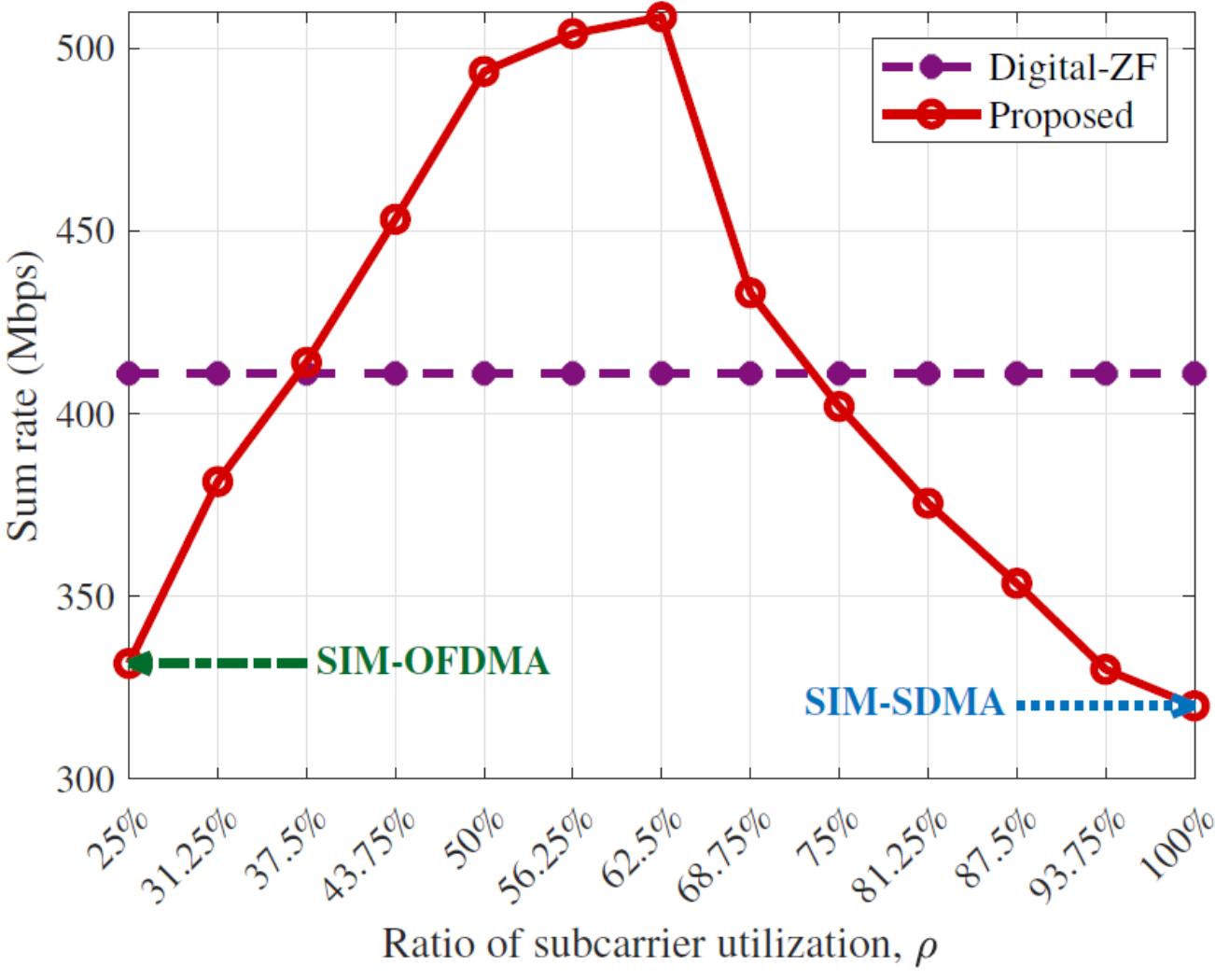}
\caption{Sum rate vs.\ subcarrier reuse~\cite{li2025fundamental}.}
\label{fig:wideband_b}
\end{subfigure}
\caption{Bandwidth trade-off. (a) With a coupling-aware wideband circuit model and the same total number of elements, a single-layer RIS is slightly better in the narrowband regime, whereas the multilayer SIM overtakes it as the bandwidth grows~\cite{ibrahim2026wideband}; the companion curve vs.\ noise variance (Fig.~9 therein) shows the same crossover from low to high SNR. (b) In SIM-assisted OFDMA (4 users, 16 subcarriers, 28~GHz, 40~MHz), the sum rate peaks at a subcarrier-reuse ratio $\rho\approx62.5\%$ and exceeds the digital-ZF baseline for $K_c\in[6,12)$ shared subcarriers; beyond the optimum the stack can no longer suppress the inter-user interference~\cite{li2025fundamental}. }
\label{fig:wideband}
\end{figure}

\textbf{Evidence:} The RIS-versus-SIM study of~\cite{ibrahim2026wideband} is the most direct (Fig.~\ref{fig:wideband}(a)): under a coupling-aware wideband circuit model and equal total element count, the single-layer RIS achieves slightly higher spectral efficiency in the narrowband regime and at low SNR, where noise rather than inter-user interference limits performance, whereas the multilayer SIM outperforms it as the bandwidth or the SNR grows, because its additional spatial degrees of freedom suppress interference and hold the gain across the band; a partially reconfigurable SIM (only one tunable layer) is near-optimal. The OFDMA trade-off of~\cite{li2025fundamental} (Fig.~\ref{fig:wideband}(b)) places the optimum reuse at about 62.5\%, above which the stack can no longer create interference-free channels for the shared subcarriers. Unfolding finds one configuration for all bands with consistent gains on unseen channels~\cite{ibrahim2026unfolding}, and the baseband-free OFDM transmitters of~\cite{li2026basebandfree,li2026fullywave} approach digital benchmarks by fitting a second stack offline to the multiport OFDM operator and modeling the residual as an effective coupling matrix.

\textbf{Design rule:} The advantage of depth is a multi-user, wideband, interference-limited advantage; in narrowband or noise-limited links a single layer is as good. Evaluate at the band edges, include the measured dispersion of the cells, and consider delay-augmented layers (Section~\ref{sec:advanced}) at large fractional bandwidth.

\subsection{Geometry, Alignment, and Calibration Sensitivity}
\label{sec:perf_geometry}

\textbf{Trade-off:} Closer layers give stronger coupling and thinner stacks but violate the forward-only assumption and amplify spacing errors; larger spacing weakens coupling and increases spillover. Focusing and nulling require the inter-layer coefficients to a fraction of a wavelength.

 \begin{figure}[t]
\centering
\begin{subfigure}[t]{0.48\columnwidth}
\centering
\includegraphics[
    width=\linewidth,
    height=3.4cm,
    keepaspectratio
]{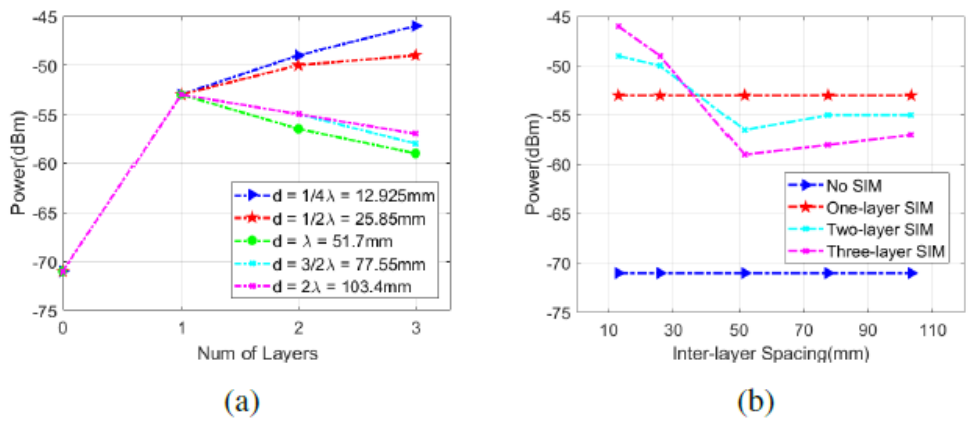}
\caption{Measured received power vs.\ number of layers and inter-layer spacing~\cite{wang2024multi}.}
\label{fig:geometry_a}
\end{subfigure}\hfill
\begin{subfigure}[t]{0.48\columnwidth}
\centering
\includegraphics[
    width=\linewidth,
    height=3.4cm,
    keepaspectratio
]{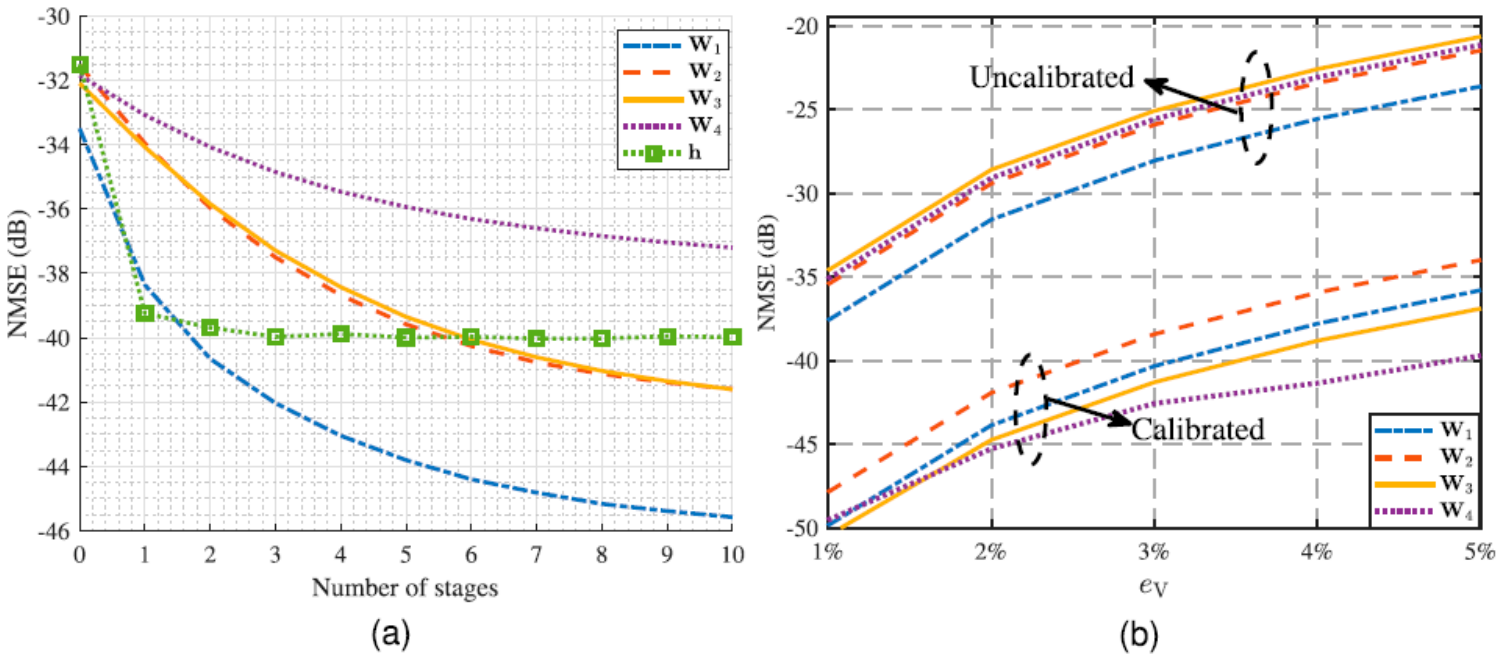}
\caption{Calibration of inter-layer coefficients~\cite{jia2026calibration}.}
\label{fig:geometry_b}
\end{subfigure}
\caption{Geometry sensitivity. (a) Three 1-bit $16\times16$ layers, TX power 10~dBm, TX--SIM 1.75~m, SIM--RX 7.2~m: received power rises from $-53$~dBm (one layer) to $-46$~dBm (three layers at $\lambda/4$) but falls to $-59$~dBm at $\lambda$ spacing~\cite{wang2024multi}. (b) Four-layer 28-GHz stack with errors of $\pm0.1\%\lambda$, $\pm1\%\lambda$, $\pm0.01^\circ$: ten-stage calibration reduces the inter-layer NMSE by up to 12.07~dB and the channel NMSE by 8.48~dB~\cite{jia2026calibration}.}
\label{fig:geometry}
\end{figure}

\textbf{Evidence:} Three layers at $\lambda/4$ improve on one layer by 7~dB, whereas the same layers at $\lambda$ spacing are 6~dB worse than one layer (Fig.~\ref{fig:geometry}(a))~\cite{wang2024multi}, a 13-dB dependence on a parameter the cascaded model treats as a constant; calibration (Fig.~\ref{fig:geometry}(b)) implies that uncalibrated inter-layer coefficients carry an error of the order of 10~dB~\cite{jia2026calibration}. The misalignment results of~\cite{mengu2020misalignment} and the 2-dB simulated-versus-measured gap of~\cite{gu2024classification} complete the picture (Section~\ref{sec:tolerance}).

\textbf{Design rule:} Treat spacing as a design variable~\cite{dong2026geometry} and the inter-layer coefficients as quantities to calibrate; include a geometry-error distribution in the optimization or training data; report residual alignment error with every measurement.

\subsection{Energy Efficiency and the Power Model}
\label{sec:perf_ee}

\textbf{Trade-off:} The spectral-efficiency gain grows sub-linearly with $L$ and $M$ while control power $LM\,P_{\rm meta}$ and loss-induced transmit power grow at least linearly, so energy efficiency has an interior optimum that depends on $P_{\rm meta}$ and on the baseline.

  \begin{figure}[t]
\centering
\begin{subfigure}[t]{0.48\columnwidth}
\centering
\includegraphics[width=\linewidth,height=3.4cm,keepaspectratio]{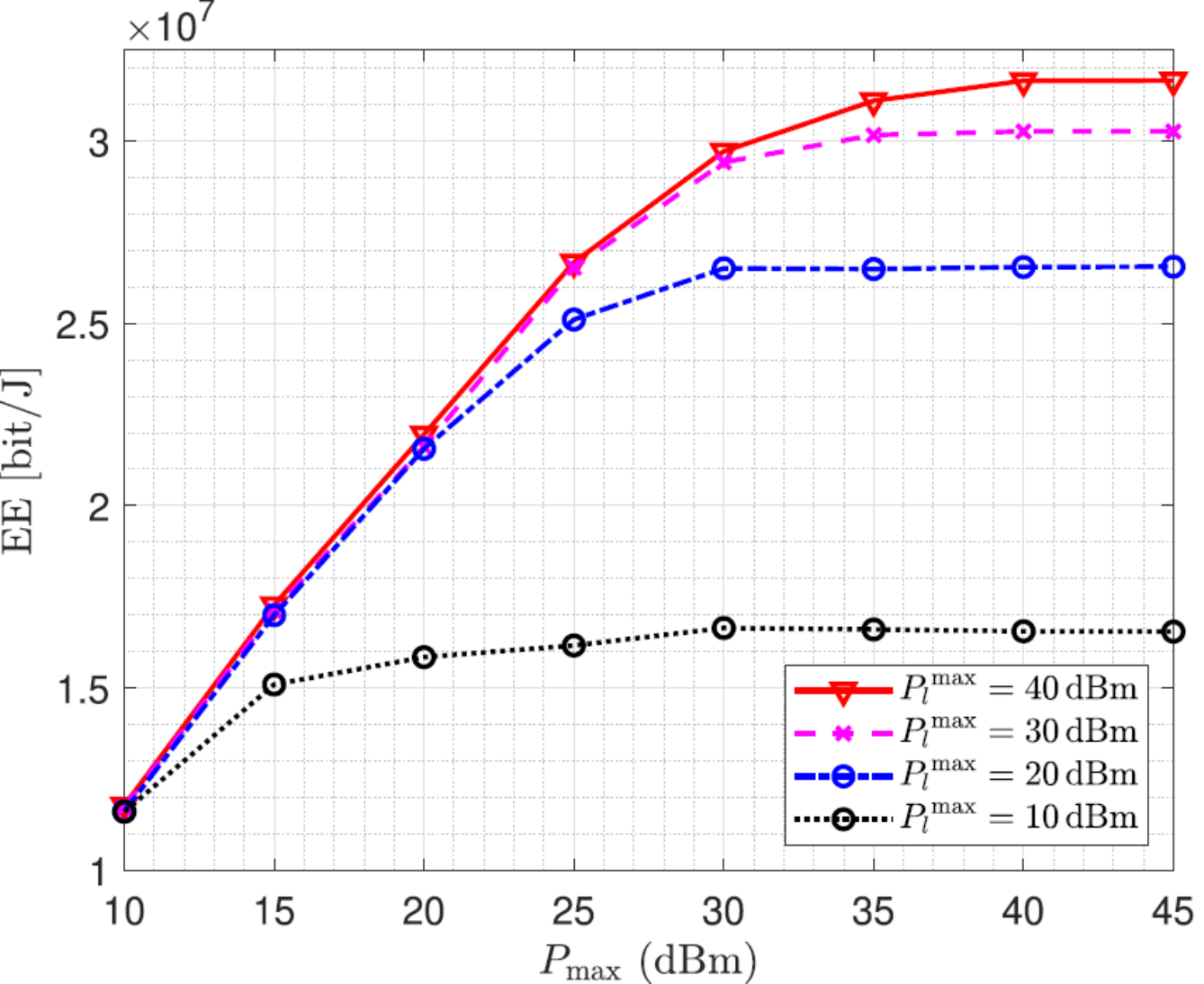}
\caption{EE vs.\ transmit-power budget $P_{\max}$ for different per-antenna limits~\cite{shi2025energy}.}
\label{fig:ee_a}
\end{subfigure}\hfill
\begin{subfigure}[t]{0.48\columnwidth}
\centering
\includegraphics[width=\linewidth,height=3.4cm,keepaspectratio]{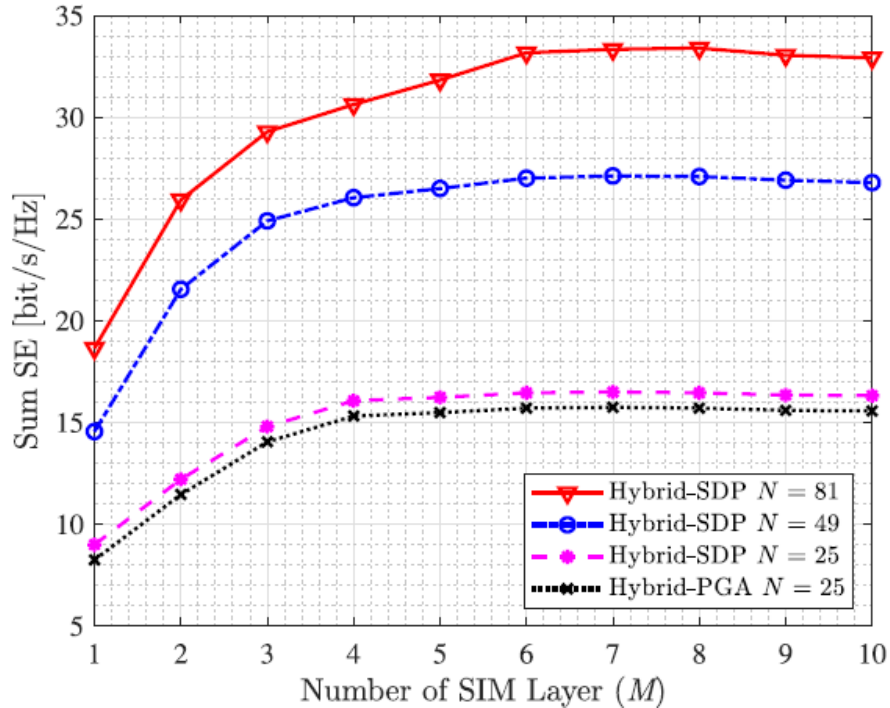}
\caption{Sum SE vs.\ number of layers for different numbers of meta-atoms per layer~\cite{shi2025energy}.}
\label{fig:ee_b}
\end{subfigure}
\caption{Energy-efficiency trade-off~\cite{shi2025energy} (28~GHz, 10~MHz, $K=4$ users, static BS power 4.5~W, 0.4~W per active RF chain, $P_{\rm meta}=10$~dBm, controller 25~dBm; hybrid wave-domain/digital precoding with RF-chain selection). (a) With a four-layer, 49-element-per-layer SIM, EE grows with $P_{\max}$ and then saturates once the per-antenna limit binds: an EE-optimal design uses only part of the budget, and a 10-dBm per-antenna limit halves the attainable EE. (b) Sum SE saturates after 4--6 layers and decreases slightly beyond eight, whereas increasing the meta-atoms per layer from 25 to 81 doubles it; the SDP and PGA designs differ by less than 1~bit/s/Hz. In~\cite{shi2025energy}, $M$ and $N$ denote the numbers of layers and of meta-atoms per layer, i.e., $L$ and $M$ here.}
\label{fig:ee}
\end{figure}

\textbf{Evidence:} With the power model~\eqref{eq:power}, energy efficiency peaks at moderate depth, hybrid precoding improves it by about 80\% over digital precoding without SIM, and the optimal $M$ decreases as $P_{\rm meta}$ grows~\cite{shi2025energy}; broadcast-MIMO analyses agree~\cite{perovic2025energy}. Active layers can still improve energy efficiency when amplifier power is included~\cite{shen2026active}, and the ISAC design of~\cite{ataeebojd2026isac} reports about 230\% gain over no SIM with fewer BS antennas. Table~\ref{tab:tuning} suggests that $P_{\rm meta}=10$~mW is realistic for PIN diodes but optimistic once the controller (about 5~W per FPGA) is included, and pessimistic for liquid crystal, whose latency precludes per-block adaptation.

\textbf{Design rule:} State the power model and the baseline; with diode-based layers the optimum is shallow and the controller power matters; with liquid-crystal layers the optimum is deeper but the design must be two-timescale; compare active layers at equal total DC power.

\subsection{Computational Cost: Optimization Versus Learning Versus Closed Forms}
\label{sec:perf_complexity}

\textbf{Trade-off:} The configuration must be computed within the coherence (or statistics-update) interval on the controller; iterative optimizers scale with $I\,LM^2K$ (Table~\ref{tab:method_compare}), closed-form updates and unfolded networks reduce the constant and the iteration count at the cost of generality or offline training.

 \begin{figure}[t]
\centering
\begin{subfigure}[t]{0.48\columnwidth}
\centering
\includegraphics[
    width=\linewidth,
    height=3.4cm,
    keepaspectratio
]{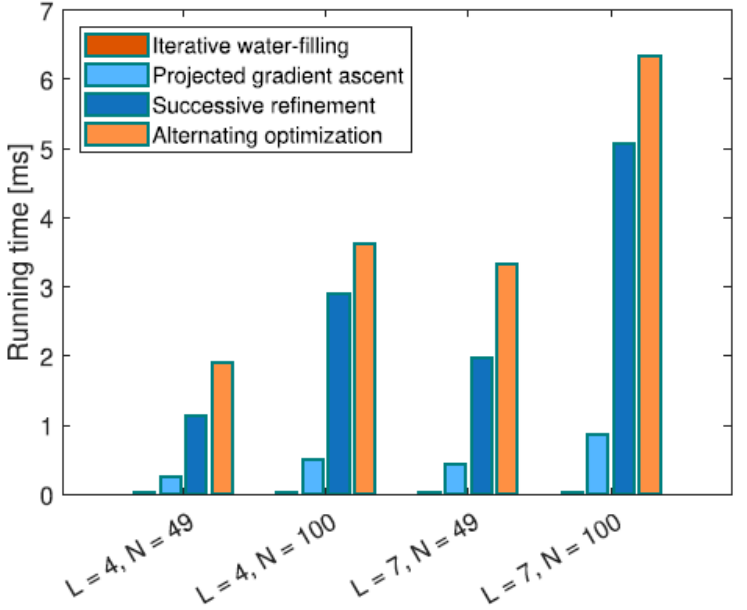}
\caption{Runtime of model-based solvers~\cite{an2025stacked}.}
\label{fig:complexity_a}
\end{subfigure}\hfill
\begin{subfigure}[t]{0.48\columnwidth}
\centering
\includegraphics[
    width=\linewidth,
    height=3.4cm,
    keepaspectratio
]{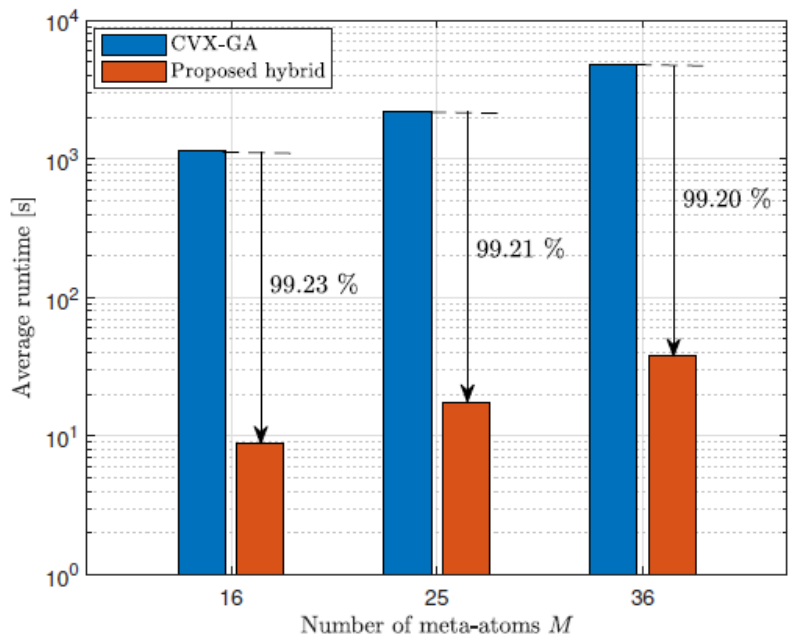}
\caption{Closed-form vs.\ convex-programming updates~\cite{park2026efficient}.}
\label{fig:complexity_b}
\end{subfigure}
\caption{Computational-cost trade-off. Closed-form per-AP and per-layer updates reduce runtime by more than 99\% at the same sum rate~\cite{park2026efficient}; unfolded networks match high-iteration benchmarks at a fraction of the runtime~\cite{wang2026unfolding} after an offline training whose cost the SIM literature does not report.}
\label{fig:complexity}
\end{figure}

\textbf{Evidence:} Fig.~\ref{fig:complexity} reproduces the runtime comparison of~\cite{an2025stacked} and the runtime-versus-$M$ result of~\cite{park2026efficient}. Unfolded networks~\cite{wang2026unfolding,ibrahim2026unfolding,xue2026unfolded} match high-iteration benchmarks at a fraction of the runtime and are reported robust to model mismatch; their training cost is unreported, and the diffractive-network literature indicates $10^4$--$10^5$ samples and hours of GPU time (Section~\ref{sec:learning}). Physically consistent models add one to two orders of magnitude per evaluation (Table~\ref{tab:complexity}) but scale linearly in $L$.

\textbf{Design rule:} For per-block reconfiguration use closed-form or unfolded updates on the cascaded model; for design and calibration use the structured $Z$ or $T$ model offline; report wall-clock time on the target controller and the training cost against the coherence interval and the hardware latency.

\subsection{Measured Versus Simulated Performance}
\label{sec:perf_measured}

\textbf{Trade-off:} Wave-domain processing delivers nanosecond latency and near-zero digital operations, but measured accuracy or gain is below the simulated one by an amount set by loss, alignment, resolution, and calibration.

 \begin{figure}[t]
\centering
\includegraphics[
    width=1\columnwidth,
    height=6.5cm
]{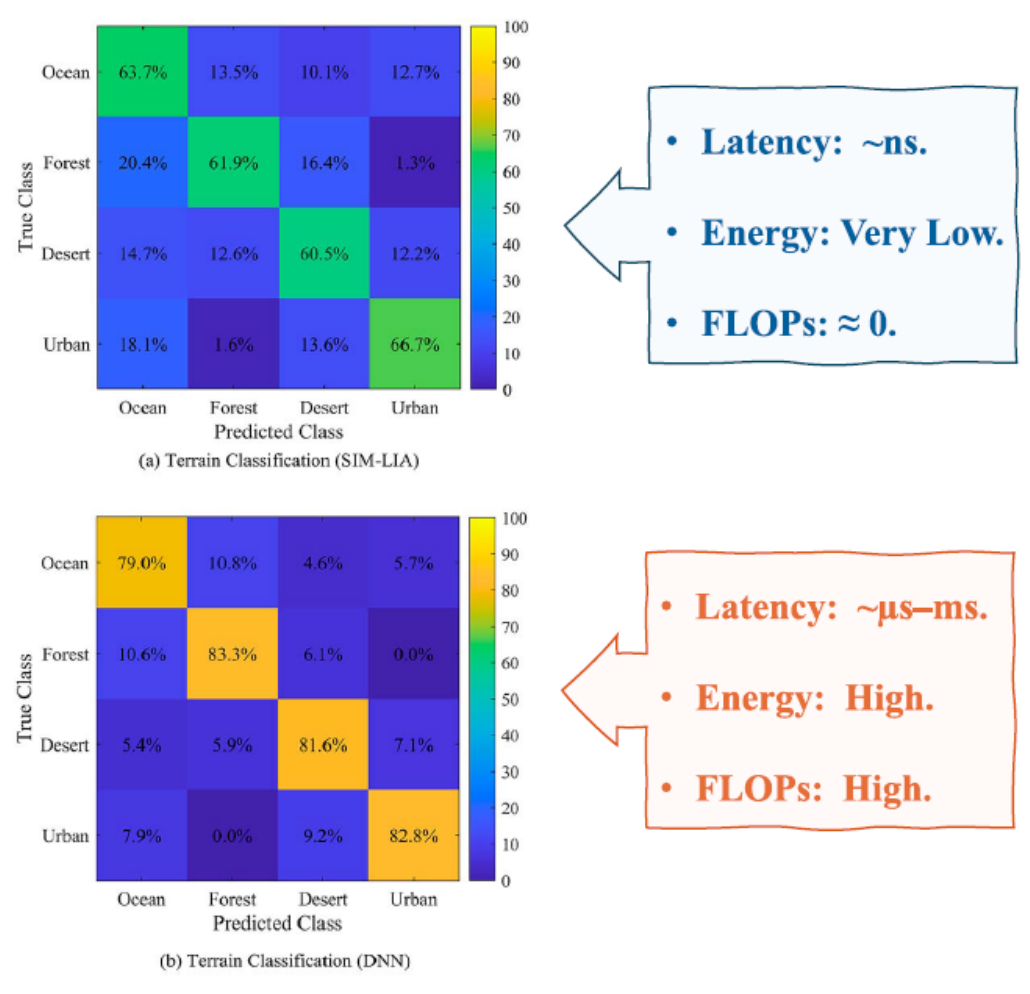}
\caption{Measured-versus-digital trade-off for in-wave inference~\cite{liu2026inwave}: confusion matrices for the SIM-based in-wave classifier (SIM-LIA) and a digital DNN for satellite terrain classification. SIM-LIA achieves 60.5--66.7\% class-wise accuracy with nanosecond-scale latency, very low energy consumption, and near-zero FLOPs, whereas the digital DNN exceeds 79\% accuracy for several classes but requires microsecond-to-millisecond latency, high energy consumption, and substantially higher computational cost.}
\label{fig:measured}
\end{figure}

\textbf{Evidence:} Beyond Fig.~\ref{fig:measured}~\cite{liu2026inwave}, the prototypes of Table~\ref{tab:prototypes} show the same pattern: 88\% experimental agreement with simulation for terahertz classifiers~\cite{lin2018all,li2021spectrally}, 92.6\% measured versus 92.8\% simulated for the nonlinear microwave network~\cite{ning2025multilayer}, 92.5\% versus 95--97\% for the movable-type network~\cite{yu2026movable}, a 2-dB focal-efficiency gap~\cite{gu2024classification}, and a 7-dB measured gain of three layers over one at the best spacing~\cite{wang2024multi}, obtained by exhaustive search rather than by the model-based algorithm proposed alongside. Calibrated optical processors reach 96--97\% at 56--70 frames per second~\cite{zhou2021reconfigurable_dpu} and 72~ns per frame at 4.4~nJ~\cite{chen2023accel}.

\textbf{Design rule:} Expect a few percentage points of accuracy and a few dB of gain between simulation and a calibrated stack, and a collapse for an uncalibrated or misaligned one; the latency and energy advantages are measured, the accuracy and gain advantages must be demonstrated per design.

\begin{table*}[t]
\centering
\caption{Quantitative trade-off results reported in the SIM literature. Level: I = ideal phase-only cascaded model; C = cascaded model with loss/quantization/Lorentzian or power model; P = physically consistent (multiport/circuit) model; M = measured.}
\label{tab:tradeoff_numbers}
\renewcommand{\arraystretch}{1.05}
\setlength{\tabcolsep}{2.5pt}
\scriptsize
\arrayrulecolor{SIMLine}
\begin{tabularx}{\textwidth}{|>{\columncolor{pastelpurple}\raggedright\arraybackslash}p{1.9cm}|>{\raggedright\arraybackslash}X|>{\centering\arraybackslash}p{1.25cm}|>{\raggedright\arraybackslash}p{4.0cm}|}
\hline
\rowcolor{pastelblue}\textbf{Trade-off} & \textbf{Result} & \textbf{Level} & \textbf{Source / setup} \\
\hline
Depth & Fitting NMSE saturates for $L\geq5$; $\approx$150\% capacity gain vs.\ MIMO/RIS; $2\times$ gain of single layer at equal total meta-atoms & I & \cite{an2023stacked}, 28~GHz, $M=100$, $\lambda/2$ pitch, $5\lambda$ thickness \\
& Sum rate peaks at $L=7$, +130\% vs.\ single layer; +200\% vs.\ ZF with equal antennas & C (2-bit) & \cite{an2025stacked}, $M=49$, $K=4$ \\
& 7-layer $<$ 4-layer at 17\% loss/layer; $<$ MIMO at 26\%; MF-SIM/FILM 2-layer: $>$11~dB / $>$7~dB less transmit power at BER $10^{-5}$ & C (loss) & \cite{niu2026twolayer}, 28~GHz, 100 meta-atoms/layer \\
& $-17$\% rate at $L=6$ from insertion loss; +140\% vs.\ no SIM & C (loss) & \cite{rostamikafaki2026loss}, URLLC \\
& Sum rate non-monotonic in $L$ and thickness; geometry-aware spacing $>$ uniform & P (near field) & \cite{dong2026geometry} \\
\hline
Aperture vs.\ depth & +8~bit/s/Hz when $M$: 25$\rightarrow$100 & I & \cite{an2023stacked} \\
& Ideal model overestimates deep stacks by up to 7.5~bit/s/Hz; aperture before depth & C (Lorentzian) & \cite{gonzalez2026lorentzian} \\
\hline
Phase resolution & 2 bits: $\approx$0.3 (max 0.5)~bit/s/Hz loss; 4 bits near continuous; $b\geq6$ indistinguishable & C & \cite{an2025stacked} \\
& 3-bit near continuous with quantum-inspired solver & C & \cite{ji2026hmimo} \\
\hline
CSI & Pilot blocks $N$ instead of $N^2$ via tensor structure & I & \cite{dealmeida2026tensim} \\
& Polar-domain SBL $4\times$ faster at equal accuracy & I & \cite{yao2026sparse}, mmWave near field \\
& Calibration: inter-layer NMSE $-12.07$~dB (max), $-9.25$~dB (avg.); channel NMSE $-8.48$~dB & C (errors) & \cite{jia2026calibration}, 4 layers, 28~GHz \\
\hline
Bandwidth & Single-layer RIS sufficient in narrowband/low SNR; SIM better in wideband multi-user & P (circuit) & \cite{ibrahim2026wideband} \\
& Optimal subcarrier reuse $\approx$62.5\% & I & \cite{li2025fundamental}, 4 users, 16 subcarriers, 40~MHz \\
\hline
Geometry & Received power $-53$ (1 layer) $\rightarrow$ $-46$~dBm (3 layers, $\lambda/4$) vs.\ $-59$~dBm (3 layers, $\lambda$); DoA error 3.57$^\circ\!\rightarrow$1.39$^\circ$ & M & \cite{wang2024multi}, 5.8~GHz, 1-bit \\
& Accuracy 97.8\%$\rightarrow$38.4\% at $\pm2.12\lambda$ lateral shift; 94.4\% with misalignment-aware training & M/C & \cite{mengu2020misalignment}, THz, 5 layers \\
& Focal efficiency 18.7\% sim.\ vs.\ 11.7\% meas. & M & \cite{gu2024classification}, 11~GHz, 3 layers \\
\hline
Energy efficiency & +80\% EE vs.\ digital precoding without SIM; best at 2--5 layers & C (power) & \cite{shi2025energy}, 28~GHz \\
& +230\% EE vs.\ no SIM with fewer BS antennas & C (power) & \cite{ataeebojd2026isac}, ISAC \\
& 12--12.6~mW per ON PIN diode; 4.8~W per FPGA; 15.7--103~W per $60\times60$ panel & M & \cite{wang2024power} \\
& $\approx$429~mW per active unit; $\approx$252~W per network; 17.7~ns per nonlinear layer & M & \cite{ning2025multilayer} \\
\hline
Complexity & $>$99\% runtime reduction with closed-form AO at equal rate & I & \cite{park2026efficient}, cell-free \\
& Structured $Z$/$T$ models: $\mathcal{O}(LM^3)$ vs.\ $\mathcal{O}(LM^2S)$ cascaded & -- & Section~\ref{sec:complexity} \\
& Training: 55k images, $\approx$8~h (5-layer THz); $\approx$80~h (3-layer spectral); $10^4$ samples, 100 epochs (mmWave DoA) & M & \cite{lin2018all,li2021spectrally,gao2024superresolution} \\
\hline
Measured vs.\ simulated & In-wave 60.5--66.7\% vs.\ digital $>$79\% (ns vs.\ $\mu$s--ms latency) & M & \cite{liu2026inwave} \\
& 88\% experiment--simulation agreement (THz); 92.6\% vs.\ 92.8\% (nonlinear microwave); 92.5\% vs.\ 95--97\% (movable-type) & M & \cite{lin2018all,li2021spectrally,ning2025multilayer,yu2026movable} \\
\hline
\end{tabularx}
\arrayrulecolor{black}
\end{table*}

\begin{lessons}
\begin{itemize}[leftmargin=*, nosep]
\item Depth saturates under the ideal model and peaks under loss; with diode-based layers the useful depth is 2--5, and two-layer designs with richer connectivity are the promising alternative.
\item Buy aperture before depth; 2--3 bits inside the optimizer suffice for rate but not for deep nulls.
\item Structured estimators cut pilots from $N^2$ to $N$; optimize training length for net rate; calibrate the device (about 10~dB of inter-layer error is at stake).
\item Depth pays in wideband, interference-limited, multi-user links, not in narrowband or noise-limited ones; evaluate at the band edges.
\item Geometry moves measured performance by 2--13~dB; energy efficiency has an interior optimum set by per-element and controller power; latency and energy advantages are measured, accuracy and gain advantages must be demonstrated per design.
\end{itemize}
\end{lessons}

 \section{Challenges and Future Research Directions}
\label{sec:challenges}

Each challenge below is tied to the evidence of Sections~\ref{sec:fund}--\ref{sec:performance}, followed by the directions that this evidence suggests and by a metric against which progress can be measured.

\subsection{Physically Consistent yet Tractable Modeling}
The cascaded model fails where the interesting regimes begin: a 13-dB measured dependence on inter-layer spacing~\cite{wang2024multi}, up to 7.5~bit/s/Hz of overestimation under state-dependent amplitude~\cite{gonzalez2026lorentzian}, non-monotonic depth behavior in the near field~\cite{dong2026geometry}, and dispersive, active, or nonlinear layers that need fixed-point or recursive formulations~\cite{abrardo2026nonlinear,abrardo2026cascade}. The consistent alternatives cost $\mathcal{O}(LM^3)$ per evaluation (Table~\ref{tab:complexity}). \emph{Directions:} reduced-order differentiable operators keeping the dominant coupling paths of the banded $Z$ model; frequency- and polarization-aware models constrained by passivity and reciprocity; physics--data calibration of the cascaded operator against multiport or measured responses~\cite{li2026fullywave}; and a shared full-wave-validated benchmark~\cite{pettanice2026modeling}. \emph{Metric:} model NMSE versus $d/\lambda$, cell $Q$-factor, and bandwidth at a cost below $\mathcal{O}(LM^2S)$.

\subsection{Scalable Joint Wave-Domain and Digital Optimization}
Per-layer projected-gradient AO needs hundreds of $\mathcal{O}(LM^2K)$ iterations; closed-form updates cut runtime by more than 99\%~\cite{park2026efficient} and unfolding gives a fixed inference cost~\cite{wang2026unfolding}, but discrete, multi-objective, and robust constraints still require heavier solvers~\cite{ji2026qubo,hassan2026robust}. \emph{Directions:} layer-wise decomposition with provable descent, discrete-aware solvers with bounded suboptimality, online and incremental updates, distributed optimization with bounded coordination for multi-SIM and cell-free settings, and joint geometry--phase design~\cite{dong2026geometry,wei2026fsim}. \emph{Metric:} configuration time on an embedded controller against the coherence interval and the hardware latency.

\subsection{CSI Acquisition, Identifiability, and Tracking with Few RF Chains}
Tensor-structured estimators cut pilots from $N^2$ to $N$ blocks~\cite{dealmeida2026tensim}, training length must be chosen for net rate~\cite{bahanshal2026training}, near-field users need range-aware dictionaries~\cite{yao2026sparse}, aging degrades mobile cell-free systems~\cite{ding2025channel,wang2026aging}, and the device itself carries about 10~dB of uncalibrated error~\cite{jia2026calibration}. \emph{Directions:} joint channel--device estimation, pilot and training-state co-design for identifiability under $S\ll M$, predictive tracking with SIM-aware observation models, and sensing-aided CSI tracking. \emph{Metric:} net rate after training versus mobility, and pilot blocks per coherence interval as a fraction of $N$.

\subsection{Wideband and Doubly Dispersive Operation}
Depth pays only in wideband interference-limited links~\cite{ibrahim2026wideband}; subcarrier reuse saturates the interference-cancellation capacity~\cite{li2025fundamental}; wave-domain OFDM needs offline fitting to a multiport operator~\cite{li2026fullywave}; delay-augmented layers remain conceptual (Table~\ref{tab:maturity}); and no prototype reports cell dispersion. \emph{Directions:} dispersion engineering with measured cells, joint optimization across subcarriers with band-edge constraints, delay--Doppler design for mobility~\cite{ranasinghe2025doubly}, and bounds linking bandwidth, depth, spacing, and focusing robustness. \emph{Metric:} band-edge gain relative to the center frequency versus $B/f_c$ and $L$.

\subsection{Hardware Non-Idealities, Loss Accumulation, and Control Overhead}
Programmable cells lose 1.2--1.9~dB each~\cite{rana2021experimental,huang2023wideband,zhang2023onebit}; beyond 17--26\% loss per layer depth hurts~\cite{niu2026twolayer}; bias costs 12~mW per diode and 5~W per controller~\cite{wang2024power}; active units draw 0.4~W each with no thermal data~\cite{ning2025multilayer}; and energy efficiency peaks at 2--5 layers~\cite{shi2025energy}. \emph{Directions:} loss- and quantization-aware design by default, stability- and thermal-aware hybrid passive--active stacks, low-power addressing (row--column biasing, non-volatile switches~\cite{matos2024thz}), two-layer rich-connectivity designs~\cite{niu2026twolayer,kim2026bdsim}, and end-to-end power models including controller and calibration overhead. \emph{Metric:} energy efficiency at equal total DC power against a digital or hybrid baseline with equal RF chains, and temperature rise of active stacks.

\subsection{Calibration, Characterization, and Reproducibility}
A single programmable testbed exists, configured by exhaustive search~\cite{wang2024multi}; simulated and measured efficiencies differ by 2~dB~\cite{gu2024classification}; diffractive networks collapse under misalignment unless trained for it~\cite{mengu2020misalignment}; calibration recovers 9--12~dB~\cite{jia2026calibration}; and no reporting standard exists (Section~\ref{sec:gap}). \emph{Directions:} in-situ self-calibration with embedded probes, over-the-air reciprocity-based calibration, compressed calibration codes exploiting the tensor structure, digital-twin-driven calibration under drift, and reference scenarios, reporting standards, and shared datasets. \emph{Metric:} residual calibrated NMSE per unit of calibration overhead, and independent reproduction of a published measured gain.

\subsection{Network-Level Integration and Orchestration}
Cell-free and non-terrestrial designs rely on two-timescale control~\cite{shi2024harnessing,lin2024stacked,zhang2026satellite}, distributed learning~\cite{zhu2025joint}, and rate splitting against aging~\cite{ding2026hsr}, while Doppler, payload, thermal, and mechanical constraints are rarely modeled and multi-surface cooperation is only beginning~\cite{liu2025sum,wang2026positionaware}. \emph{Directions:} hierarchical control planes separating fast local tuning from slow global management, fronthaul-efficient and group-based control, interoperability with RIS/STAR-RIS and advanced MIMO front ends, NTN designs with Doppler, intermittent-link, payload, and safety constraints, and network-calculus delay guarantees~\cite{chen2025stacked_vtol}. \emph{Metric:} control-plane overhead (bits per update per SIM) and end-to-end latency versus the number of SIMs and users.

\subsection{Trustworthy AI Control and Task-Oriented Operation}
Learned controllers are fast online but their training cost is unreported (Section~\ref{sec:learning}); in-wave inference trails digital accuracy by more than ten points~\cite{liu2026inwave}; policies trained on the ideal model inherit its optimism; and EM and actuator constraints are rarely enforced. \emph{Directions:} physics-constrained architectures (unfolding on the multiport model, passivity- and reciprocity-preserving layers), measurement-calibrated data with sim-to-real fine-tuning, uncertainty quantification and safe fallback configurations, continual and meta-learning under drift, and joint task metrics (accuracy, latency, energy, reliability). \emph{Metric:} measured task performance after sim-to-real transfer relative to simulation, and constraint-violation rate on hardware.

The eight strands are coupled: modeling fidelity bounds optimization, CSI acquisition, and learning; hardware loss bounds depth; and calibration bounds every focusing and nulling gain. The most valuable near-term contributions are therefore a second and third programmable testbed reporting loss, latency, power, and alignment error, a validated reduced-order model, and calibration-aware training, because these unlock the rest.

\begin{lessons}
\begin{itemize}[leftmargin=*, nosep]
\item The eight challenges are coupled: modeling fidelity bounds optimization, CSI, and learning; hardware loss bounds depth; calibration bounds every focusing and nulling gain.
\item The most valuable near-term contributions are not new applications but a second and third programmable testbed with reported loss, latency, power, and alignment error, a validated reduced-order model, and calibration-aware training.
\item Each direction comes with a metric; proposals that report it can be compared, proposals that do not cannot.
\end{itemize}
\end{lessons}

 \section{Conclusions}
\label{sec:conclusion}

This article has surveyed SIMs from a model-to-hardware perspective that complements the function-oriented overviews. Five conclusions emerge. First, the cascaded diffraction operator used by almost all SIM papers is a first-order approximation of the multiport $Z$, $S$, and $T$ models, valid under four explicit assumptions: it suits algorithm design and like-for-like comparisons, but in closely spaced, resonant, near-field, wideband, or active stacks it changes trends rather than numbers, while the physically consistent models cost $\mathcal{O}(LM^3)$ and are best used offline. Second, the literature reduces to five comparable methodological families, and most apparent contradictions stem from differing EM models, hardware assumptions, or baselines. Third, gains are largest for interference-limited wave-domain tasks under RF-chain, payload, or power constraints, and most fragile when they rely on focusing or nulling. Fourth, the hardware record is thin: a single programmable SIM testbed with wireless metrics, two programmable-active networks, and static diffractive stacks, none reporting loss, latency, and thermal behavior together; most advanced architectures remain conceptual. Fifth, the published trade-offs converge: depth peaks at 2--5 layers under realistic loss and power, aperture should precede depth, 2--3 phase bits suffice for rate but not for nulls, structured estimators cut pilots quadratically, depth pays in wideband interference-limited links, and geometry errors shift measured performance by 2--13~dB; the latency and energy advantages of wave-domain processing are measured, whereas its accuracy and gain advantages must still be shown per design. The path to deployable SIM front ends thus runs less through new applications than through validated reduced-order models, calibration-aware training, hardware-aware optimization, and prototypes that report what today's omit.


\end{document}